\documentclass[a4paper,11pt]{article}
\usepackage{jheppub}
\pdfoutput=1
\usepackage[T1]{fontenc}

\usepackage{amsmath,amssymb,amsfonts, bm, natbib}
\usepackage{diagbox}
\usepackage{epsfig}
\usepackage{mathtools}
\usepackage{graphicx}
\usepackage{slashed}
\usepackage{physics}
\usepackage{multirow}
\usepackage{subcaption}
\usepackage{adjustbox}
\usepackage{slashed} 
\usepackage{mathrsfs}
\usepackage{soul}
\usepackage{soul}
\usepackage{braket}
\usepackage{hyperref}
\usepackage{calc}
\usepackage{xcolor}
\usepackage{tablefootnote}
\usepackage{float}

\usepackage[font=small]{caption}

\newcommand{\bea}{\begin{eqnarray}}
\newcommand{\eea}{\end{eqnarray}}
\newcommand{\beq}{\begin{equation}}
\newcommand{\eeq}{\end{equation}}
\newcommand{\ec}{\end{center}}
\newcommand{\bc}{\begin{center}}

\newcommand{\gev}{{\rm GeV}}

\newcommand{\pdir}{p\kern -5.2pt\raise 0.2ex\hbox {/}}

\newcommand{\vdir}{v\kern -5.75pt\raise 0.15ex\hbox {/}}
\newcommand{\kdir}{k\kern -5.75pt\raise 0.15ex\hbox {/}}
\newcommand{\epsdir}{\epsilon\kern -5.0pt\raise 0.15ex\hbox {/}}
\newcommand{\bvdir}{\bar{v}\kern -5.75pt\raise 0.15ex\hbox {/}}
\newcommand{\Ddir}{D\kern -7.75pt\raise 0.20ex\hbox {/}}
\newcommand{\Adir}{A\kern -7.75pt\raise 0.20ex\hbox {/}}
\newcommand{\ldir}{l\kern -5.0pt\raise 0.2ex\hbox{/}}
\newcommand{\varepsdir}{\varepsilon\kern -5.5pt\raise 0.15ex\hbox{/}}

\newcommand{\ii}{\mathrm{i}}

\makeatother

\definecolor{niceblue}{rgb}{0.15,0.15,0.6}
\definecolor{nicegreen}{rgb}{0.1,0.5,0.1}
\definecolor{Red}{rgb}{1.,0.,0.}

\definecolor{Green}{rgb}{0.2,.7,0.2}

\makeatletter
\newcommand*{\rom}[1]{\expandafter\@slowromancap\romannumeral #1@}
\makeatother

\definecolor{gray}{rgb}{0.4,0.4,0.4}

\definecolor{darkblue}{rgb}{0.0,0.0,0.3}
\hypersetup{colorlinks,breaklinks,
            linkcolor=darkblue,urlcolor=darkblue,
            anchorcolor=darkblue,citecolor=darkblue}

\title{\boldmath $|V_{ub}|$ determination and testing of lepton flavour universality in semileptonic $B_c \rightarrow D^{(\ast)}$ decays}
\author{Domagoj Leljak,} 
\author{Bla\v zenka Meli\'c}

\affiliation{Rudjer Bo\v skovi\'c Institute, Division of Theoretical Physics, Bijeni\v cka 54, HR-10000 Zagreb, Croatia}

\emailAdd{dleljak@irb.hr}
\emailAdd{melic@irb.hr}

\abstract{
In light of prospects for measurements of $B_c \rightarrow D^{(\ast)} l \nu$ decays in the upcoming Upgrade II of the LHC,  we show that by using calculated $B_c \rightarrow D^{(\ast)}$ form factors a competitive extraction of the $|V_{ub}|$ CKM matrix element from the $B_c \to D \mu \bar{\nu}_{\mu}$ decay might be possible. 
To minimize experimental and theoretical uncertainties we provide the ratio $|V_{ub}|/|V_{cb}|$ by normalizing the $B_c \rightarrow D^{(\ast)} \mu \bar{\nu}_{\mu}$ to $B_c \to J/\psi \mu \bar{\nu}_{\mu}$ decay. 
We also briefly examine the suggestion to extract $|V_{ub}|/|V_{cs}|$ from the theoretically interesting ratio of $B_c \rightarrow D^0 e \bar{\nu}_{e}$  and $B_c \rightarrow B_s e \bar{\nu}_{e}$  decay rates in the zero-recoil limit.
With the present average value of $|V_{ub}|$, the predicted branching ratios are estimated to be $BR(B_c \to D^0 \mu \bar{\nu}_{\mu}) = (2.4\pm 0.4)\cdot 10^{-5} $ and $BR(B_c \to D^{\ast} \mu \bar{\nu}_{\mu}) = (7\pm3)\cdot 10^{-5} $, and the semileptonic ratios for testing  the lepton flavour universality in these $B_c$ decays are $R_c(D^0) =0.64 \pm 0.05$ and $R_c(D^{\ast}) = 0.55 \pm 0.05$. We also provide $q^2$ distributions and various angular observables of $B_c \rightarrow D^{(\ast)} l \nu$ decays.
}

\begin{document}

\maketitle

\setcounter{page}{1}
\setcounter{footnote}{0}
\setcounter{equation}{0}
\noindent
\section{Introduction} 

Precise determination of the CKM matrix elements, in particular of $|V_{ub}|$ and $|V_{cb}|$ are crucial for studies of flavor physics and CP violation in the quark sector. There is a tremendous experimental progress in the extraction of these CKM matrix elements through both inclusive and exclusive $b$-quark decay measurements and the values have already achieved impressive precision \cite{PDG}: 
\begin{eqnarray}
\arraycolsep=1.4pt\def\arraystretch{2.2}
\label{eq:CKM1}
&&|V_{ub}|^{\rm incl} = (4.49 \pm  0.16 \substack{+0.16 \\-0.17} \pm 0.17)\cdot 10^{-3}, 
\quad\quad   |V_{cb}|^{\rm incl} = (42.2 \pm 0.8)\cdot 10^{-3},
\nonumber \\
&&|V_{ub}|^{\rm exc} = (3.67 \pm 0.09 \pm  0.12)\cdot 10^{-3} 
,\qquad\qquad  |V_{cb}|^{\rm exc} =(41.9 \pm 2.0)\cdot 10^{-3}, 
\nonumber \\
\nonumber \\
&&\qquad\qquad\qquad\qquad\qquad\qquad|V_{ub}|/|V_{cb}|^{\rm inc} =0.107 \pm 0.007\,,
\nonumber \\ 
&&\qquad\qquad\qquad\qquad\qquad\qquad|V_{ub}|/|V_{cb}|^{\rm excl} = 0.088 \pm 0.006 \,,
\end{eqnarray}
with the present averaged values
\begin{eqnarray}
\label{eq:CKM2}
&&|V_{ub}|^{\rm aver} = (3.94 \pm 0.36)\cdot 10^{-3}, \quad\quad  |V_{cb}|^{\rm aver} =(42.2 \pm 0.8)\cdot 10^{-3} , 
\\
\nonumber \\
\label{eq:CKM3}
&&\qquad\qquad\qquad\qquad|V_{ub}|/|V_{cb}|^{\rm aver} = 0.092 \pm 0.008\,.
\label{eq:excinc}
\end{eqnarray}
The problem with existing determinations of $|V_{ub}|$ and  $|V_{cb}|$   
is a persistent discrepancy between the values obtained from inclusive and exclusive $b$-hadron decays. While the inclusive $|V_{cb}|$ determination already achieved the precision of $1-2\%$ and it was shown recently \cite{BigiGambino,Bernlochner:2019ldg} that $|V_{cb}|^{\rm excl}$  may be shifted towards the inclusive value by using more sophisticated the BGL parametrization of the form factors \cite{Boyd1,Boyd2} (although care must be taken in the interpretation of the results), we can see that the values for $|V_{ub}|$
obtained from inclusive and exclusive decays still differ by approximately 3.5 standard deviations. 

The main source for the extraction of $|V_{ub}|$ from the exclusive decays is the semileptonic $B \to \pi l \nu$ decay, which is precisely measured and also relatively precisely determined theoretically. Theoretical uncertainties are mainly connected with the hadronic non-perturbative uncertainties hidden in the transition $B \to \pi$ form factors form factors. Fortunately, there are nowadays precise theoretical calculation of $B \to \pi$ form factors in the framework of the light-cone sum rules, at $q^2 \le 12-15 \, \gev$ \cite{BallZwicky,DuplancicKMM,Imsong:2014oqa} and on the lattice at 
$q^2 \ge 15 \, \gev$ \cite{latticeBpi1, latticeBpi2}, which combined, by using constraints from unitarity and analyticity, enable the form factor determination in a full $q^2$ range and very precise determination of $|V_{ub}|$ \cite{Gonzalez-Solis:2018ooo}:
\begin{eqnarray}
|V_{ub}|^{B \to \pi} = (3.53\pm 0.08_{\rm stat}\pm 0.06_{\rm syst} )\cdot 10^{−3}\,. 
\end{eqnarray}
With the accumulation of a big sample of $\Lambda_b^0$ data at LHCb, it became possible to study also the semileptonic $\Lambda_b^0$ decays for extraction of the $|V_{ub}|/|V_{cb}|$ ratio  \cite{Aaij:2015bfa} using the QCD lattice results for the form factors \cite{Detmold:2015aaa}:
\begin{eqnarray}
|V_{ub}|/|V_{cb}|^{\Lambda_b \to \Lambda_c} = 0.084 \pm 0.004_{\rm exp} \pm 0.004_{\rm lattice}  \,,
\end{eqnarray}
which is again somewhat lower than the inclusive determination of this ratio in (\ref{eq:CKM1}).

The extraction of $|V_{ub}|$ (and $|V_{cb}|$) from measured inclusive or exclusive semileptonic $B$ meson decays rely on different experimental techniques to isolate the signal and on different theoretical descriptions of QCD contributions to the underlying weak decay processes. Therefore, in the future there will be need for more information from various $b \to u l \nu$ decays to extract $V_{ub}$ \cite{expHE}. The most promising exclusive decays are $B \to (\eta,\eta^\prime,\omega,\rho) l \nu$ and in the near future also $B_c \to D^{(\ast)} l \nu$ decay which we discuss here.
LHCb plans to go for rare $B_c \to D^0 l \nu$ decays in the Upgrade II \cite{LHCb_book}. As stated for the LHCb Upgrade II, approximately 30,000 reconstructed $B_c \to D^0 l \nu$ decays can be expected with the 300 $fb^{-1}$ Upgrade II dataset, which could lead to a competitive measurement of $|V_{ub}|$ from these decays too. 

Also there is an extensive research in testing of the lepton flavor universality in various semileptonic decays, probing the ratios
\begin{eqnarray}
R(H_{b,c}) = \frac{BR(H^\prime_{b,c} \to H_{b,c}\tau \nu)}{BR(H^\prime_{b,c} \to H_{b,c} (\mu,e) \nu)},
\end{eqnarray}
which for $B \to K^{\ast}$, $B \to D^{\ast}$ consistently show $2 - 3\, \sigma$ lower values than predicted in the Standard Model (SM). Recently, also the potential sign of the lepton-flavour non-universality was observed for semileptonic $B_c \to J/\psi$  decay \cite{RjpsiEX, CohenLammLebed,RjpsiOUR}. With upgraded detectors in the next run of LHC, it would be possible to sample enough data in LHCb experiment for the analysis of other semileptonic $B_c$ decays.
 

Therefore, in this paper we address the calculation of $B_c \to D^{(\ast)}$ form factors and analyse the semileptonic $B_c\to D^{(\ast)}$ decays. For the calculation of the form factors we employ the three-point sum rule (3ptSR) method \cite{Va78}. Although the method itself has some general limitations, for such a type of heavy-to-heavy decays only the method of QCD sum rules seems to be applicable. It is known that the 3ptSR method has some problems with the description of heavy-to-light transitions, in particular at the end-point region of momenta when almost whole of the final state meson momentum is carried by one of the constituents, which can then cause a strange behaviour of some of the form factors in the heavy-quark limit~\cite{Ball:1997rj}. But, this behaviour mainly concerns heavy-to-light transitions, while in $B_c \to D^{(*)}$ decays, considered here, the final state $D^{(*)}$ meson is somewhere in-between to be described as a light or a heavy particle. This is the main reason why neither the use of heavy-quark symmetries, nor the description of the mesons in terms of light-cone distribution amplitudes are apriori trustworthy approaches of analyzing $B_c \to D^{(\ast)}$ semileptonic decays.
Namely, as we will discuss briefly in Sec.3.3, one can parametrize $B_c \to D^{(*)}$ matrix elements in the heavy quark limit in terms of two form factors which can be further expressed with the help of the heavy-quark symmetry as integrals over the $B_c$ meson wave function, Eqs.(\ref{eq:fits}-\ref{eq:eqf0}). However, there are no fully reliable and controllable models for calculating the $B_c$ wave function without further approximations being involved, such as various non-relativistic or heavy-quark approximations at zero-recoil, or the use of constituent quark models  related to a quark potential and/or relativistic quark kinematics, as for example the ones used in~\cite{Colangelo:1999zn}. Also, the impact of the deviation from the infinite heavy quark mass limit is then difficult to judge upon and incorporate into systematical uncertainties. 
On the other hand, the light-cone sum rule method relies on the known description of the final (light) meson or the decaying heavy-meson distribution amplitudes (DAs) of increasing twist, which is hardly applicable for our $B_c \to D^{(*)}$ transitions, since the $B_c$ meson DA's aren't known, nor are the $D^{(*)}$ mesons light enough that their DAs could be systematically expanded near the light-cone. There exists the Brodsky-Huang-Lepage (BHL) prescription~\cite{Brodsky:1981jv} on how to (for a relativistic two particle state) approximately connect the wave function with the light-cone functions, used in~\cite{Huang:2007kb}. But, this approach involves models with constituent quark masses and arbitrary phenomenological parameters which are hardly numerically controllable. On the other hand, OPE expansion in the 3ptSR is under control, the non-perturbative vacuum condensates are universal and known also from sum rule calculations. 
There are no available lattice QCD form factor predictions for $B_c \to D^{(*)}$ decays.  Alternative methods used in the estimation of $B_c \to D^{\ast}$ form factors, like various relativistic quark models provide form factors with a precision that cannot be systematically controlled and therefore calculated values for the form factors differ in a wide range, see Table \ref{tab:ffactor}. 

The 3ptSR method was developed a long time ago, and was since applied successfully to the calculation of the pion electromagnetic form factor at intermediate momentum transfer. Soon afterwards it was applied for the first time in the description of the semileptonic decay $D \to K e \nu$, where it was used in the calculation of weak $B$-meson decays. New insights in the application of the 3ptSR in weak decays were given in \cite{BallBraunDosch}. In this paper the authors discussed in detail the validity of the approach and possible issues of the application to the 3ptSR model. We will follow the discussion and extend the estimation of the form factors by including the non-local condensate contributions to the non-perturbative part, apart from the leading small local gluon-condensate contribution discussed in the literature. 

The structure of the paper is as follows. In the Sec.2 we discuss the 3ptSR calculation of the form factors, their $q^2$ dependence and extrapolation to high $q^2$ transition momenta by using different $z$-parametrizations and present our results. In the Sec.3 we give our predictions for decay rates and several asymmetries in $B_c \to D^{(\ast)} l \nu$ decays. In the same section we also analyze the determination of the $|V_{ub}|$ and the $|V_{ub}|/|V_{cb}|$ ratio and discuss the usefulness of the $\Gamma(B_c \to D^0 e \bar{\nu}_e)/\Gamma(B_c \to B_s e \bar{\nu}_e)$ ratio at the zero recoil to extract $|V_{ub}|/|V_{cs}|$ from the experiment. We conclude in Sec.4. The analytical results for various 3ptSR perturbative and non-perturbative form factor contributions  are given in Appendix A, while numerical parameters used  and the details of form factor high $q^2$ parametrizations are summarized in Appendix B. In Appendix C we give expressions relevant for the calculation of various $q^2$ and angular distributions, as well as binned $q^2$ distributions to be compared with the future measurements. 


\section{Calculation of $B_c \to D^{(\ast)}$ form factors}
We choose to parametrize $B_c \to D^{(\ast)}$ matrix elements in terms of form factors according to BSW parametrization \cite{Wirbel:1985ji}. For the $B_c\rightarrow D^0$ we use
    	\begin{equation}
			\bra{D^0(p_2)}V_{\mu}\ket{B_c(p_1)} = f_+(q^2)\big[P_{\mu} - \frac{m_{B_c}^2-m_{D^0}^2}{q^2}q_{\mu}\big] + f_0(q^2)\frac{m_{B_c}^2-m_{D^0}^2}{q^2}q_{\mu}\,,
		\end{equation}
whereas for $B_c\rightarrow D^{*}$ the parametrization is
\begin{equation}
        \label{eq:ffdef}
		\begin{split}
			-i\bra{D^{*}(p_2)}(V-A)_{\mu}\ket{B_c(p_1)}  = &\,\, i\frac{2 \, V(q^2)}{m_{B_c}+m_{D^*}}\varepsilon_{\mu\nu\alpha\beta}\epsilon^{*\nu}p_2^{\alpha}p_1^{\beta}
			- (m_{B_c}+m_{D^*})A_1(q^2)\epsilon^*_{\mu}\\
	  & \hspace*{-3cm} +\frac{A_2(q^2)}{m_{B_c}+m_{D^*}}(\epsilon^*\cdot q)(p_1+p_2)_{\mu}
			+ \frac{2 m_{D^*}}{q^2}(A_3(q^2)-A_0(q^2))(\epsilon^*\cdot q)q_{\mu}\,,		\end{split}
\end{equation}
where $q=p_1-p_2$ and $P=p_1+p_2$ and it should be noted that there is the identity to be satisfied, namely, $A_3(0)=A_0(0)$. Also, then
\begin{equation}
    A_3(q^2)=\frac{m_{B_c}+m_{D^*}}{2m_{D^*}}A_1(q^2)-\frac{m_{B_c}-m_{D^*}}{2m_{D^*}}A_2(q^2) \,.
\end{equation}
In the case of pseudoscalar final state one has $f_+(0)=f_0(0)$. Decay constants are defined as 	
    \begin{equation}
		\begin{split}
			\bra{0}\bar{c}i\gamma_5 b \ket{B_c} & = f_{B_c}\frac{m_{B_c}^2}{m_c+m_b} \,,\\
			\bra{0}\bar{c}i\gamma_5 u \ket{D^0} & = f_{D^0}\frac{m_{D^0}^2}{m_u+m_c} \,,\\
			\bra{0}\bar{c}\gamma_{\nu}c\ket{D^*} & =  f_{D^{*}}m_{D^{*}}\epsilon_{\nu}\,.
		\end{split}
	\end{equation}
The standard procedure for the evaluation of form factors in the framework of 3ptSR starts by considering the three-point functions 
\begin{eqnarray}
\Pi_P^{\mu}(p_1,p_2) &=& i^2 \int \dd^4 x \,\dd^4 y \,\mathrm{e}^{-i(p_1 x - p_2 y)}\langle 0|\mathcal{T}\big\{j_P(y)V^{\mu}j^{\dag}_{B_c}(x)\big\}|0  \rangle \,, \nonumber \\
\Pi_V^{\mu\nu}(p_1,p_2) &=& i^2 \int \dd^4 x \,\dd^4 y\,\mathrm{e}^{-i(p_1 x - p_2 y)}\langle 0| \mathcal{T}\big\{j_{V}^{\nu}(y)(V-A)^{\mu}j^{\dag}_{B_c}(x)\big\}|0  \rangle \,,
\end{eqnarray}
where $B_c$ and $P=D^0,V=D^{\ast}$ states are interpolated by the currents
	\begin{equation}
		\begin{split}
			j_{B_c}(x) & = \bar{c}(x)i\gamma_5 b(x),\\
			j_P(x) & = \bar{q}_1(x)i\gamma_5 q_2(x),\\
			j_V^{\nu}(x) & = \bar{q}_1(x)\gamma^{\nu}q_2(x).\\
		\end{split}
	\end{equation}
We use here $P$ and $V$ to generally denote cases with pseudoscalar and vector mesons in  the final state, respectively. This is useful in order to describe in the same way also $B_c\to J/\psi$ and $B_c \to B_s$ processes later in the paper. By writing the correlation function for the $B_c \to D$ transition as 
\begin{equation}
\Pi_P^{\mu}(p_1,p_2) = \Pi_{P,1} p_1^{\mu} + \Pi_{P,2}p_2^{\mu}\,,
\label{eq:corr1}
\end{equation}
and the one for $B_c \to D^{\ast}$ like 
\begin{equation}
    \begin{split}
	    -\ii\Pi_{V}^{\mu\nu}(p_1,p_2) & = g^{\mu\nu}\,\Pi_{V,0} + p_{2}^{\mu} p_{1}^{\nu\vphantom{\mu}}\,\Pi_{V,1} + p_{1}^{\mu}p_{1}^{\nu\vphantom{\mu}} \,\Pi_{V,2}+ p_{2}^{\mu}p_{2}^{\nu\vphantom{\mu}}\,\Pi_{V,3} + p_{1}^{\mu}p_{2}^{\nu\vphantom{\mu}}\,\Pi_{V,4} - \mathrm{i}\epsilon^{\mu\nu\alpha\beta}p_{2\alpha} p_{1\beta}\,\Pi_{V,v}\,,
    \end{split}
    \label{eq:corr2}
\end{equation}
where $\Pi_{P,V}$ are functions of $q^2$, $p_1^2$ and $p_2^2$, and after the Borel transforming them one gets the final expressions for the pseudoscalar form factors
\begin{equation}
    \begin{split}
        & f_+(q^2) = \frac{(m_b+m_c)m_c}{2m_{B_c}^2m_{D^*}^2f_{B_c}f_{D^0}}\mathrm{e}^{\frac{m_{B_c}^2}{M_1^2}+\frac{m_{D^*}^2}{M_2^2}}M_1^2M_2^2 \bigg[\mathcal{B}_{-p_1^2}(M_1^2)\mathcal{B}_{-p_2^2}(M_2^2)\bigg(\Pi_{P,1}(q^2)+\Pi_{P,2}(q^2) \bigg)\bigg], \\
        & f_0(q^2) = \frac{(m_b+m_c)m_c}{2m_{B_c}^2m_{D^*}^2f_{B_c}f_{D^0}}\mathrm{e}^{\frac{m_{B_c}^2}{M_1^2}+\frac{m_{D^*}^2}{M_2^2}}M_1^2M_2^2\cross\\
        & \qquad\qquad\qquad\quad\,\,\cross\bigg[\mathcal{B}_{-p_1^2}(M_1^2)\mathcal{B}_{-p_2^2}(M_2^2)\bigg(\frac{\Pi_{P,1}(q^2)-\Pi_{P,2}(q^2)}{m_{B_c}^2-m_{D^*}^2}q^2+\Pi_{P,1}(q^2)+\Pi_{P,2}(q^2)  \bigg)\bigg], \\
    \end{split}
\end{equation}
and with complete analogy for the vector case
\begin{equation}
	\begin{split}
		V(q^2) & = -\frac{(m_{B_c}+m_{D^*})(m_b+m_c)}{2f_{B_c}f_{D^*}m_{B_c}^2m_{D^*}}\exp{\frac{m_{B_c}^2}{M_1^2}+\frac{m_{D^*}^2}{M_2^2}} M_1^2 M_2^2\bigg[\mathcal{B}_{-p_1^2}(M_1^2)\mathcal{B}_{-p_2^2}(M_2^2)\Pi_{V,v}\bigg],\\
		A_1(q^2) & = -\frac{m_b+m_c}{(m_{B_c}+m_{D^*})f_{B_c}f_{D^*}m_{B_c}^2m_{D^*}}\exp{\frac{m_{B_c}^2}{M_1^2}+\frac{m_{D^*}^2}{M_2^2}} M_1^2 M_2^2\cross\\
		&\qquad\qquad\qquad\qquad\qquad\qquad\qquad\qquad\qquad\qquad\qquad\qquad \cross \bigg[\mathcal{B}_{-p_1^2}(M_1^2)\mathcal{B}_{-p_2^2}(M_2^2)\Pi_{V,0}\bigg], \\
		A_2(q^2) & = +\frac{(m_{B_c}+m_{D^*})(m_b+m_c)}{2f_{B_c}f_{D^*}m_{B_c}^2m_{D^*}}\exp{\frac{m_{B_c}^2}{M_1^2}+\frac{m_{D^*}^2}{M_2^2}} M_1^2 M_2^2 \cross\\
		& \qquad\qquad\qquad\qquad\qquad\qquad\qquad\qquad\qquad\quad\,\, \cross \bigg[\mathcal{B}_{-p_1^2}(M_1^2)\mathcal{B}_{-p_2^2}(M_2^2)\big(\Pi_{V,1}+\Pi_{V,2}\big)\bigg],\\
		A_0(q^2) & = -\frac{m_b+m_c}{2f_{B_c}f_{D^*}m_{B_c}^2m_{D^*}^2}\exp{\frac{m_{B_c}^2}{M_1^2}+\frac{m_{D^*}^2}{M_2^2}} M_1^2 M_2^2\,\cross\\ 
		& \qquad \cross \bigg[\mathcal{B}_{-p_1^2}(M_1^2)\mathcal{B}_{-p_2^2}(M_2^2)\bigg(\Pi_{V,0} + (m_{B_c}^2-m_{D^*}^2) \frac{\Pi_{V,1}+\Pi_{V,2}}{2}-q^2\frac{\Pi_{V,1}-\Pi_{V,2}}{2}\bigg)\bigg],\\
	\end{split}
\end{equation}
where $M_{1}^2, M_{2}^2$ are the Borel parameters in the $B_c$ and $D^{(\ast)}$ channels, respectively.

\begin{figure}[H]
    \centering
    \includegraphics[scale=0.4]{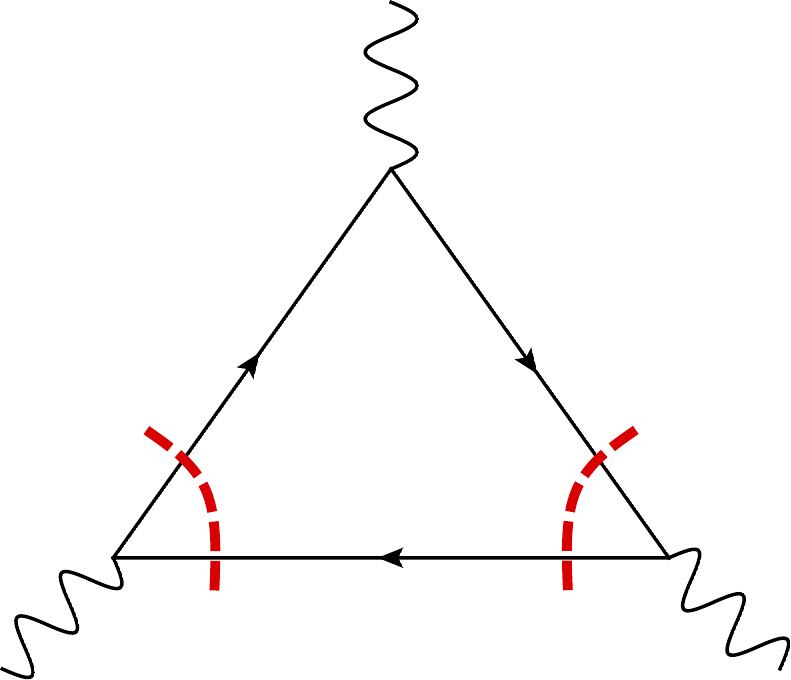}
    \includegraphics[scale=0.4]{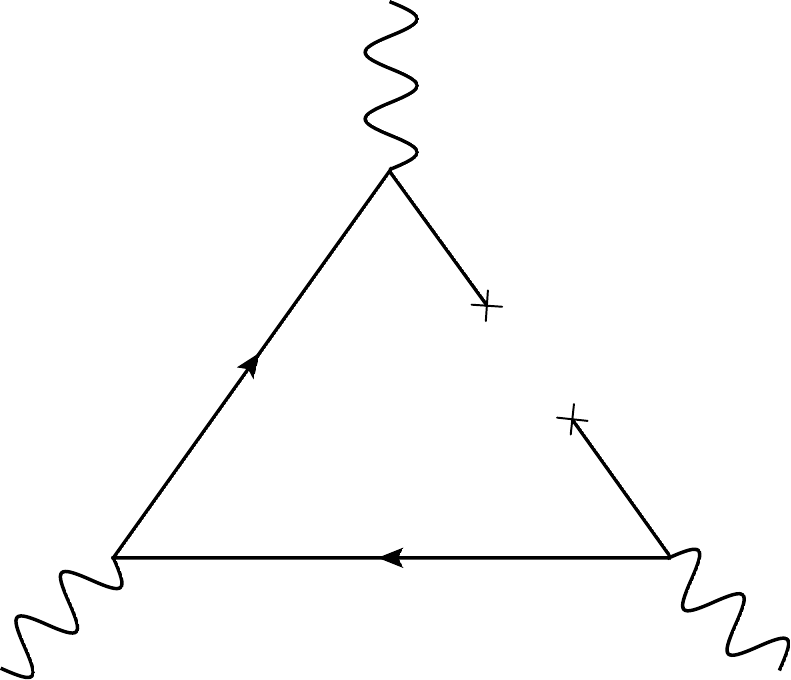}
    \includegraphics[scale=0.4]{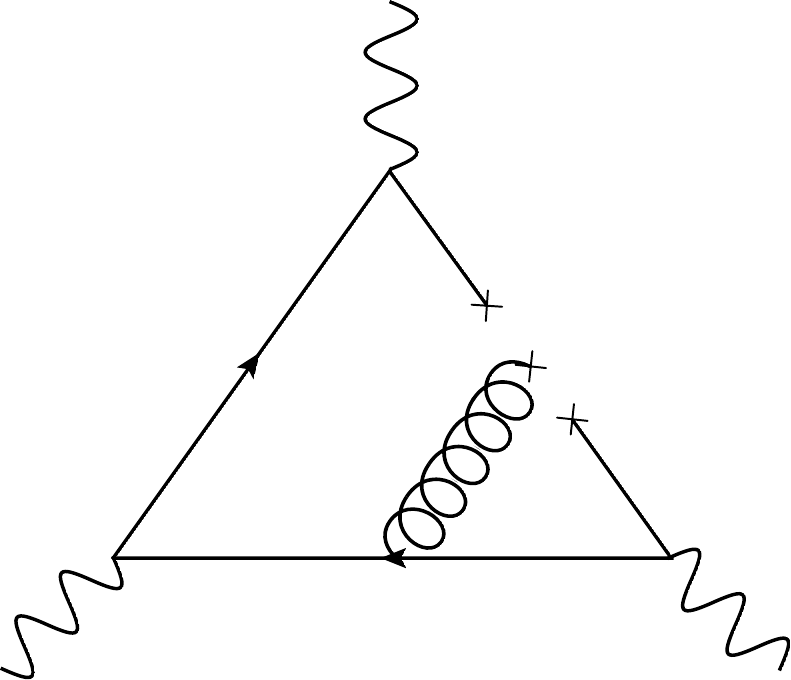}
    \includegraphics[scale=0.4]{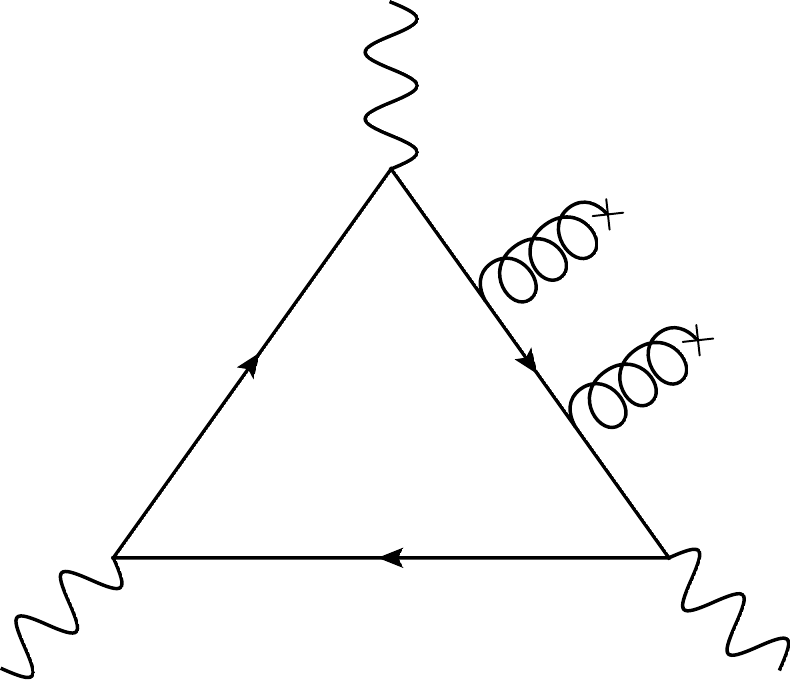}
    \caption{ Contributions to correlation function from left to right: \textbf{(a)} perturbative diagram,\textbf{(b)} nonlocal quark condensate diagram, \textbf{(c)} an example of a quark-gluon condensate diagram, \textbf{(d)} an example of a gluon condensate diagram.}
    \label{fig:diags}
\end{figure}

In the OPE expansion in QCD we will take the perturbative part and non-perturbative terms up to dimension 5:
\begin{eqnarray}
\label{qe:CFsum}
\Pi_i = \Pi_i^{\rm pert} + \Pi_i^{\rm (3)} +\Pi_i^{\rm (4)} + \Pi_i^{\rm (5)}+\dots \,,
\end{eqnarray}
with contributions shown in Fig.~\ref{fig:diags}. The dim=3 terms are proportional to to the quark condensates $\langle \bar{q} q \rangle \, (q = u, d, s)$, dim=4 terms are proportional to the gluon condensate $\langle G G \rangle$, while the dim=5 terms come from the contribution of the mixed quark-gluon condensate $\langle \bar{q} \sigma G q \rangle$. 
The perturbative part is calculated by standard methods imposing the Cutkosky rules to calculate simultaneously discontinuities in $p_1^2$ and  $p_2^2$ of amplitudes (Figure \ref{fig:diags}a),
\\$(-4){\rm Im}_{s_1,s_2}\Pi^{\rm pert}_{i}(s_1,s_2,q^2) =  \rho_{i}(s_1,s_2,q^2)$, and then by using the double dispersion relation
\begin{equation}
\label{eq:DR}
\Pi_{i}^{\rm pert}(p_1^2,p_2^2,q^2) = -\frac{1}{(2\pi)^2}\int\!\!\!\int\frac{\rho_{i}(s_1,s_2,q^2)}{(s_1-p^2_1)(s_2-p_2^2)}ds_1 ds_2\,.
\end{equation}
The explicit results for $\rho_{i}(s_1,s_2,q^2)$ for various form factors are given in Appendix A.1. 

It is easy to see that the quark- and mixed quark-gluon condensate contributions (Figure \ref{fig:diags}b,c) vanish after the Borel transformations in both variables $p_1^2 =p_{B_c}^2$ and $p_2^2 = p_{D^{(\ast)}}^2$. However, this is only true if one considers local condensates. To improve the picture we examine the influence of non-local quark condensates $\langle \bar{q}(x) [x,0] q(0) \rangle$ \cite{MikhailovRady}  in $B_c \to D^{(\ast)} l \nu$ decays. 
The non-local quark condensates are usually introduced as   
\begin{eqnarray}
 \langle \bar{q}(x) q(0) \rangle = \langle \bar{q} q  \rangle \int\displaylimits_0^{\infty}\mathrm{d}\nu\, \mathrm{e}^{\nu\frac{x^2}{4}} f(\nu)
\end{eqnarray}
with the model-dependent function \cite{BraunNLC}
\begin{eqnarray}
f(\nu) = \frac{\lambda^{a-2}}{\Gamma(a-2)} \nu^{1-a} e^{-\lambda/\nu}\,, \quad a-3 = \frac{4 \lambda}{m_0^2}
\label{eq:fnu1}
\end{eqnarray}
or in the simpler version \cite{MikhailovRady}, which we use in this paper
\begin{eqnarray}
f(\nu) = \delta \left ( \nu - \frac{m_0^2}{4} \right ).
\end{eqnarray}
The first two moments of the model function $f(\nu)$ are fixed by the OPE as 
\begin{equation}
	\begin{split}
		& \int\displaylimits_0^{\infty}\mathrm{d}\nu\, f(\nu) = 1 , \qquad  \int\displaylimits_0^{\infty}\mathrm{d}\nu\, \nu\, f(\nu) = \frac{m_0^2}{4} \,. \\
	\end{split}
\end{equation}
Here $m_0^2$ is the standard OPE parameter \cite{Ba94,Br04} connected with the average quark virtuality, and is defined as a ratio of quark and quark-gluon condensates
\begin{equation}
g \langle\bar{q}(x)(\sigma\!\cdot\! G)q(0) \rangle \approx m_0^2 
\langle \bar{q}(x)q(0)\rangle \,.
\end{equation}
We have checked explicitly that the use of the more sophisticated function $f(\nu)$ given in (\ref{eq:fnu1}) does not change anything in the conclusion. 
Namely, numerically the non-local quark condensates $\Pi_i^{(3)}= \Pi_i^{\langle \bar{q}q \rangle}$ and the mixed quark-condensate $\Pi_i^{(5)}= \Pi_i^{\langle \bar{q} \sigma.G q \rangle}$ are contributing just up to $1\%$ to the result. The analytic expressions are given in Appendix A.2 and A.3, respectively. 

The gluon condensate contributions $\Pi_i^{(4)}= \Pi_i^{\langle G G\rangle}$ to $B_c \to D^{(\ast)}$ form factors are calculated from diagrams in Figure \ref{fig:diags}d. The procedure is well known and the expressions are very lengthy and cumbersome and similar to those already published in the literature for $B_c \to J/\psi, \eta_c$ transitions \cite{Kiselev1} and will not be given explicitly here. Some subtleties of the calculation are given in Appendix A.3. Numerically the gluon condensate contributions do not exceed $\mathcal{O}(1\%)$ in determined window of Borel parameters, but somewhat stabilize the sum rules at smaller values of these parameters.

We deduce that all non-perturbative parts are numerically negligible, and can be safely neglected. The main contribution to the correlation function (\ref{qe:CFsum}) comes from the perturbative parts, Eqs.(\ref{eq:appA1}-\ref{eq:appA2}).

To calculate the form factors in QCD sum rules, in which the correlation function is written as a sum of perturbative and non-perturbative contributions as in Eq.~(\ref{qe:CFsum}), the perturbative part is calculated by the usual expansion in the coupling constant, while the non-perturbative part is described by the manner of Wilson's operator product expansion as a sum of expectation values of operators of increasing dimension. Since it is known that when using the Borel-transformed sum rules in calculating heavy meson decay constants higher orders of perturbation series can contribute as much as 30-40\%, depending on the scheme (heavy-light decay constants are known to NNLO \cite{Gelhausen:2013wia}), whereas the QCD 3-point function is only known to LO, here we parametrize the 3-point function with the same  threshold parameters $s_0^{\mathrm{eff}}$ that, at LO in QCDSR reproduce the meson decay constants obtained from the lattice QCD calculations, listed in Table~\ref{tab:dconst}, whereas the 3ptSR Borel mass parameters $M^2$ are taken in the region where the stability is achieved in the sense of appearance of the so called Borel plateau.

\begin{center}
\addtolength{\tabcolsep}{-3pt}
\renewcommand{\arraystretch}{1.5}
 \begin{tabular}{|| c | c c c c ||}
 \hline
 Meson & lattice [MeV] & our value [MeV] & $s_0^{\mathrm{eff}}$ [GeV$^2$] & $M^2$ [GeV$^2$]\\ [0.5ex]
 \hline\hline
 $f_{B_c}$ & 427$\pm$8~\cite{McNeile:2012qf}, 434$\pm 5$~\cite{Colquhoun:2015oha} & 425$\pm$25 & 53-55 & 30-50 \\
 \hline
 $f_{D^0}$ & 213$\pm$2 \cite{Bazavov:2014wgs}, 207$\pm$4 \cite{Carrasco:2014poa} & 212$\pm$16 & 7-7.5 & 4-6\\
 \hline
 $f_{D^*}$ & 278$\pm$23 \cite{Becirevic:2012ti}, 224$\pm$9 \cite{Lubicz:2017asp} & 258$\pm$40 & 6-8 & 6-8\\
 \hline
\end{tabular}
\captionof{table}{Decay constants of mesons with 3ptSR parameters.}
\label{tab:dconst} 
\end{center}
An approximate relation connecting the Borel mass parameters of different meson decay constants  noticed by authors in~\cite{Ball:1991bs}
\begin{equation}
    \label{eq:borelrat}
    \frac{M_1^2}{M_2^2}\approx \frac{m_{\mathrm{M_1}}^2-m_{Q_1}^2}{m_{\mathrm{M_2}}^2-m_{Q_2}^2}
\end{equation}
where $m_{\mathrm{M}_i}$ is mass of the meson, and $m_{Q_i}$ is the mass of its heavier quark, is found to hold here too, and, as will be shown later, in 3-point calculations as well. Note however, that the uncertainties of decay constants arising in our calculation are connected with our specific method of calculation, since the threshold parameters are actually fixed so that they reproduce the lattice values, along with their uncertainties. Venturing into the 3-point calculation, as mentioned above, we use the same Borel thresholds paired with the decay constants that are reproduced by them.
It is important to notice that, when estimating uncertainties in the parameters of the 3ptSR calculation we do not vary the decay constants and thresholds independently, but rather we always use decay constants values together with the corresponding thresholds fixed by the decay constants calculation.
The hope is that all the higher order/higher dimension operator contributions are reproduced through the threshold modification in the 3-point calculation as well. Otherwise, for the $b$ quark we use the so called "potential subtracted" mass~\cite{Beneke:1998rk}, which is coincidentally very close to both the $\Upsilon(1S)$ scheme mass~\cite{Hoang:1998uv} and the kinetic scheme mass \cite{Bigi:1996si,Neubert:2004sp}, whereas the $c$-quark the mass is then given by varying the ratio $Z$ of the two masses given by the QCD lattice calculation \cite{Aoki:2016frl,Bazavov:2018omf}, keeping in mind that we do not use the $\overline{\mathrm{MS}}$ masses, and this ratio for pole masses tends to be lower as higher order corrections are included - which is why we choose to use a somewhat lower value of $Z$. The same method described above was already used for calculating $B_c \to \eta_c, J/\psi$ transition form factors in \cite{RjpsiOUR}. All parameters used are listed in Table \ref{tab:pars}. 
\begin{table}[H]
	    \centering
		\begin{tabular}{ |c|c|c| }
			\hline
			$m_b = 4.6^{+0.1}_{-0.1}$ GeV & $m_{B_c}=6.275$ GeV & $M^2_{B_c}=60-90$ GeV$^2$\\ [0.75ex] 
			$m_c = Zm_b$, $Z\approx 0.29^{+0.1}_{-0.1}$ & $m_{D^0} = 1.865$ GeV & $M^2_{D^0}=8-12$ GeV$^2$ \\[0.75ex]
            & $m_{D^*} = 2.007$ GeV & $M^2_{D^*}=12-16$ GeV$^2$ \\[0.75ex] 
			\hline\hline 
			$\expval{\frac{\alpha_s}{\pi}GG}=0.009 \pm 0.007$ GeV$^4$ \cite{Ioffe:2002be} & $m_0 = 0.8-1.05$ GeV & $\tau_{B_c}=0.507\pm 0.009$ ps\\ [0.75ex] 
			\hline
		\end{tabular}
	\captionof{table}{Parameters used in the QCDSR calculation.}
	\label{tab:pars}
\end{table}
As for the Borel mass parameters, it is found that stability in the sense of appearance of the Borel plateau is achieved for approximately twice of the values of the Borel mass parameters used in the 2-point sum rule, so that 
\begin{equation}
    \label{eq:3pt2ptrel}
    \frac{M^{2}_\mathrm{2pt.}}{M^{2}_\mathrm{3pt.}}\approx\frac{1}{2}, 
\end{equation}
which is a heuristic finding also confirmed by prior QCDSR studies. We additionally demand that heavier hadronic states contribute less than $50\%$ of the ground-state $B_c$ meson contribution to 3ptSR in the $B_c$ channel. This condition (and the one arising from the $m_{B_c}$ reproduction) yields an upper limit in ${}^{\uparrow}M_{B_c}^2 = 90$ GeV$^2$, and a lower limit in ${}^{\downarrow}M_{D^{(*)}}^2 = 8$ GeV$^2$. To get the upper constraint on $M_{D^{(*)}}^2$ we would need to know more precisely the three-point correlation function, since obviously we are missing $\mathcal{O}(\alpha_s)$ corrections which we model by adjusting the effective threshold $s_0^{eff}$ parameters of continuum contributions from the 2ptSR as described above. The sum rules for observables should in principle be independent of the auxiliary Borel parameters $M^2$. In practice, however, this is only approximately true because of the various approximations made. Therefore it is important to pick the right 'Borel window' where all above requirements are satisfied, by checking  the stability of the sum rule against variation of the Borel and other parameters and including the errors into analysis. Our form factor uncertainties reflect also these parameter variations, and the form factors do not change by more than $\sim$5\% in the acceptable range of the Borel mass parameters.
The estimation of the systematic errors is hard here, since we lack higher perturbative corrections in the 3ptSR, which could be large and would significantly stabilize the sum rules. The extracted values of sum rule parameters described above are in the acceptable range, expected by some general considerations, such as for example the relation among Borel parameters of two different meson constants (\ref{eq:borelrat}), as well as among Borel parameters from 2ptSR and 3ptSR calculations, given in (\ref{eq:3pt2ptrel}).

In Table \ref{tab:ffactor} we give our prediction for the form factors at $q^2 = 0$ together with an extensive list of the earlier results found in literature. Note that the form factors predicted in previous QCDSR calculation \cite{Kiselev:2002vz} are significantly larger then our results and those obtained from other model calculations. The reason can be found in the fact that the authors there renormalize their perturbative spectral densities using Coulomb-like gluon exchange corrections which usually results in multiplying the bare values by a factor of three or more. Another remark is that this kind of renormalization should work at $q^2_{\mathrm{max}}$, leaving the scaling with $q^2$ somewhat ambiguous, especially having in mind that the QCDSR are supposed to be reliable in the maximum recoil region. The authors thus claim that this implies that their results can be considered to represent the upper bounds in the QCDSR approach.
\begin{table}[H]
\begin{center}
\addtolength{\tabcolsep}{-3pt}
\renewcommand{\arraystretch}{1.5}
    \begin{adjustbox}{center}
 \begin{tabular}{||c | c c c c c c c c c c c ||}
 \hline
 Form Factor & \begin{tabular}{@{}c@{}}QCDSR \\ this work \end{tabular} & \begin{tabular}{@{}c@{}}QCDSR\\ ~\cite{Kiselev:2002vz}\end{tabular} &
 \begin{tabular}{@{}c@{}}LCSR\\ ~\cite{Huang:2007kb}\end{tabular} & \begin{tabular}{@{}c@{}}CCQM\\ ~\cite{Dubnicka:2017job,Issadykov:2017wlb}\end{tabular} & \begin{tabular}{@{}c@{}}pQCD\\ ~\cite{Wang:2014yia}\end{tabular} & \begin{tabular}{@{}c@{}}RQM\\ ~\cite{Ebert:2003cn}\end{tabular} & \begin{tabular}{@{}c@{}}RQM\\ ~\cite{Nobes:2000pm}\end{tabular} & \begin{tabular}{@{}c@{}}LFQM\\ ~\cite{Wang:2008xt}\end{tabular} & \begin{tabular}{@{}c@{}}SMD \\ ~\cite{Du:1988ws}\end{tabular} & \begin{tabular}{@{}c@{}}QCDSR\\ ~\cite{Colangelo:1992cx}\end{tabular} & \begin{tabular}{@{}c@{}}BSW \\ ~\cite{Dhir:2008hh}\end{tabular}\\ [0.5ex]
 \hline\hline
 $f^{D^{0}}_{+,0}(q^2=0)$ & $0.16\pm0.02$ & 0.32 & 0.35 & 0.19 & 0.19(7) & 0.14 & 0.14 & 0.16(4) & 0.15 & 0.13(5) & 0.08\\
 \hline\hline
 $V^{D^{*}}(q^2=0)$ & $0.27\pm0.04$ & 1.66 & 0.57 & 0.23 & 0.25(11) & 0.18 & 0.17 & 0.13(3) & 0.22 & 0.25(8) & 0.16\\
 \hline
 $A_1^{D^{*}}(q^2=0)$ & $0.17\pm0.03$ & 0.43 & 0.32 & 0.14 & 0.18(8) & 0.17 & 0.10 & 0.08(2) & 0.15 & 0.11(4) & 0.10\\
 \hline
 $A_2^{D^{*}}(q^2=0)$ & $0.17\pm0.03$ & 0.51 & 0.57 & 0.15 & 0.20(8) & 0.19 & 0.11 & 0.07(2) & 0.13 & 0.17(8) & 0.11\\
 \hline
 $A_0^{D^{*}}(q^2=0)$ & $0.17\pm0.04$ & 0.35 & - & 0.13 & 0.17(7) & 0.14 & 0.09 & 0.09(2) & 0.16 & - & 0.08\\
 \hline
\end{tabular}
\end{adjustbox}
\captionof{table}{Form factor predictions at $q^2 =0$ in various models.} \label{tab:ffactor} 
\end{center}
\end{table}
\subsection{High $q^2$ extrapolation of the form factors}
 Considering the fact that the QCDSR method is reliable only in the low $q^2$ region, we calculate the form factors for $ 0 \leq q^2 \leq 10$ GeV$^2$, extrapolate them to high $q^2$ region using the Bourrely-Caprini-Lellouch (BCL) approach \cite{Bourrely:2008za} and then compare it to the Boyd-Grinstein-Lebed (BGL) \cite{Boyd1,Boyd2} one. Both methods use unitarity by defining a conformal variable
 \begin{equation}
     z\equiv z(q^2,t_0) = \frac{\sqrt{t_* - q^2}-\sqrt{t_* - t_0}}{\sqrt{t_* - q^2}+\sqrt{t_* - t_0}}, 
 \end{equation}
 which maps the $q^2$-plane for $q^2>t_*$ onto a unit disk in the complex plane, with $z(t_*,t_0)=-1$, $z(\infty,t_0)=1$, and $z(t_0,t_0)=0$. The parameter $t_0$ therefore determines the $q^2$ point which is mapped to the origin of the disk, and all the poles beneath $t_*$ need to be compensated for by the fitting  function. Traditionally, one would choose $t_*=t_+=(m_{B_c}+m_{D^{(*)}})^2$ and then compensate for all the resonances beneath $t_*$. Then, one would assume that the two-particle contributions to the form factor (seen as branch cuts beneath $t_*$) are negligible, even though there might be plenty. In our case, however, it turns out to be beneficial to use $t_* = (m_{B^{(*)}}+m_\eta)^2$ for $B_c \to D$ semileptonic decays and $t_* = (m_{B^{(*)}} + m_{\rho})^2$ for $B_c \to D^*$ semileptonic decays, for then there is a maximum of two resonances contributing to a single form factor which are then removed by the Blaschke factor in the BGL parametrization or by the standard pole dependence of form factors in the BCL
 \footnote{As noticed in \cite{CohenLammLebed} in earlier works the threshold was always taken as $t^{\ast} = t_+$, which might introduce a sub-threshold branch cut in the region $|z| < 1$. 
 Similarly as used in \cite{CohenLammLebed} for their case, since we are interested only in the semileptonic decays at  $m_l^2  \leq t \leq t_{-}$, we can take the smaller thresholds:  $t_* = (m_{B}^{(\ast)} + m_{\eta})^2$ for $B_c \to D$ semileptonic decays and $t_* = (m_{B}^{(\ast)} + m_{\rho})^2$ for $B_c \to D^*$ semileptonic decays, which are always larger than allowed $t$ in these decays. It should be noticed however that these thresholds are not the lowest possible ones. The lowest threshold for $B_c \to D^{(*)}$ semileptonic decays would be $t_* = (m_{B}^{(*)} + m_{\pi})^2$. However, in that case some of the form factors would not have expected pole behavior and would show a slight instability at higher $t$ which we avoid with above choices of the threshold parameters $t_*$.}. 
 
 The fitting function is then expanded in a power series in $z$ multiplied by a function compensating for the poles, which is for the two cases inspected here given as
 \begin{equation}
    \label{eq:fits}
    F_i^{\textrm{BGL}}(z)=\frac{1}{B_i(z)\phi_i(z)}\sum_{k=0}^{\infty}a_k z^k \,, \qquad\qquad F_i^{\textrm{BCL}}(z)=\frac{1}{P_i(z)}\sum_{k=0}^{\infty}b_k z^k \,, 
 \end{equation}
 where $F_i^{\mathrm{BGL}}(z)$ are the helicity form factors defined as
 \begin{equation}
    \label{eq:fitsF}
    \begin{split}
        g(z) & = \frac{2}{m_{B_c}+m_{D^*}}V(z); \quad f(z) = (m_{B_c}+m_{D^*})A_1(z); \quad \mathcal{F}_2(z) = 2A_0(z);\\
        \mathcal{F}_1(z) & = \frac{1}{m_{D^*}}\bigg[-\frac{\lambda(m_{B_c},m_{D^*},q^2)}{m_{B_c}+m_{D^*}}A_2(z)-\frac{1}{2}(q^2-m_{B_c}^2+m_{D^*}^2)(m_{B_c}+m_{D^*})A_1(z)\bigg]\\
    \end{split}
 \end{equation}
 for the vector particle and 
 \begin{equation}
    \label{eq:eqf0}
    f_0^{\mathrm{BGL}}(q^2) = (M_{B_c}^2-M_{D}^2)f_0(q^2)
 \end{equation}
 for the pseudoscalar one, whereas for $F_i^{\mathrm{BCL}}(z)$ the standard form factor basis from Eq.~(\ref{eq:ffdef}) is used and the functions
 \begin{equation}
     B_i(z)=\prod_{R=1}^n\frac{z-z(m_R^2,t_0)}{1- z z(m_R^2,t_0)}\quad\mathrm{and}\quad P_i(z)=\prod_{R=1}^{n}\left (1-\frac{q^2(z)}{m_R^2} \right )
 \end{equation} account for $n$ resonances of masses $m_R$ below the threshold. One can notice that in the original BCL paper authors used the fact that the derivative of the form factor vanishes at $q^2=t_+$, which is a consequence of angular momentum conservation, and expression-wise relies on the fact that $z(q^2=t_+)=-1$, which isn't the case here, since we chose $t_*\neq t_+$. Therefore, we do not utilize this fact  and keep the parametrization in its more simple form. The form factors in the helicity basis are used in the case of fitting to the BGL function, since in this basis unitarity relations are diagonalized and the $\phi_i(z)$ functions are readily available. The latter are calculable perturbatively and have been known for a long time now. We list them in Appendix B.3. 
 
 A final comment concerns the parameter $t_0$. Here we have chosen the value that optimizes the fit in the sense that it reduces the possible error originating from truncating the series in (\ref{eq:fits}). This is achieved for $z(0,t_0)=-z(q^2_{\mathrm{max}},t_0)$, which lowers the overall maximum value of $z$, and thus $|z_{\mathrm{max}}|\approx 0.106$. 
 
 For $m_{B^{(*)}}$ we use the experimentally well established values, while for the other resonances we use values obtained by a recently updated quark model \cite{Godfrey:2016nwn}, all listed in Table \ref{tabular:res}. 
\begin{center}
    \addtolength{\tabcolsep}{-3pt}
    \renewcommand{\arraystretch}{1.5}
     \begin{tabular}{|| c c c || c c c | c || c c c | c ||}
     \hline
        $J^P$ & threshold & $m_R$  [GeV] & BGL: & $a_0$ & $a_1$ & $\chi^2[10^{-2}]$ & BCL: & $b_0$ & $b_1$ & $\chi^2[10^{-2}]$ \\ [0.5ex]
        \hline
        $1^-$ & $B \eta$ & 5.32 & $f_+$ & 0.0087 & -0.032 & 0.1& $f_+$ & 0.23 & -0.7 & 4\\
        $0^+$ & $B \eta$ & 5.76 & $f_0^{\mathrm{BGL}}$ & 0.019 & -0.07 & 5 & $f_0$ & 0.18 & -0.2 & 0.8\\
        \hline
        \hline
        $1^-$ & $B^*\rho $ & 5.32, 5.93 & $g$ & 0.019 & -0.04 & 1 & $V$ & 0.25 & 0.2 & 0.9 \\
        $1^+$ & $B\rho$ & 5.78, 5.78 & $f$ & 0.0055 & -0.01 & 2 & $A_1$ & 0.11 & 0.5 & 4\\
        $1^+$ & $B\rho$ & 5.78, 5.78 & $\mathcal{F}_1$ & 0.0010 & -0.003 & 0.6 & $A_2$ & 0.16 & -0.1 & 6\\
        $0^-$ & $B\rho$ & 5.28, 5.91 & $\mathcal{F}_2$ & 0.022 & -0.05 & 0.5 & $A_0$ & 0.12 & 0.3 & 20\\
        \hline
    \end{tabular}
        \captionof{table}{Summary of the fits for $B_c \to D$ and $B_c \to D^{*}$ form factors.}
    \label{tabular:res}
\end{center}

\begin{figure}[h!]
    \centering
    \includegraphics[width=0.47\textwidth]{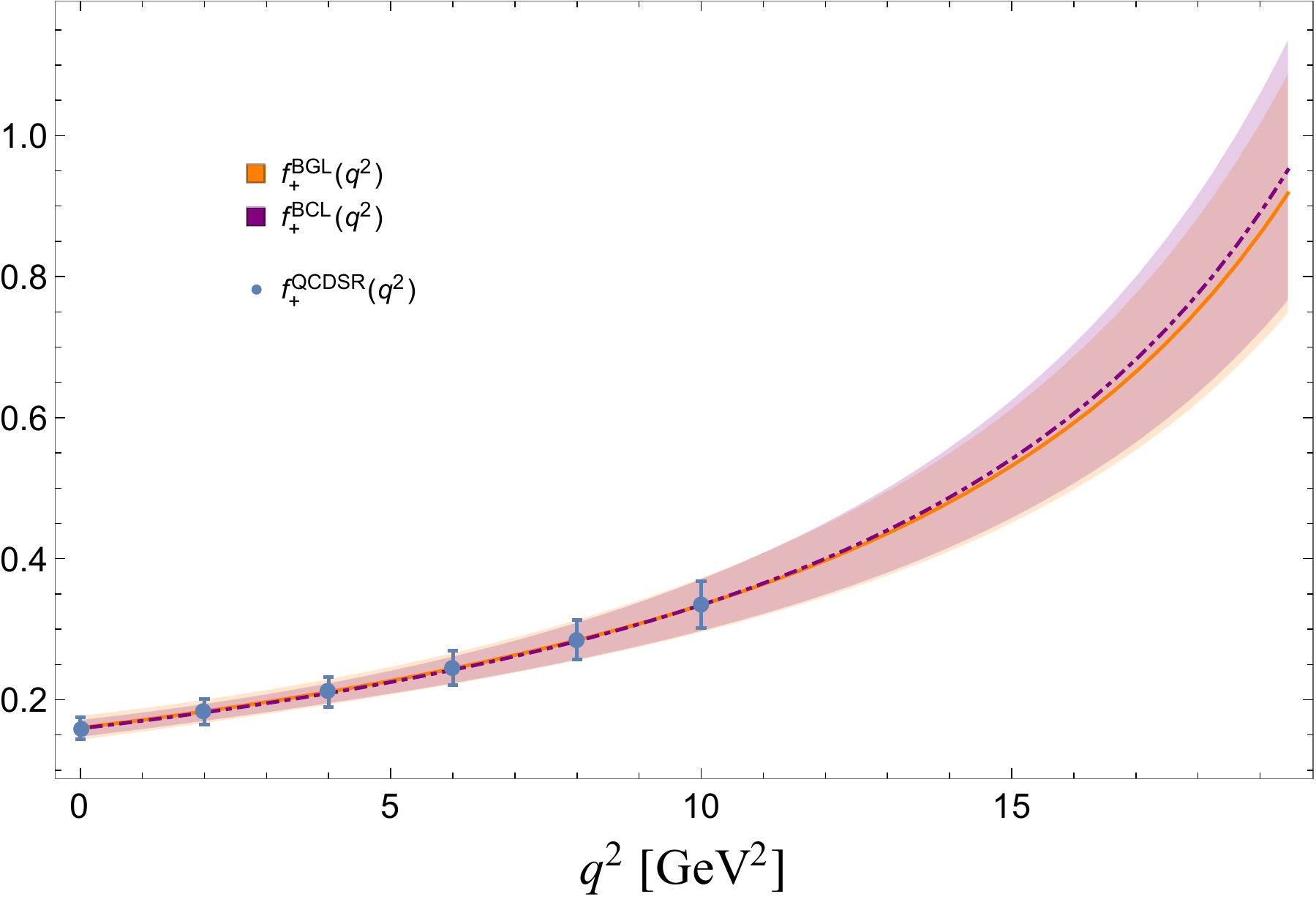}
    \hspace*{0.3cm}
    \includegraphics[width=0.47\textwidth]{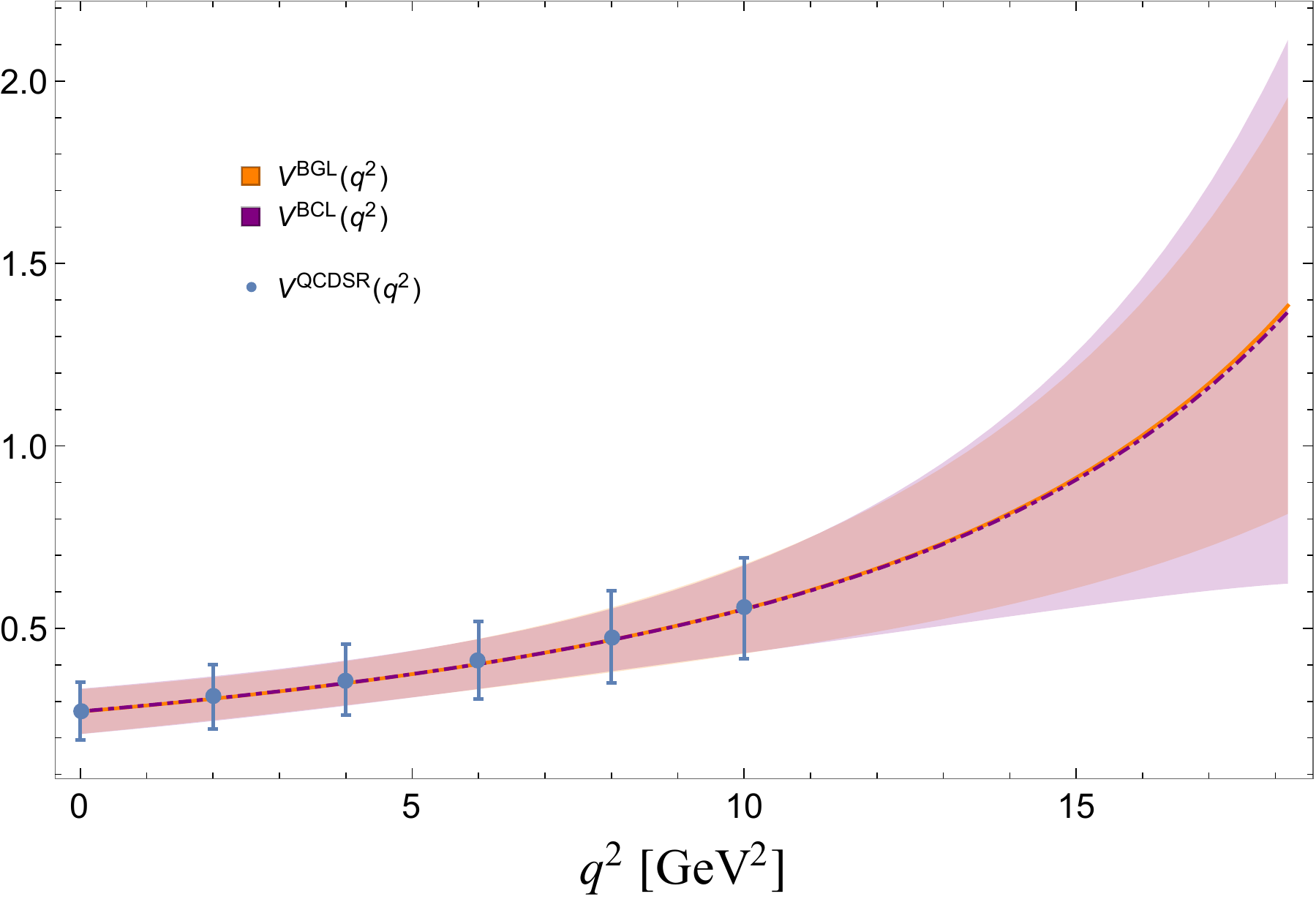}
    \caption{The BGL (purple, dot-dashed line) and BCL (orange, full line) fits to the QCDSR values of $f_+(q^2)$ and $V(q^2)$.}
    \label{fig:BCLBGL}
\end{figure}

Errors of the fitting parameters from Table~\ref{tabular:res} and their correlations are given in Appendix B.2. 
In $B_c \to D^{(*)}$ decays, being $b\to u$ transitions, the parameter $|z_{\mathrm{max}}|$ in the form factor $q^2$-expansion is somewhat larger, and the functions $|\phi_i(z)_{\mathrm{min}}|$ are smaller than in typical $b\to c$ transitions, and one would need to go to higher order in $z$ to reduce the truncation error - which would be unusable here, since we have no high-$q^2$ points to impose bounds on parameters of the fit multiplying higher orders in $z$. Actually adding higher orders of $z$ to the fit function only marginally changes its central shape, which is mostly witnessed through the central value of $f^{\mathrm{fit}}_+(q^2_{\mathrm{max}})$, which changes at most by $\sim 10\%$ and always stays inside the uncertainties of the linear $z$ fit. Using $f_+(q^2)$ as a benchmark, the difference of the two fitting procedures is made obvious in Figure \ref{fig:BCLBGL}, where one can see that compensating multiple resonances using a multipole function $P_i(q^2)$ can be a bit more violent, driving the fit towards higher form factor values.
\newline
Knowing that traditionally (in $B\to \pi$, $B\to D^{(*)}$ decays) sum rules undervalue the form factors at zero-recoil, one is tempted to use the fit that reproduces higher values of zero-recoil form factors, even if this is somewhat less faithful to our QCDSR results in terms of $\chi^2$, defined for the $i-$th form factor as
\begin{equation}
    \chi^2_i=\sum_{j}\frac{[F_i^{\mathrm{fit}}(q^2_j)-F_i^{\mathrm{QCDSR}}(q^2_j)]^2}{[\sigma_{F_i}^2(q^2_j)]^2}.
\end{equation}
In our fits this turns out to be the case when the more simple BCL choice of parametrization is adopted, which is consequently the one we use in our phenomenological analysis from now on. The difference anyways turns out to be almost negligible for all of the form factors. 

\begin{figure}[h!]
    \centering
    \includegraphics[width=0.47\textwidth]{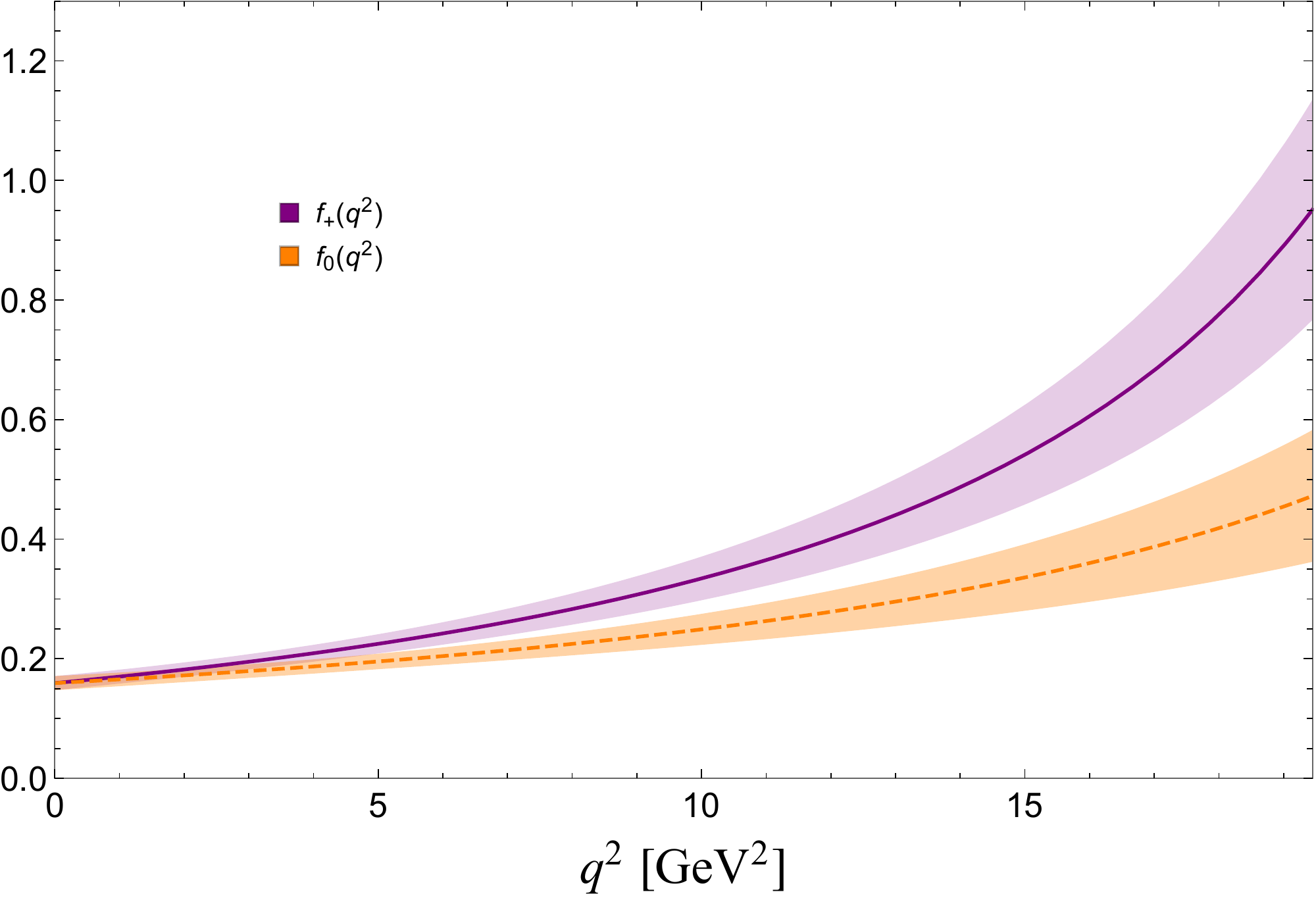}\\
    \includegraphics[width=0.47\textwidth]{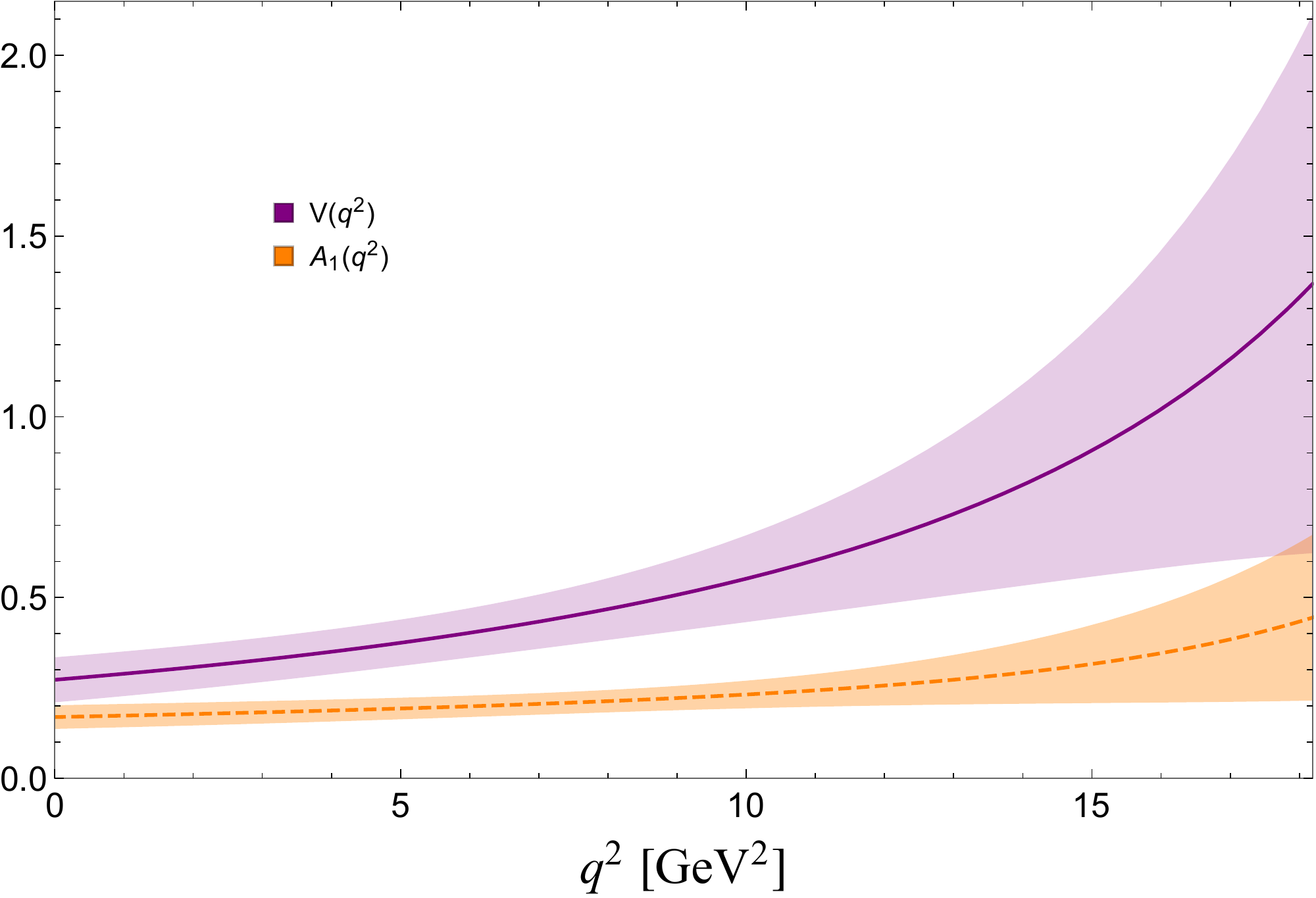}
        \hspace*{0.2cm}
    \includegraphics[width=0.47\textwidth]{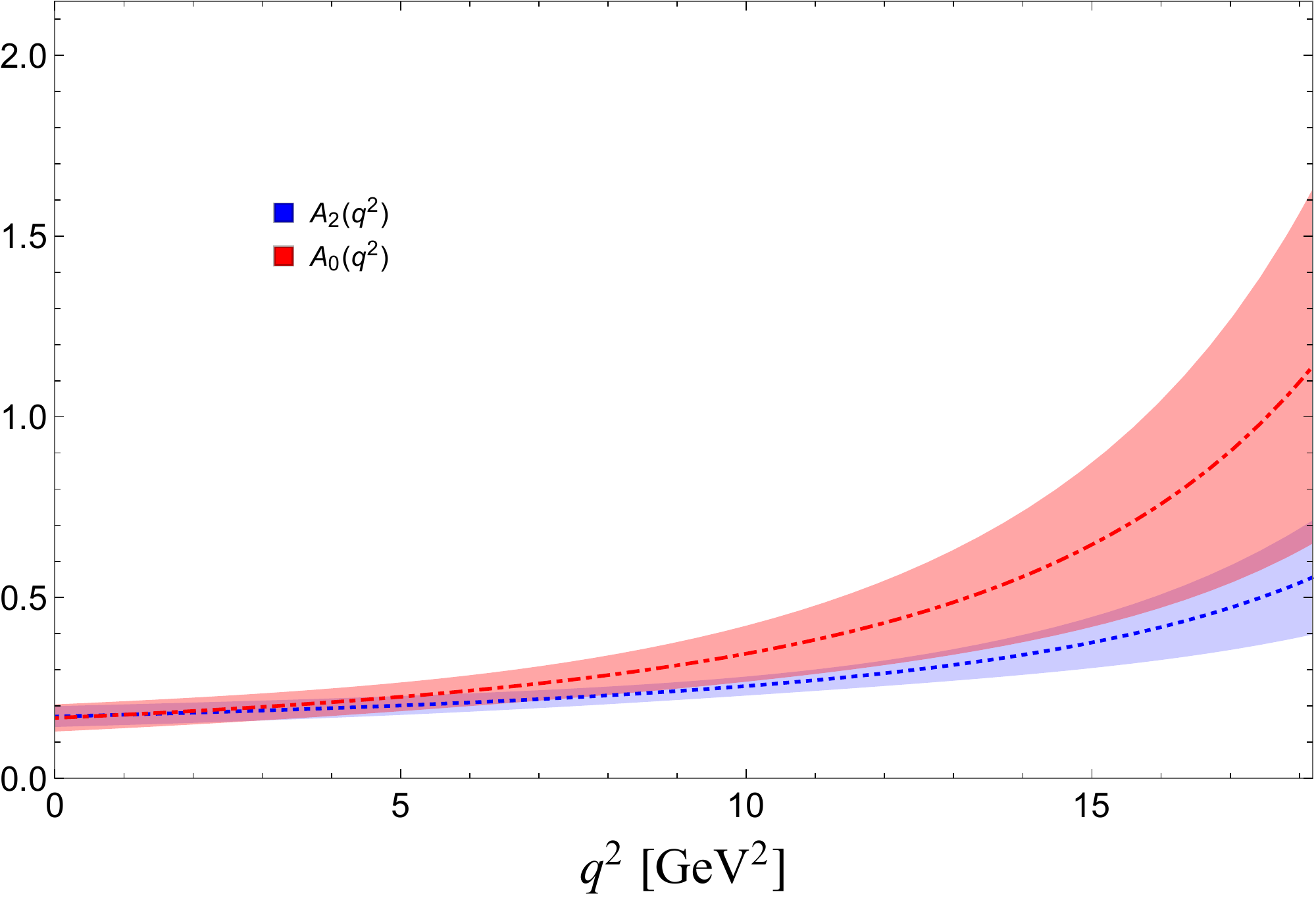}
    \caption{Final predictions for $B_c \to D^{(\ast)}$ form factors obtained by extrapolating 3ptSR results to higher $q^2$ regions by using the BCL parametrizations.}
    \label{fig:BCLBGL2}
\end{figure}

\begin{figure}[h!]
    \centering
    \includegraphics[width=0.49\textwidth]{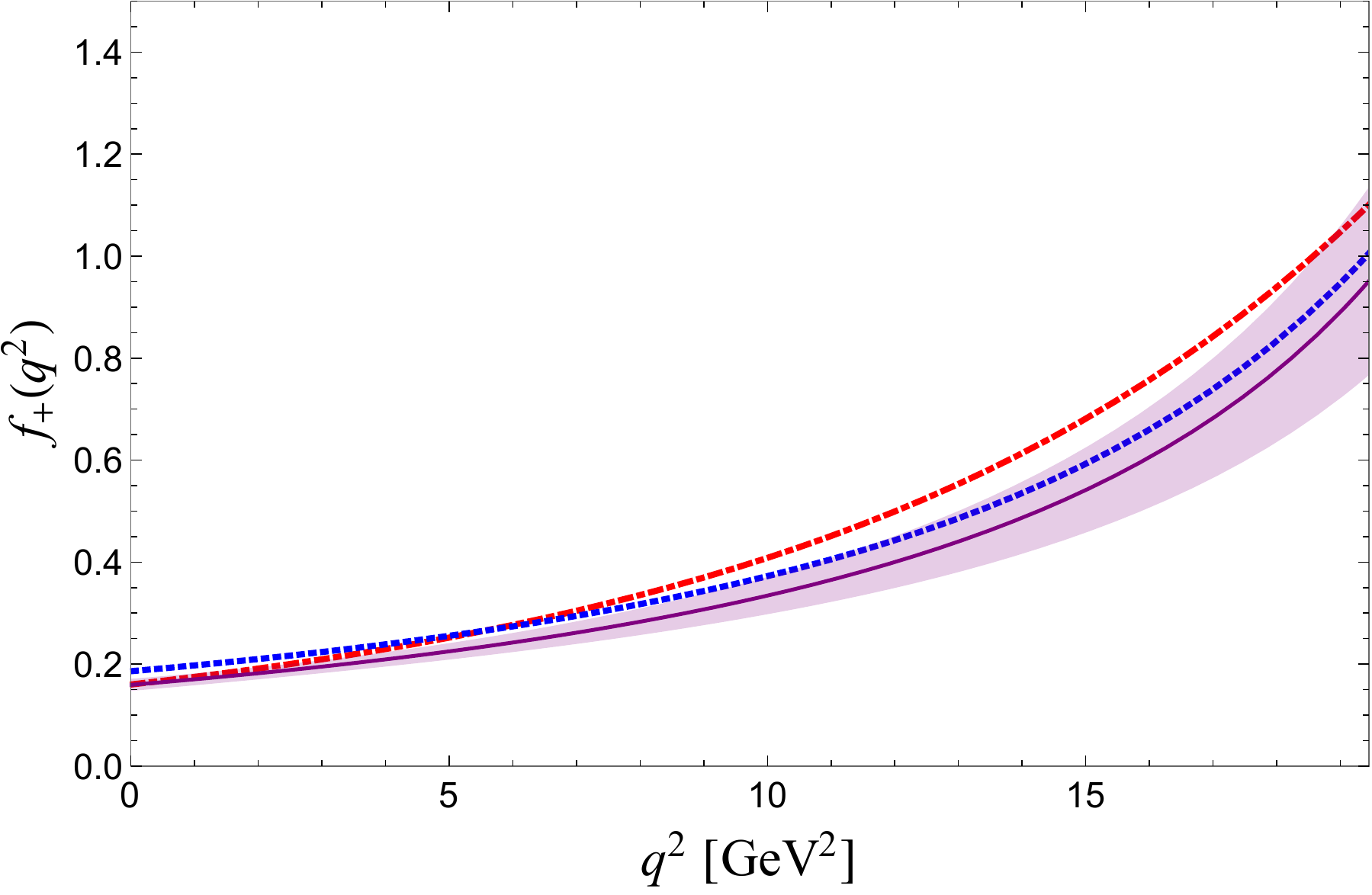}
    \caption{The purple solid line and the area represents our result for $f_+(q^2)$ form factor with errors; the blue dashed-line is the result of CCQM \cite{Dubnicka:2017job} and the red dash-dotted one is the LFQM prediction from \cite{Wang:2008xt}.}
    \label{fig:BCLBG3L}
\end{figure}

In Figure \ref{fig:BCLBGL2} we present the $q^2$-dependence of form factors which we use further in the analysis. 
One can notice that the $B_c\rightarrow D^*$ form factors come with a larger uncertainty, which stems from the fact that the value of $f_{D^*}$ decay constant is more uncertain than of $f_D$, in both lattice and our fitted results, see Table \ref{tab:dconst}.
In Figure \ref{fig:BCLBG3L} we show a comparison of our prediction for $f_+(q^2)$ to two quark models, namely the constituent quark model, CCQM~\cite{Dubnicka:2017job} and the light quark model, LFQM~\cite{Wang:2008xt}, where a good agreement among results can be noticed, despite the difference in approaches when obtaining them.


\subsection{A comment on correlations between pseudo-data points}
\label{sec:corr}

The reader might notice that we chose our pseudo-data points to be uncorrelated. This is a rather strong assumption, and stems from the fact that the correlations are hard to be determined exactly. In order to estimate the effect the correlation might have on the uncertainties of our form factors and observables, first we define a new $\tilde{\chi}^2$ to be minimized, namely,
\begin{equation}
    \tilde{\chi}_i^2 \equiv [F_i(q_a^2;\vec{\theta})-F^{\mathrm{QCDSR}}_i(q_a^2)]\big(\Sigma^{-1}\big)_{ab}[F_i(q_b^2;\vec{\theta})-F^{\mathrm{QCDSR}}_i(q_b^2)].
\end{equation}

For the first estimate of $\tilde{\chi}_i^2$ we correlate the points rather crudely by introducing the covariance
\begin{equation}
    \Sigma_{ij} \equiv (1-x)\cdot s_i s_j \delta_{ij} + x\cdot s_i s_j\,,
\end{equation}
where $x$ describes the correlation, varied between $50\%-90\%$, while $s_i$ are the errors of the parameters arising from the QCDSR calculation. The effect this arbitrary correlation has on the goodness of fit is that (as expected) $\tilde{\chi}^2$ is  now much larger, $\tilde{\chi}^2 \approx [2-10]\cdot \chi^2$. On the side of the uncertainties of the observables and  the integrated decay rates have the same uncertainty.  The observables defined as ratios have much smaller uncertainties (as expected, since they cancel to a larger extent in the ratio), while the form factors and differential decay rates have larger uncertainties in the $q^2$ region below $\sim 8$ GeV$^2$, and somewhat smaller uncertainties in the upper $q^2$ region.

As for our preferable estimate of the correlated errors we calculate the Jacobian for each form factor and correlate the data points using it. We estimate the Jacobian $\mathcal{J}$, which is a $11\cross 6$ matrix (11 pseudo-data points and 6 parameters) at central values of parameters and estimate a "raw" covariance matrix by
\begin{equation}
    \Sigma = \mathcal{J}^T P \mathcal{J},
\end{equation}
where $P$ is a $6\cross 6$ diagonal matrix containing the errors in the parameters (we assume the parameters themselves are uncorrelated). The covariance matrix we use when minimizing $\tilde{\chi}^2$ is estimated then by
\begin{equation}
    \tilde{\Sigma} = \sigma^T\!\! \rho\, \sigma,
\end{equation}
where $\rho$ is the correlation matrix obtained from $\Sigma$, and $\sigma$ is a vector containing the QCDSR errors of the pseudo-data points. 


Globally, now the $\tilde{\chi}^2$ grows to an even larger value,  $\tilde{\chi}^2 \approx 20\cdot \chi^2$. The effect this has on the net uncertainties of the quantities presented in this paper is the same as described above, only to a larger extent. 

\begin{figure}[H]
    \centering
    \includegraphics[scale=0.33]{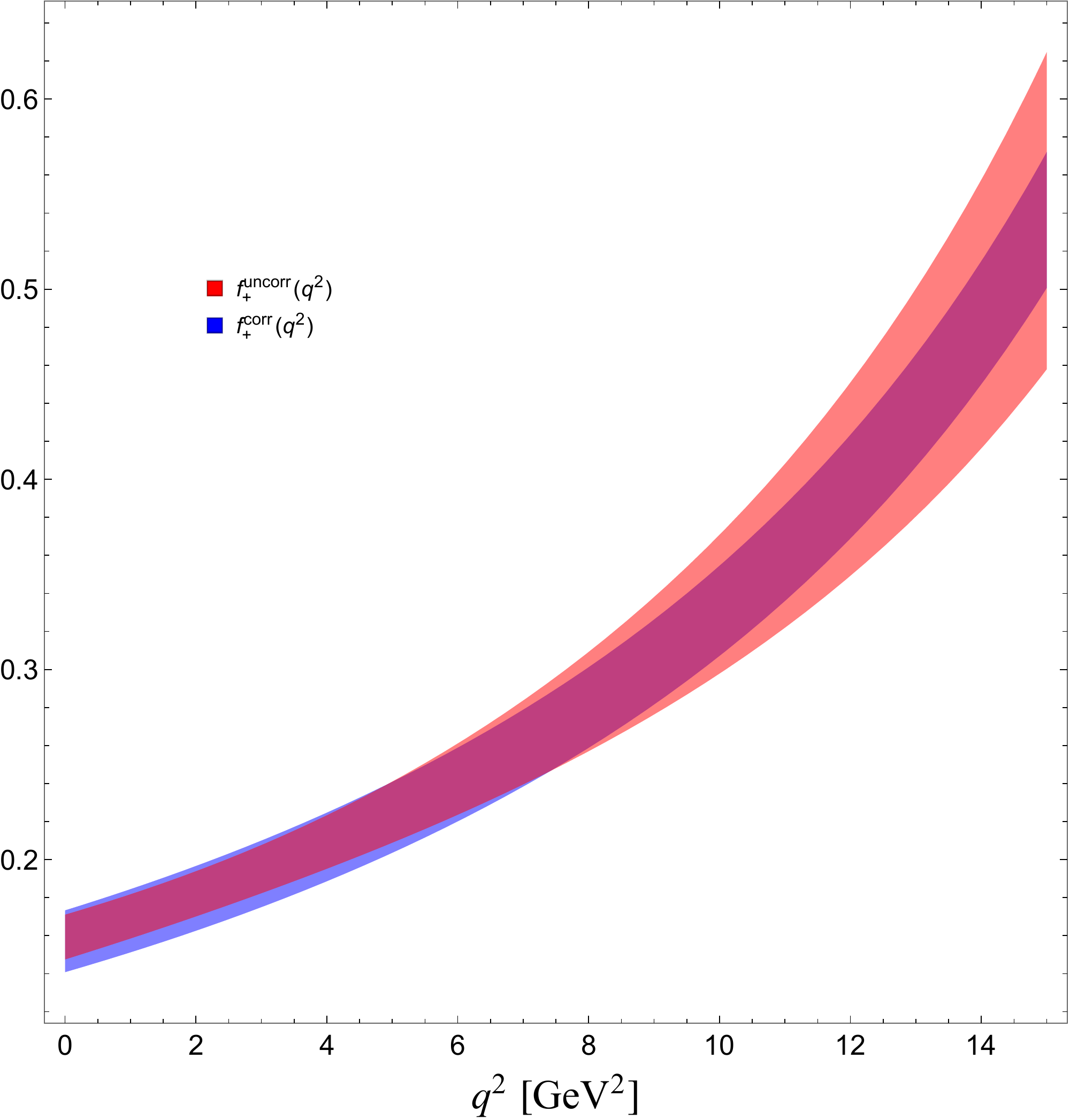}
    \caption{Comparison of the uncorrelated fit with the correlated one.}
    \label{fig:cor}
\end{figure}

In particular, for the correlated cases the lower $q^2$ region have now larger uncertainties, while in the upper $q^2$ region the errors become smaller, as compared with the uncorrelated case, as is witnessed in the Figure \ref{fig:cor}.
This is taken into consideration when we present our $|V_{ub}|$ determination prospect both in Sec.~\ref{Vub-deter} and Appendix~\ref{binneddel}. 

It should be stressed that all the central values of the fit parameters   remain the same.
\section{Semileptonic $B_c\rightarrow D^{(*)}$ decays and the $|V_{ub}|$ determination}
\subsection{Predictions for the decay rates and angular observables} 
Having in  hands calculated form factors in the full $q^2$ range we now turn towards the predictions of $B_c\rightarrow D^{(*)} l \nu$ branching ratios and asymmetries. 

The general expressions for the double differential distributions can be given as 
\begin{equation}
    \frac{\dd^2\Gamma(B_c\to D^{(*)} l \bar{\nu}_l)}{\dd q^2 \dd \cos{\theta_l}} = a^{D^{(*)}}_{\theta_l}(q^2) + b^{D^{(*)}}_{\theta_l}(q^2)\cos{\theta_l} + c^{D^{(*)}}_{\theta_l}(q^2)\cos^2{\theta_l}\,,
    \label{eq:doubleGamma}
\end{equation}
where $\theta_l$ is the angle between the lepton and the final state meson in the center-of-momentum frame of the leptonic pair, and the exact expressions for the $a_{\theta_l}(q^2), b_{\theta_l}(q^2)$ and $c_{\theta_l}(q^2)$ coefficients are given in the Appendix C.1. The partial $q^2$ differential decay widths is then given in each case by integrating over $\cos{\theta_l}$, or, specifically in the case of $B_c\rightarrow D^0 l \bar{\nu}_l$, 
\begin{equation}
    \begin{split}
        &\frac{\dd\Gamma(B_c\rightarrow D^0 l \bar{\nu}_l)}{\dd q^2}=\frac{G_F^2 |V_{ub}|^2q^2}{192\pi^3m_{B_c}^3}\sqrt{\lambda(m_{B_c}^2,m_{D^0}^2,q^2)}\bigg(1-\frac{m_l^2}{q^2}\bigg)^2\cross\\
        &\qquad\qquad\qquad\qquad\qquad\qquad\,\,\cross\bigg[\bigg(1+\frac{m_l^2}{2q^2}\bigg)|h_0(q^2)|^2+\frac{3m_l^2}{2q^2}|h_t(q^2)|^2\bigg],
    \end{split}
\end{equation}
whereas for the case of $B_c\rightarrow D^* l \bar{\nu}_l$ we have
\begin{equation}
    \begin{split}
    &\frac{\dd\Gamma(B_c\rightarrow D^* l \bar{\nu}_l)}{\dd q^2}=\frac{G_F^2 |V_{ub}|^2q^2}{192\pi^3m_{B_c}^3}\sqrt{\lambda(m_{B_c}^2,m_{D^*}^2,q^2)}\bigg(1-\frac{m_l^2}{q^2}\bigg)^2\cross\\
    &\cross\bigg[\bigg(1+\frac{m_l^2}{2q^2}\bigg)(|H_+(q^2)|^2+|H_-(q^2)|^2+|H_0(q^2)|^2)+\frac{3}{2}\frac{m_l^2}{q^2}|H_t(q^2)|^2\bigg],
    \end{split}
\end{equation}
with a new set of helicity form factors, defined as 
\begin{equation}
    h_0(q^2)=\sqrt{\frac{\lambda(m_{B_c}^2,m_{D^0}^2,q^2)}{q^2}}f_+(q^2)\,,\qquad h_t(q^2)=\frac{m_{B_c}^2-m_{D^0}^2}{\sqrt{q^2}}f_0(q^2),
\end{equation}
and 
\begin{equation}
    \begin{split}
        H_{\pm}(q^2)&=-i\bigg[\pm \frac{\sqrt{\lambda(m_{B_c}^2,m_{D^*}^2,q^2)}}{m_{B_c}+m_{D^*}}V(q^2)+(m_{B_c}+m_{D^*})A_1(q^2) \bigg],\\
        H_0(q^2)&=-\frac{i}{2m_{D^*}\sqrt{q^2}}\bigg[(m_{B_c}+m_{D^*})(m_{B_c}^2-m_{D^*}^2-q^2)A_1(q^2)\\
        &\qquad\qquad\qquad\,\,\,\,-\frac{\lambda(m_{B_c}^2,m_{D^*}^2,q^2)}{m_{B_c}+m_{D^*}}A_2(q^2)\bigg],\\
        H_t(q^2)&=-i\frac{\sqrt{\lambda(m_{B_c}^2,m_{D^*}^2,q^2)}}{\sqrt{q^2}}A_0(q^2).\\
    \end{split}
\end{equation}
When dealing with helicity amplitudes one needs to remember that they are defined through the specific choice of the virtual vector boson polarization, and the seminal paper dealing with such treatment in details is~\cite{Korner:1989qb}, so we choose not to elaborate on these further. In both of these cases it is also beneficial to define the differential decay width as a sum of contributions of left and right lepton helicity projections along the $z$-axis
\begin{equation}
    \begin{split}
        &\frac{\dd\Gamma^-(B_c\rightarrow D^0 l \bar{\nu}_l)}{\dd q^2}=\frac{G_F^2 |V_{ub}|^2q^2}{192\pi^3m_{B_c}^3}\sqrt{\lambda(m_{B_c}^2,m_{D^0}^2,q^2)}\bigg(1-\frac{m_l^2}{q^2}\bigg)^2|h_0(q^2)|^2 \,,\\
        &\frac{\dd\Gamma^+(B_c\rightarrow D^0 l \bar{\nu}_l)}{\dd q^2}=\frac{G_F^2 |V_{ub}|^2q^2}{192\pi^3m_{B_c}^3}\sqrt{\lambda(m_{B_c}^2,m_{D^0}^2,q^2)}\bigg(1-\frac{m_l^2}{q^2}\bigg)^2\frac{m_l^2}{2q^2}\big[|h_0(q^2)|^2+3|h_t(q^2)|^2\big]\,,\\
    \end{split}
\end{equation}
and
\begin{equation}
    \begin{split}
    \frac{\dd\Gamma^-(B_c\rightarrow D^* l \bar{\nu}_l)}{\dd q^2}=\frac{G_F^2 |V_{ub}|^2q^2}{192\pi^3m_{B_c}^3}&\sqrt{\lambda(m_{B_c}^2,m_{D^*}^2,q^2)}\bigg(1-\frac{m_l^2}{q^2}\bigg)^2\cross\\
    &\cross\big[|H_+(q^2)|^2+|H_-(q^2)|^2+|H_0(q^2)|^2\big]\,,\\
    \frac{\dd\Gamma^+(B_c\rightarrow D^* l \bar{\nu}_l)}{\dd q^2}=\frac{G_F^2 |V_{ub}|^2q^2}{192\pi^3m_{B_c}^3}&\sqrt{\lambda(m_{B_c}^2,m_{D^*}^2,q^2)}\bigg(1-\frac{m_l^2}{q^2}\bigg)^2\frac{m_l^2}{2q^2}\cross\\
    &\cross\big[|H_+(q^2)|^2+|H_-(q^2)|^2+|H_0(q^2)|^2+3|H_t(q^2)|^2\big]\,,\\
    \end{split}
\end{equation}
so that it's obvious that $\Gamma = \Gamma^+ + \Gamma^-$. In the case of the $D^*$ in the final state, one can look at both, the longitudinal and the transverse $D^*$ polarization contribution, 
\begin{equation}
    \begin{split}
    \frac{\dd\Gamma_L(B_c\rightarrow D^* l \bar{\nu}_l)}{\dd q^2}=\frac{G_F^2 |V_{ub}|^2q^2}{192\pi^3m_{B_c}^3}&\sqrt{\lambda(m_{B_c}^2,m_{D^*}^2,q^2)}\bigg(1-\frac{m_l^2}{q^2}\bigg)^2\cross\\
    &\cross\bigg[\bigg(1+\frac{m_l^2}{2q^2}\bigg)|H_0(q^2)|^2+\frac{3}{2}\frac{m_l^2}{q^2}|H_t(q^2)|^2\bigg]\,,\\
    \frac{\dd\Gamma_T(B_c\rightarrow D^* l \bar{\nu}_l)}{\dd q^2}=\frac{G_F^2 |V_{ub}|^2q^2}{192\pi^3m_{B_c}^3}&\sqrt{\lambda(m_{B_c}^2,m_{D^*}^2,q^2)}\bigg(1-\frac{m_l^2}{q^2}\bigg)^2\cross\\
    &\cross\bigg(1+\frac{m_l^2}{2q^2}\bigg)\big[|H_+(q^2)|^2+|H_-(q^2)|^2\big]\,,\\
    \end{split}
\end{equation}
respectively, where again $\Gamma = \Gamma_L + \Gamma_T$. 

Our predictions for integrated decay rates of both decays are given in Table \ref{tab:dwid}.

\begin{center}
\addtolength{\tabcolsep}{-3pt}
\renewcommand{\arraystretch}{1.5}
    \begin{tabular}{||c | c c c c c c c c c c c ||}
    \hline
    Mode & This work & \cite{Ivanov:2006ni} & \cite{Kiselev:2002vz} & \cite{Huang:2007kb} & \cite{Issadykov:2017wlb} & \cite{Wang:2008xt} & \cite{Wang:2014yia} & \cite{Ebert:2003cn} &\cite{Scora:1995ty} & \cite{Nobes:2000pm} & \cite{AbdElHady:1999xh} \\ [0.5ex]
    \hline\hline
    $\Gamma(B_c\rightarrow D^0 l \bar{\nu}_{l})$ & $31\pm 6$ & 51 & 59 & 293 & 43 & 43 & 46 & 19 & 26 & 20 & 49 \\
    \hline
    $\Gamma(B_c\rightarrow D^0 \tau \bar{\nu}_{\tau})$ & $21\pm 4$ & 31 & 32 & 219 & 27 & 30 & 32 & - & - & - & - \\
    \hline
     $\Gamma(B_c\rightarrow D^* l \bar{\nu}_{l})$ & $85\pm 33$ & 56 & 270 & 512 & 78 & 64 & 160 & 110 & 53 & 34 & 192 \\
    \hline
    $\Gamma(B_c\rightarrow D^* \tau \bar{\nu}_{\tau})$ & $46\pm 20$ & 32 & 120 & 293 & 44 & 39 & 94 & -  & - & - & - \\
    \hline
\end{tabular}
\captionof{table}{Decay widths of $B_c\rightarrow D^{0,*}$ decays, given in $10^{-18}$ GeV using the PDG average value of $V_{ub}$ from Eq.(\ref{eq:CKM2}).}
\label{tab:dwid} 
\end{center}

\begin{figure}[H]
        \centering
        \includegraphics[width=0.46\textwidth]{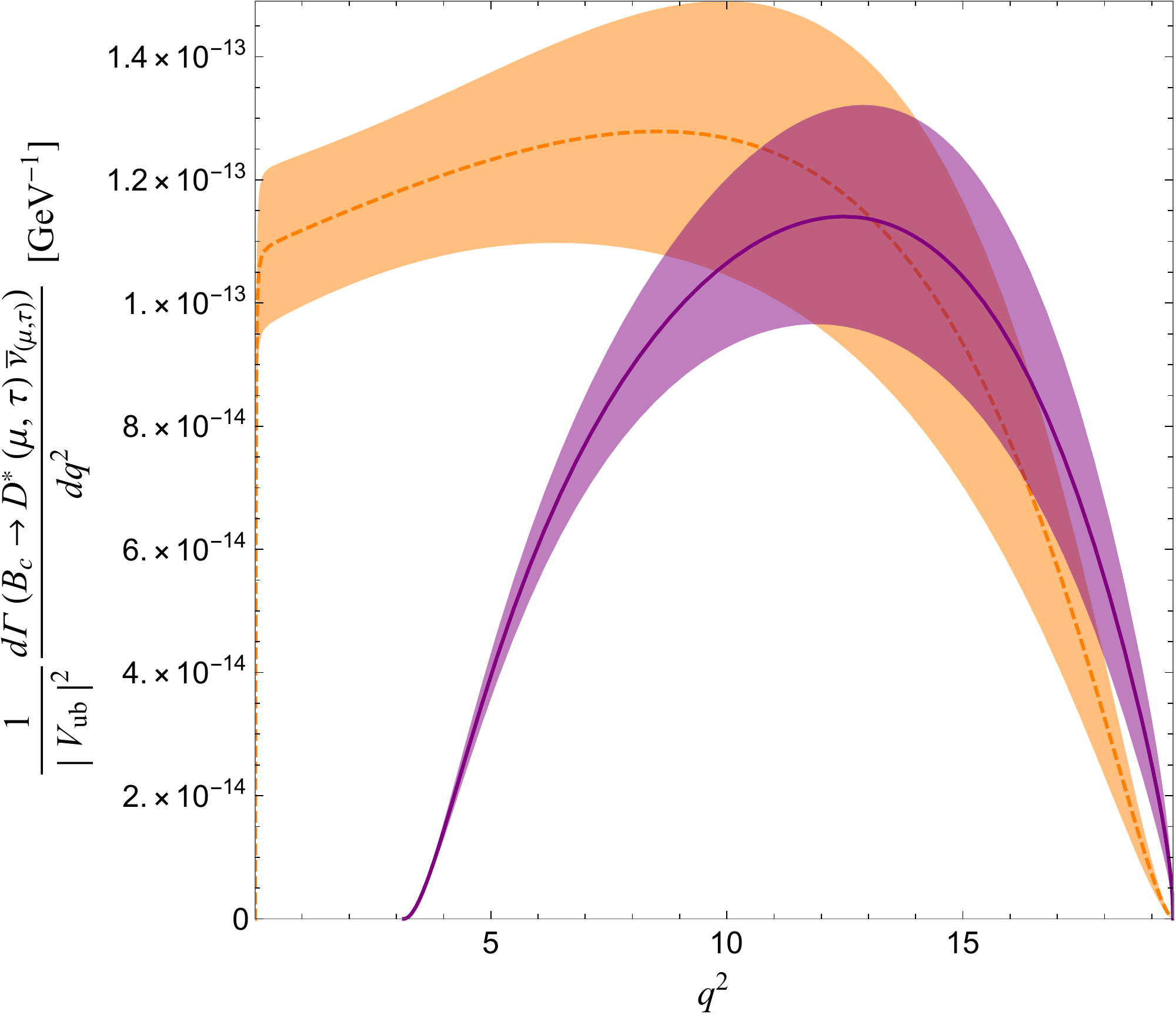}
        \hspace*{0.5cm}
        \includegraphics[width=0.46\textwidth]{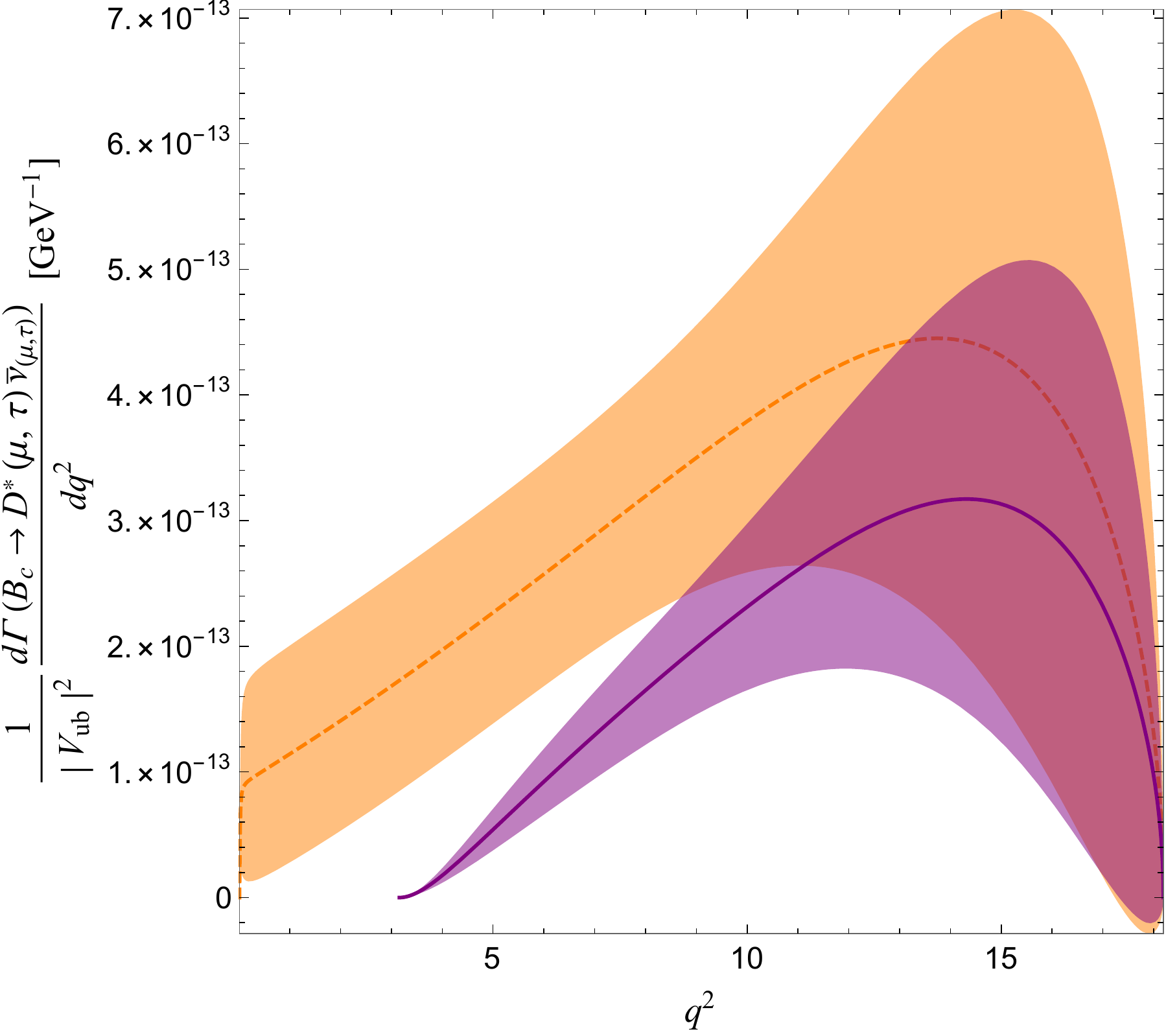}
        \caption{Partial differential decay rates with errors of $B_c$ semileptonic transitions to $D$ (left), and to $D^*$ (right), where the orange solid-line/area corresponds to $\mu$ in the final state, and the purple dashed-line/area to the case with $\tau$ final state.}
        \label{fig:ddw}
\end{figure}

We plot in Fig.\ref{fig:ddw} the partial differential decay rates in units of $|V_{ub}|$ GeV$^{-1}$. It is obvious that one can achieve the satisfactory precision for $B_c \to D l \bar{\nu}_l$ decays, while the theoretical errors in the $B_c \to D^{\ast}$ form factors and uncertainties in the $B_c$ and $D^{\ast}$ decay constants drive predictions for $B_c \to D^{\ast} l \bar{\nu}_l$ to be quite uncertain. 

Both experimentally and theoretically, due to the cancellations of systematic hadronic uncertainties, it is preferable to extract the ratios
\begin{equation}
    \label{eq:ratios}
    \begin{split}
        R_c(D^0) & \equiv\frac{{\cal B}(B_c\rightarrow D^0 \tau \bar{\nu}_{\tau})}{{\cal B}(B_c\rightarrow D^0 \mu \bar{\nu}_{\mu})} = 0.64\pm 0.05 ,\\
        R_c(D^*) & \equiv\frac{{\cal B}(B_c\rightarrow D^* \tau \bar{\nu}_{\tau})}{{\cal B}(B_c\rightarrow D^* \mu \bar{\nu}_{\mu})} = 0.55\pm 0.05, 
    \end{split}
\end{equation}
that is - ratios of branching fractions of semileptonic decays including a $\tau$ lepton in a final state to the branching fractions including a muon in a final state. Once measured, the ratios in~(\ref{eq:ratios}) will serve as a an additional test of the lepton flavour universality in $B_c$ decays. 
%
%
Considering this ratio, we note that although the models discussed in e.g. \cite{Huang:2007kb}, \cite{Kiselev:2002vz}, \cite{Dubnicka:2017job, Issadykov:2017wlb, Ivanov:2006ni}, and \cite{Wang:2014yia} apply different approaches in the calculation of the form factors,
we agree well with the predictions of \cite{Dubnicka:2017job, Issadykov:2017wlb, Wang:2014yia, Ivanov:2006ni} for the ratios $R_c(D^0)$ and $R_c(D^*)$, and quite disagree with \cite{Kiselev:2002vz}. One can also notice that, in spite of the huge difference between our form factors and decay width values and the ones reported in~\cite{Huang:2007kb} we still agree quite well on the $R_c(D^*)$ value, while their $R_c(D^0)$ seems to be somewhat larger. The main disagreement with our results is visible when we compare with the previous 3ptSR calculation of \cite{Kiselev:2002vz}. There, the authors accounted for Coulomb interactions which were modeled to be very large and consequently have driven the $B_c \to D^{(*)}$ form factors to large values, hardly compatible with any of the models above. The origin of the discrepancy was already discussed on p. 9, above Table~\ref{tab:ffactor}. Here, we just remind the reader that these corrections, aside from enlarging the form factor magnitudes, might also alter their $q^2$ scaling - which in turn might impact the ratios significantly. Also, the decay constants used in \cite{Kiselev:2002vz} (known at that time) are significantly smaller, which additionally increased their results.

\begin{figure}[t]
        \centering
        \includegraphics[width=0.46\textwidth]{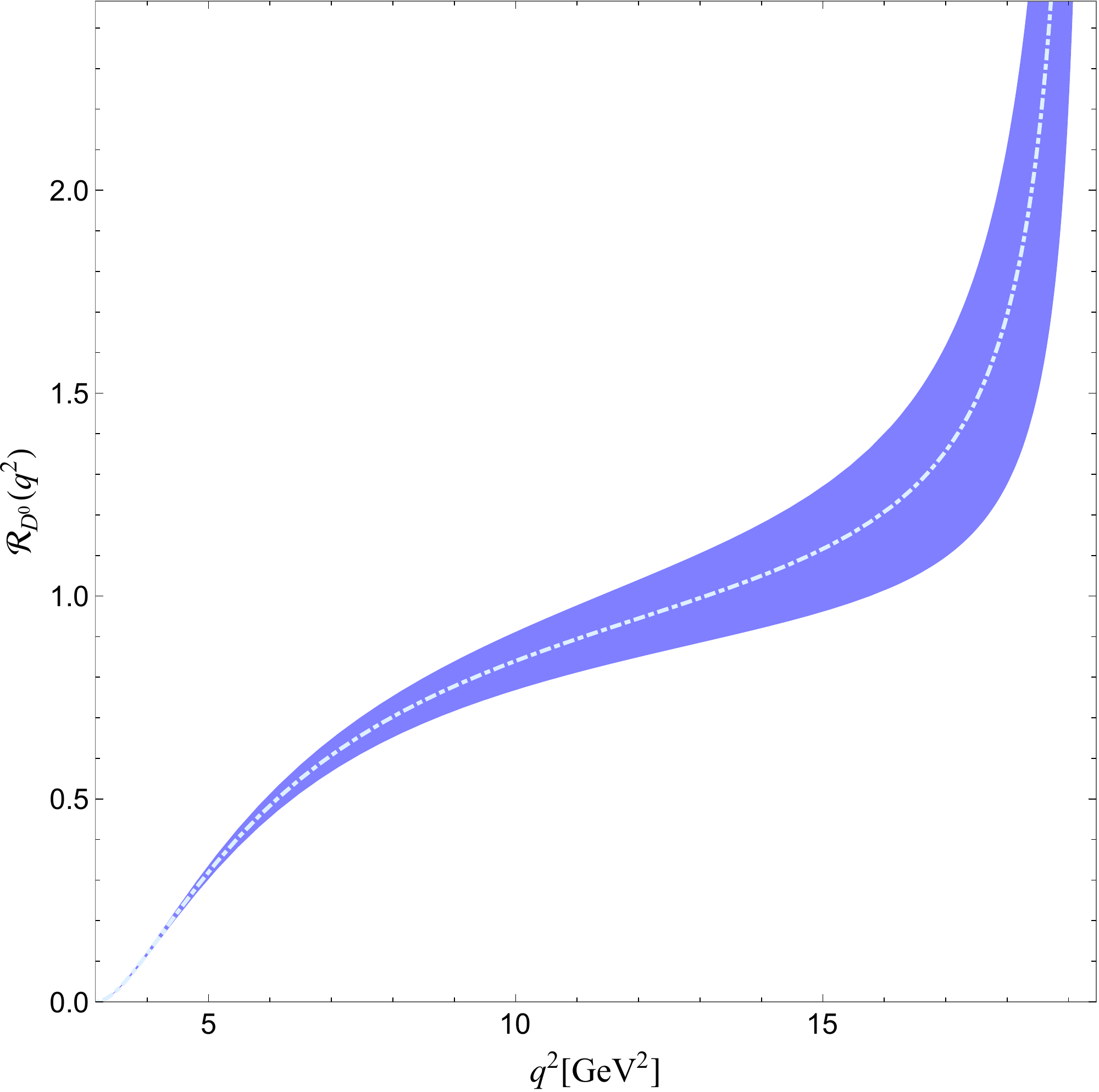}
        \hspace*{0.5cm}
        \includegraphics[width=0.46\textwidth]{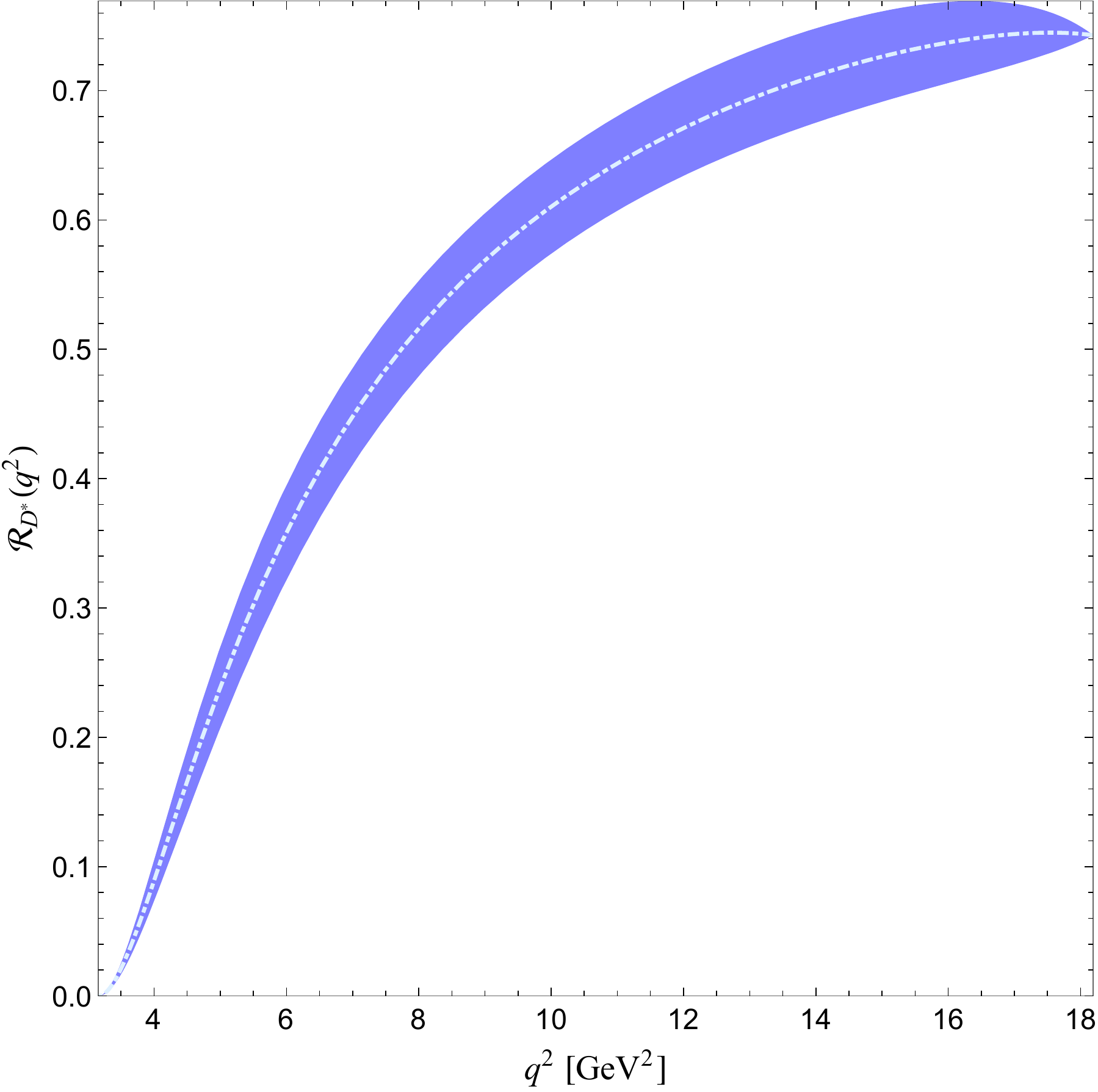}
        \caption{Ratios of differential partial decay rates $B_c$ semileptonic transitions to $D$ (left), and to $D^*$ (right) with tau in the final state to the case with muon in the final state.}
        \label{fig:ddR}
\end{figure}

The $q^2$ distributions of differential forms of $\dd R_c(D^0)$ and $\dd R_c(D^*)$ (which are just ratios of partial differential decay rates, as opposed to integrated rates) are shown in Figure \ref{fig:ddR}.
%

Further on, we define three angular observables for the decay $B_c\to D^0 l \bar{\nu}_{l}$, namely the forward-backward asymmetry  $A^{D^0\!,\,l}_{\mathrm{FB}}(q^2)$, the polarization asymmetry of the lepton $l$,  $P^{D^0\!,\,l}(q^2)$, and  the so-called convexity parameter $C_{F}^{D^0\!,\,l}(q^2)$ as:

\begin{equation}
    \begin{split}
        A^{D^0\!,\,l}_{\mathrm{FB}}(q^2) & =\frac{\bigg(\int\displaylimits_{0}^1-\int\displaylimits_{-1}^0\bigg)\dd\cos\theta \frac{\dd^2\Gamma(B_c\rightarrow D^0 l \bar{\nu}_{l})}{\dd q^2\,\dd \cos\theta}}{\frac{\dd \Gamma(B_c\rightarrow D^0 l \bar{\nu}_{l})}{\dd q^2}} = \frac{3m_l^2}{2q^2}\frac{\mathrm{Re}\big[h_0(q^2)h_t^*(q^2)\big]}{\big(1+\frac{m_l^2}{2q^2}\big)|h_0(q^2)|^2+\frac{3m_l^2}{2q^2}|h_t(q^2)|^2},
        \\
        P^{D^0\!,\,l}(q^2) & =\frac{\frac{\dd \Gamma^+(B_c\rightarrow D^0 l \bar{\nu}_{l})}{\dd q^2}-\frac{\dd \Gamma^-(B_c\rightarrow D^0 l \bar{\nu}_{l})}{\dd q^2}}{\frac{\dd \Gamma(B_c\rightarrow D^0 l \bar{\nu}_{l})}{\dd q^2}} = \frac{\frac{m_l^2}{2q^2}\big[|h_0(q^2)|^2+3|h_t(q^2)|^2\big]-|h_0(q^2)|^2}{\frac{m_l^2}{2q^2}\big[|h_0(q^2)|^2+3|h_t(q^2)|^2\big]+|h_0(q^2)|^2},
        \\
        C_{F}^{D^0\!,\,l}(q^2) = & \frac{1}{\frac{\dd \Gamma(B_c\rightarrow D^0 l \bar{\nu}_{l})}{\dd q^2}}\frac{\dd^2}{\dd(\cos\theta)^2}\bigg[\frac{\dd^2\Gamma(B_c\rightarrow D^0 l \bar{\nu}_{l})}{\dd q^2\,\dd \cos\theta}\bigg]=\frac{3}{2}\frac{|h_0(q^2)|^2\big(\frac{m_l^2}{q^2}-1\big)}{\big(1+\frac{m_l^2}{2q^2}\big)|h_0(q^2)|^2+\frac{3m_l^2}{2q^2}|h_t(q^2)|^2}.
    \end{split}
    \label{eq:observD}
\end{equation}

\begin{figure}[H]
        \captionsetup{margin=0.8cm}
        \centering
        \includegraphics[width=0.32\textwidth]{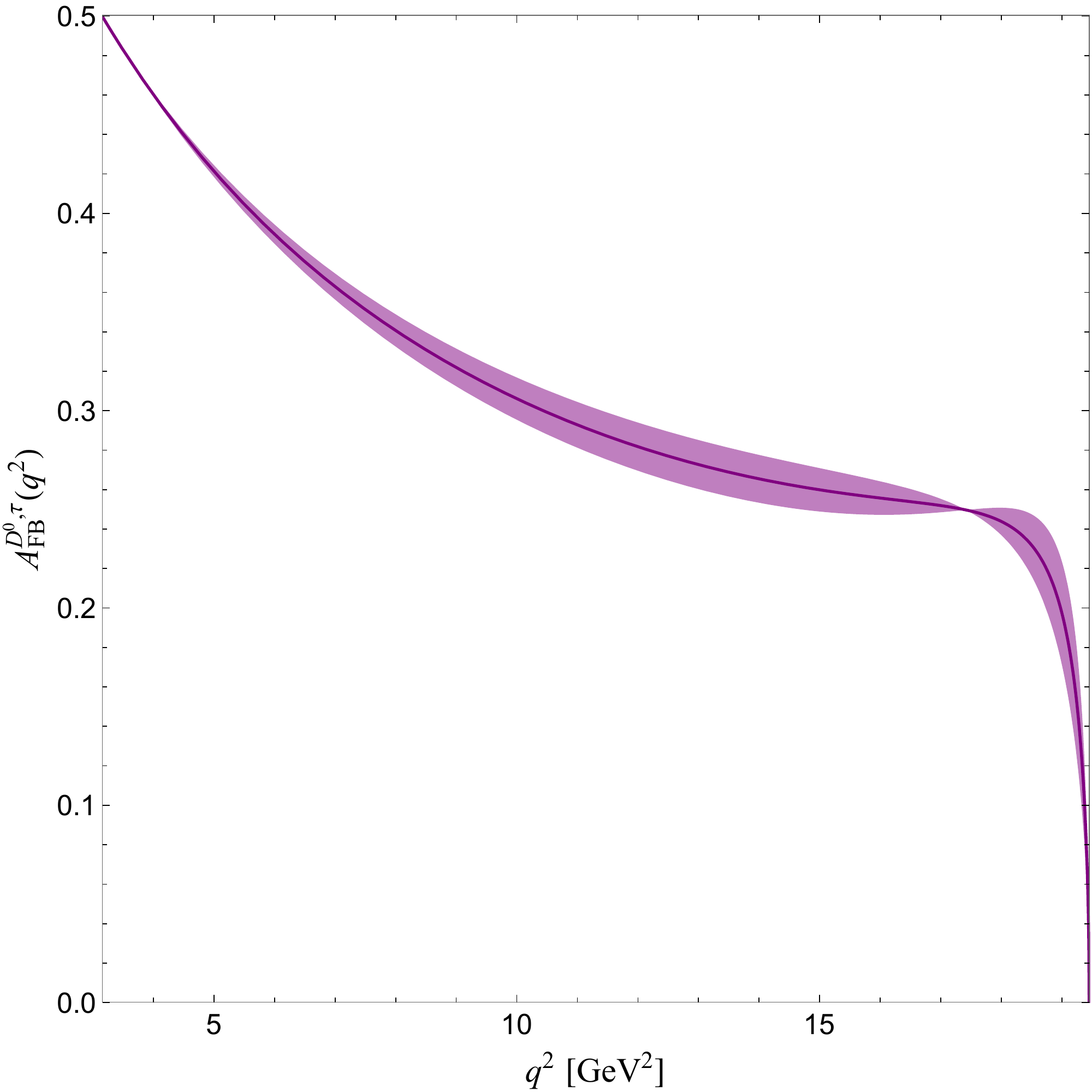}
        \includegraphics[width=0.32\textwidth]{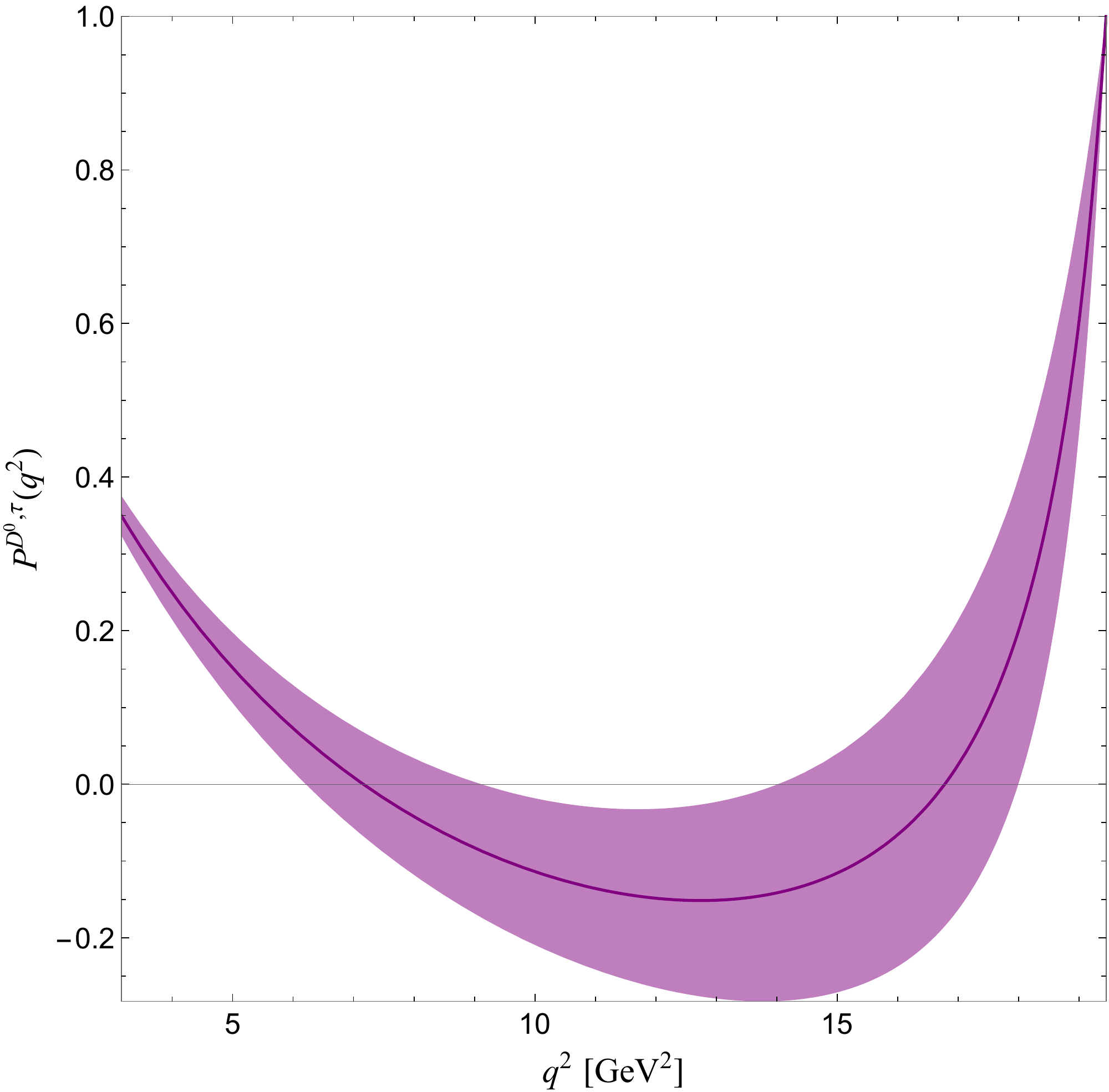}
        \includegraphics[width=0.32\textwidth]{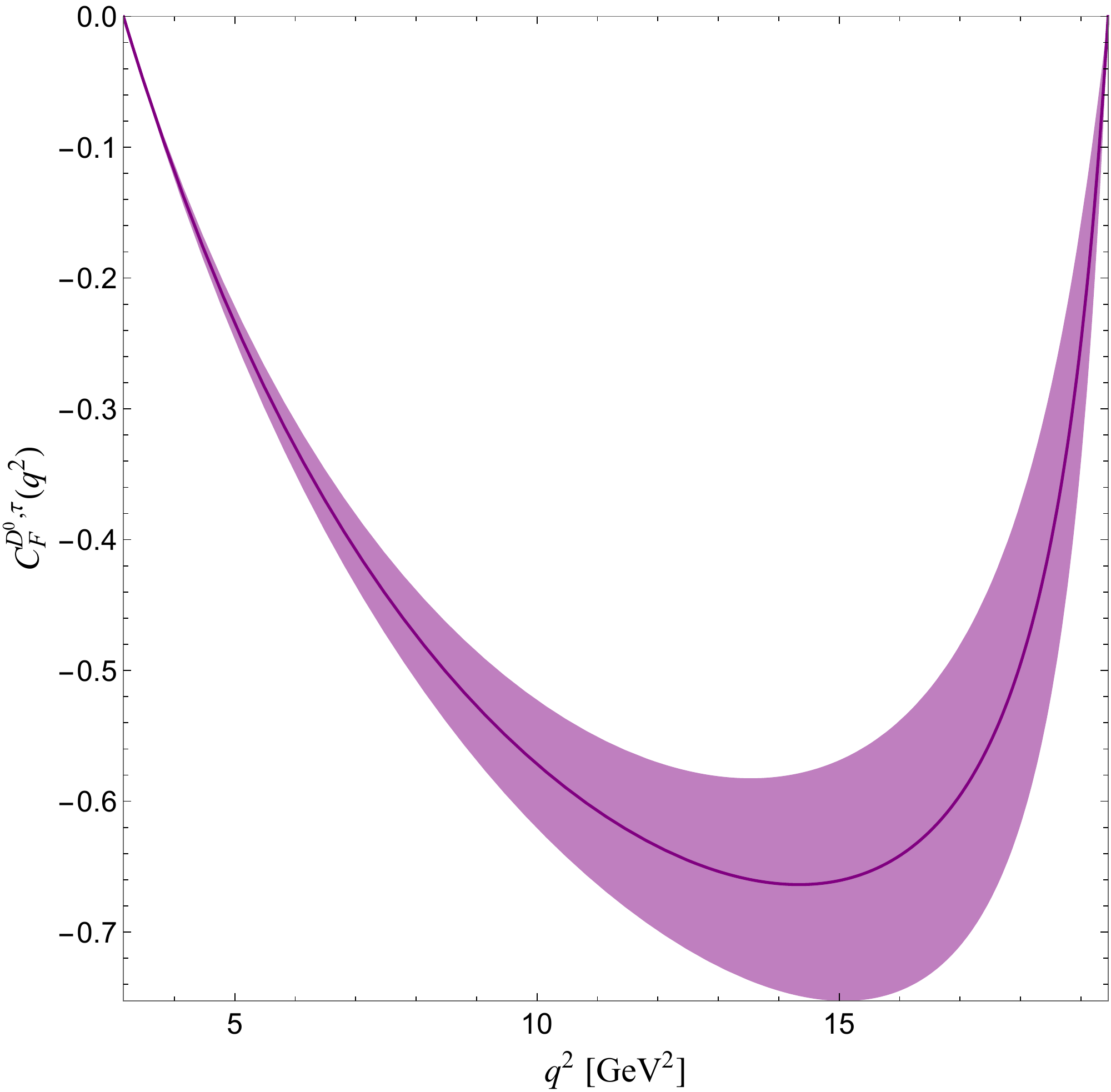}
        \caption{Angular observables defined in (\ref{eq:observD}) for $B_c\to D^0\tau\bar{\nu}_{\tau}$.}
        \label{fig:angularD0}
    \end{figure}

\begin{figure}[H]
    \centering
    \includegraphics[width=0.45\textwidth]{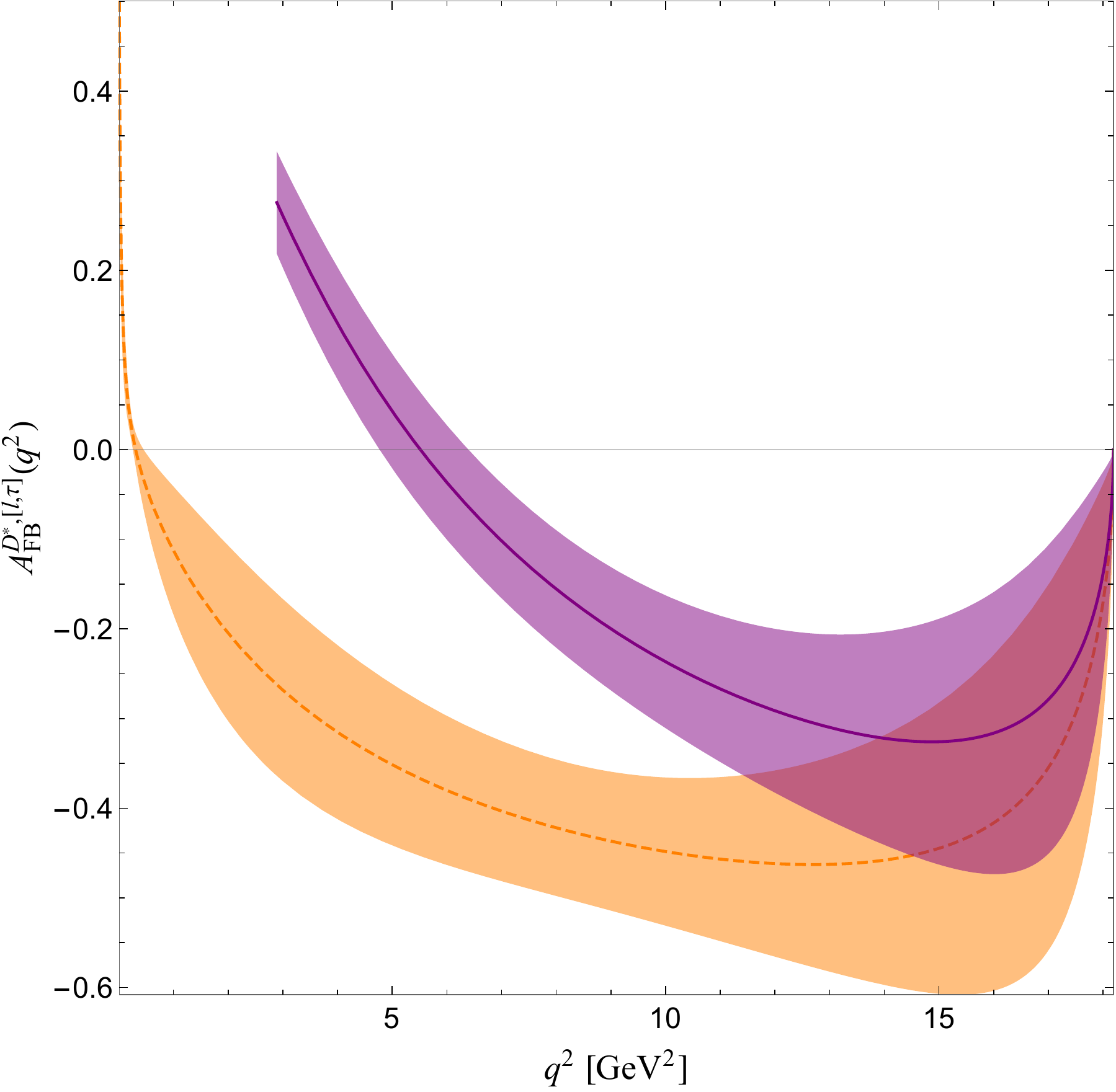}
        \hspace*{0.2cm}
    \includegraphics[width=0.45\textwidth]{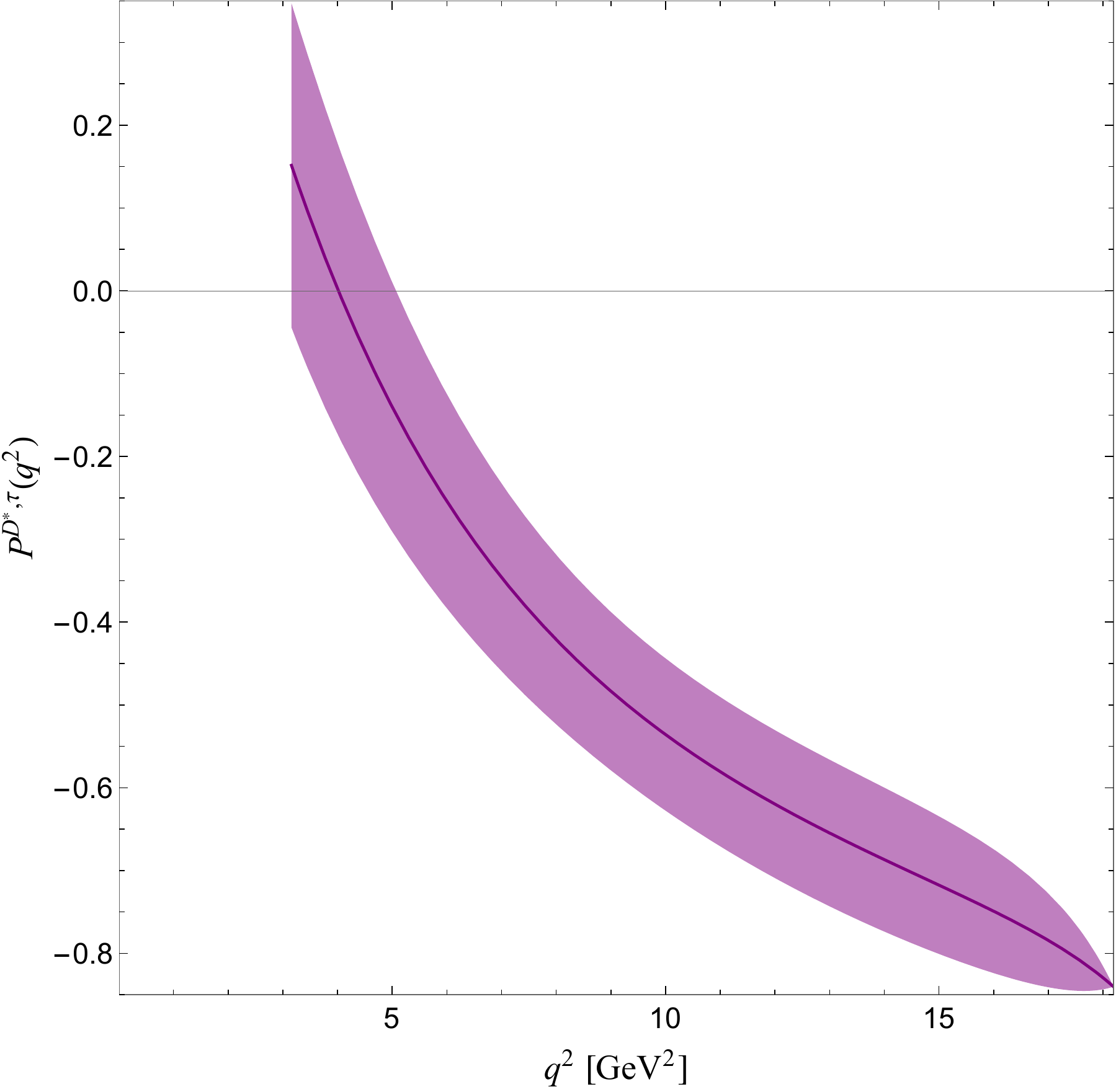}
    \includegraphics[width=0.45\textwidth]{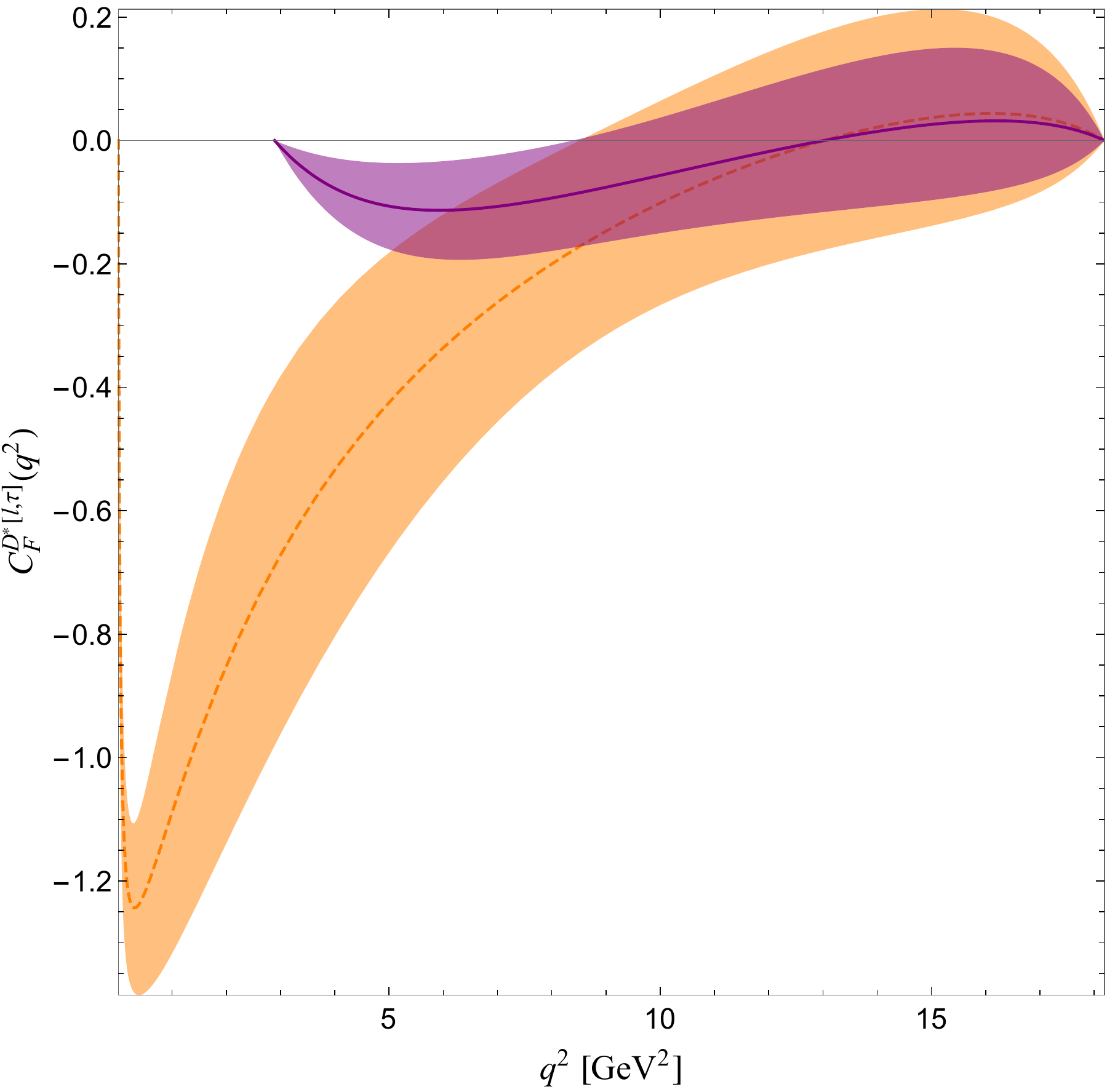}
        \hspace*{0.2cm}
    \includegraphics[width=0.45\textwidth]{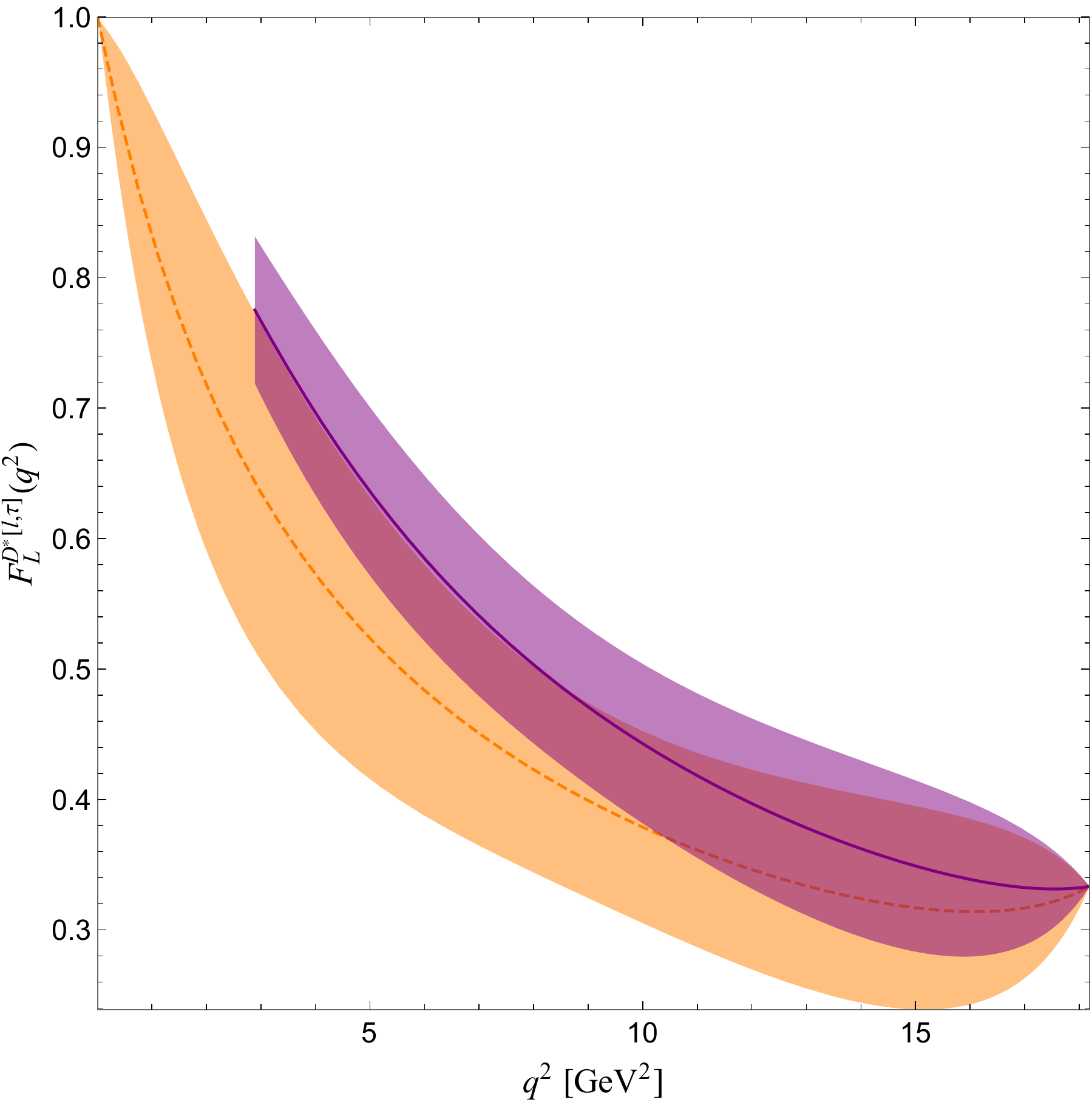}
\caption{Angular observables defined in (\ref{eq:observDstar}) with errors for $B_c\to D^*(l,\tau)\bar{\nu}_{(l,\tau)}$, where now $l$ stands for light leptons, a case depicted by a dashed-line/area in orange, whereas the case with the final $\tau$ state is depicted in purple.}
    \label{fig:angularDstar}
\end{figure}

We stress that the arbitrary choice of the lepton angle can change the value of the forward-backward asymmetry, so one should be careful when referring to its definition. In Figure \ref{fig:angularD0} we plot the observables just for the case with the $\tau$ lepton in the final state, since the asymmetries with light leptons in the final state are basically constant in the entire $q^2$ range (with the exception of extreme upper and lower kinematical limits):
\begin{equation}
    A^{D^0,\,\mu/e}_{\mathrm{FB}}(q^2) \approx 0\,, \qquad P^{D^0,\,\mu/e}(q^2) \approx - 1\,, \qquad C_{F}^{D^0,\,\mu}(q^2) \approx - \frac{3}{2} \,.
\end{equation}
For the case of $B_c\to D^* l \bar{\nu}_{l}$ we similarly have  
\begin{equation}
    \begin{split}
        A^{D^*\!,\,l}_{\mathrm{FB}}(q^2) & =\frac{\bigg(\int\displaylimits_{0}^1-\int\displaylimits_{-1}^0\bigg)\dd\cos\theta \frac{\dd^2\Gamma(B_c\rightarrow D^* l \bar{\nu}_{l})}{\dd q^2\,\dd \cos\theta}}{\frac{\dd \Gamma(B_c\rightarrow D^* l \bar{\nu}_{l})}{\dd q^2}} \\ & = -\frac{3}{4}\frac{|H_+(q^2)|^2-|H_-(q^2)|^2-2\frac{m_l^2}{q^2}\mathrm{Re}\big[H_0(q^2)H_t^*(q^2)\big]}{(|H_+(q^2)|^2+|H_-(q^2)|^2+|H_0(q^2)|^2)\big(1+\frac{m_l^2}{2q^2}\big)+\frac{3}{2}\frac{m_l^2}{q^2}|H_t(q^2)|^2}\, ,\\
        P^{D^*\!,\,l}(q^2) & =\frac{\frac{\dd \Gamma^+(B_c\rightarrow D^* l \bar{\nu}_{l})}{\dd q^2}-\frac{\dd \Gamma^-(B_c\rightarrow D^* l \bar{\nu}_{l})}{\dd q^2}}{\frac{\dd \Gamma(B_c\rightarrow D^* l \bar{\nu}_{l})}{\dd q^2}} \\
        & = -1+\frac{m_l^2}{q^2}\frac{|H_+(q^2)|^2+|H_-(q^2)|^2+|H_0(q^2)|^2+3|H_t(q^2)|^2}{(|H_+(q^2)|^2+|H_-(q^2)|^2+|H_0(q^2)|^2)\big(1+\frac{m_l^2}{2q^2}\big)+\frac{3}{2}\frac{m_l^2}{q^2}|H_t(q^2)|^2} \, ,\\
        C^{D^*\!,\,l}_{F}(q^2) & = \frac{1}{\frac{\dd \Gamma(B_c\rightarrow D^* l \bar{\nu}_{l})}{\dd q^2}}\frac{\dd^2}{\dd(\cos\theta)^2}\bigg[\frac{\dd^2\Gamma(B_c\rightarrow D^* l \bar{\nu}_{l})}{\dd q^2\,\dd \cos\theta}\bigg]\\
        & = \frac{3}{4}\bigg(1-\frac{m_l^2}{q^2}\bigg)\frac{|H_+(q^2)|^2+|H_-(q^2)|^2-2|H_0(q^2)|^2}{(|H_+(q^2)|^2+|H_-(q^2)|^2+|H_0(q^2)|^2)\big(1+\frac{m_l^2}{2q^2}\big)+\frac{3}{2}\frac{m_l^2}{q^2}|H_t(q^2)|^2}\,, \\
        F_{L}^{D^*\!,\,l}(q^2) & = \frac{\frac{\dd\Gamma_L(B_c\rightarrow D^* l \bar{\nu}_{l})}{\dd q^2}}{\frac{\dd\Gamma(B_c\rightarrow D^* l \bar{\nu}_{l})}{\dd q^2}} = \frac{|H_0(q^2)|^2+3|H_t(q^2)|^2/(1 + \frac{2q^2}{m_l^2})}{|H_+(q^2)|^2+|H_-(q^2)|^2+|H_0(q^2)|^2+3|H_t(q^2)|^2/(1 + \frac{2q^2}{m_l^2})} \, ,\\
    \end{split}
    \label{eq:observDstar}
\end{equation}
where in addition we compute the longitudinal polarization fraction of $D^*$, $F_L^{D^*}$, in the decay. The results for these observables are shown in Figure \ref{fig:angularDstar}. 
Similarly to the prior case one observable is approximately constant
\begin{equation}
    P^{D^*\!,\,\mu/e}(q^2)\approx - 1 
\end{equation}
and it it not shown there.  Integrated values of $B_c \to D^*$ angular observables are given in Table \ref{tab:iang}. Again, proving that these observables are relatively independent of the hadronic form factors, good agreement with the recent analysis provided in~\cite{Dutta:2018vgu}, where the LFQM form factors from~\cite{Wang:2008xt} are used, is found.
\begin{center}
\addtolength{\tabcolsep}{-3pt}
\renewcommand{\arraystretch}{1.5}
    \begin{tabular}{||c | c  c || c | c c ||}
        \hline
                & $l=\mu$ &\hspace*{0.5cm} $l=\tau$ &  & $l=\mu$ &\hspace*{0.4cm} $l=\tau$\\
        \hline\hline
            $ A^{D^*\!,\,l}_{\mathrm{FB}}$ & $-0.4\pm 0.2$ &\hspace*{0.4cm} $-0.3\pm 0.2$ & $ A^{D^0\!,\,l}_{\mathrm{FB}}$ & $ \approx 0$ &\hspace*{0.4cm} $0.30\pm 0.06$ \\
            $P^{D^*\!,\,l}$ & $\approx-1$ &\hspace*{0.4cm} $-0.6\pm 0.4$ & $P^{D^0\!,\,l}$ & $\approx -1$ &\hspace*{0.4cm} $-0.1\pm 0.1$ \\
            $C^{D^*\!,\,l}_{F}$ & $-0.2\pm 0.2$ &\hspace*{0.4cm} $-0.02\pm 0.08$ &  $C^{D^0\!,\,l}_{F}$ & $\approx -\frac{3}{2}$ &\hspace*{0.4cm} $-0.6\pm 0.1$ \\
            $F_{L}^{D^*\!,\,l}$ & $0.4\pm 0.2$ &\hspace*{0.4cm} $0.4\pm 0.3$ & - & - & \hspace*{0.5cm}- \\
        \hline
\end{tabular}
\captionof{table}{Angular observables integrated over the entire kinematic region.}
\label{tab:iang} 
\end{center}
\subsection{$|V_{ub}|$ determination and the $|V_{ub}|/|V_{cb}|$ ratio} \label{Vub-deter}

We propose to determine $|V_{ub}|$ by measuring the decay width of $B_c \to D^{} \mu \bar{\nu}_{\mu}$.  Here we give our estimates for 
\begin{equation}
    \zeta_{D^0}|V_{ub}|^2\equiv \Gamma(B_c\rightarrow D^0 \mu \bar{\nu}_{\mu})
\end{equation}
as:
\begin{eqnarray}
        \zeta_{D^0} = (2.0 \pm 0.3)\cross 10^{-3} \,\mathrm{eV}\,.
        \label{eq:zeta}
\end{eqnarray}
We see that combining our predictions from Table \ref{tab:dwid} with the future experimental data  the $|V_{ub}|$ can be determined from $B_c \to D^0 \mu \nu$  with the theoretical uncertainty of 7.5$\%$. By calculating the same for the semileptonic $B_c \to D^{\ast} \mu \nu$ decay, 
\begin{eqnarray}
    \zeta_{D^*}|V_{ub}|^2 &\equiv& \Gamma(B_c\rightarrow D^* \mu \bar{\nu}_{\mu}), 
    \nonumber \\
    \zeta_{D^*} &=& (5 \pm 2)\cross 10^{-3} \,\mathrm{eV}\,, 
\end{eqnarray}
we see that there the error are much larger there and amount to 20$\%$, which makes this decay at present less suitable  for the $|V_{ub}|$ determination.  

\begin{figure}[H]
        \centering
        \includegraphics[width=0.55\textwidth]{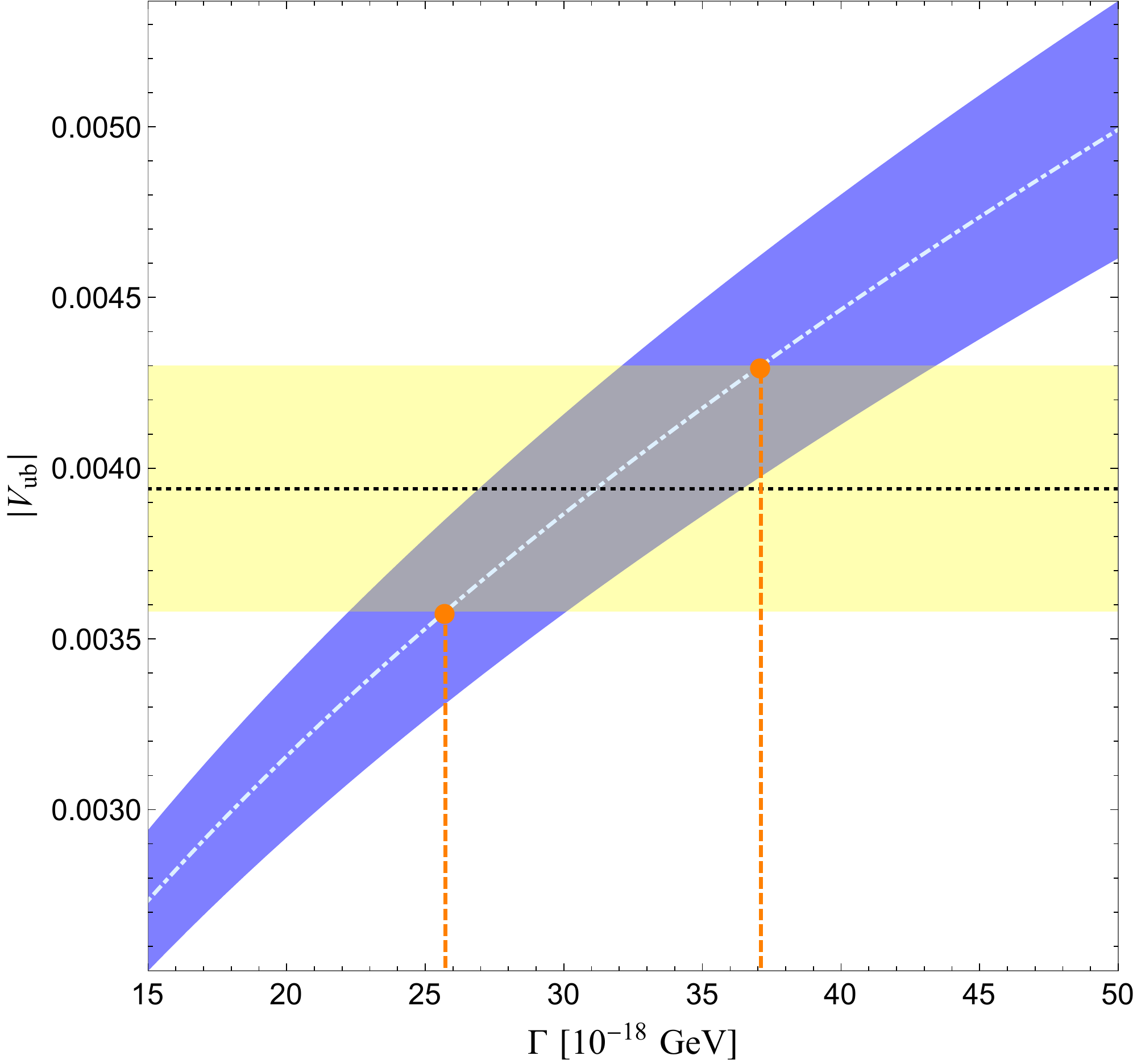}
        \caption{ The prospect for $|V_{ub}|$ determination from  $\Gamma(B_c \to D^{} \mu \bar{\nu}_{\mu})$; the yellow band represents the PDG value of $|V_{ub}|$, Eq.(\ref{eq:CKM2}).}
\label{fig:VubD}
\end{figure}

In Fig.\ref{fig:VubD} we present $|V_{ub}|$ dependence on the decay rate using our calculated value of $\zeta_D$. It is clear that if the decay rate can be measured with $10-20\%$ accuracy, as expected in the LHCb Run II \cite{Marta}, then the extraction of the $V_{ub}$ would be possible at the same level and even more precise.

We also give here the value for 
\begin{eqnarray}
\Delta \zeta_{D^0}(q_1^2,q_2^2) \equiv \frac{1}{|V_{ub}|^2} \int_{q_1^2}^{q_2^2} dq^2 \frac{d\Gamma(B_c\rightarrow D^0 \mu \bar{\nu}_{\mu})}{dq^2}
\label{eq:zetaD0}
\end{eqnarray}
which, employing our predicted $B_c \to D^0$ form factor $f^+(q^2)$ from the 3ptSR at $m_{\mu^2} \le q^2 \le 10$ GeV$^2$, Figure \ref{fig:BCLBGL}, amounts to  
\begin{eqnarray}
\Delta \zeta_{D^0}(m_{\mu}^2,10\; {\rm GeV}^2) = (1.2 \pm 0.1 \pm 0.1)\cross 10^{-3} \,\mathrm{eV}\,,
\end{eqnarray}
where an additional 10\% error has been added, which is our error estimate for the correlation between the QCDSR pseudo-data points in the low $q^2$ region, while we also present the other bins of $\Delta \zeta_{D^0}(q_1^2,q_2^2)$  which can be used together with future experimental data to determine $V_{ub}$ from $B_c\rightarrow D^0 \mu \bar{\nu}_{\mu}$ decays in Table~\ref{tab:deltabin}, Appendix C.2.
The theoretical error of 7.5$\%$ in (\ref{eq:zeta}) might be improved by explicitly adding the $\alpha_s$-corrections to the 3ptSR, which would certainly reduce the main systematic theoretical uncertainty of adjusting $s_0$ sum rule parameter.
In order to suppress the unknown systematic uncertainty in the estimation of $|V_{ub}|$ arising from the method itself, here we define the ratio of branching fractions
\begin{equation}
       \mathcal{R}_{D^0 J/\psi}\equiv \frac{\zeta_{D^0}(q_{\mathrm{min}_1}^2, q_{\mathrm{max}_1}^2)}{\zeta_{ J/\psi}(q_{\mathrm{min}_2}^2, q_{\mathrm{max}_2}^2)} = \frac{1}{     \frac{|V_{ub}|^2}{|V_{cb}|^2}} \frac{\int\displaylimits_{q_{\mathrm{min}_1}^2}^{q_{\mathrm{max}_1}^2}\frac{\dd\Gamma(B_c \to D^0 \mu \bar{\nu}_{\mu})}{\dd q^2}\dd q^2}{\int\displaylimits_{q_{\mathrm{min}_2}^2}^{q_{\mathrm{max}_2}^2}\frac{\dd\Gamma(B_c \to J/\psi \mu \bar{\nu}_{\mu})}{\dd q^2}\dd q^2}\,,
       \label{eq:ratiodef}
\end{equation}
where the form factors entering $\dd\Gamma(B_c\to J/\psi \mu \bar{\nu}_{\mu})/\dd q^2$ are known to some extent from lattice calculation \cite{Colquhoun:2016osw}, and are reproduced to a satisfactory precision by the QCDSR method explained in detail in Sec.2. The form factors for $B_c \to J/\psi$ transition were already briefly presented in \cite{RjpsiOUR}. Here the used parameters differ a little from ones used there, due to a necessary update. The specific values of parameters used in the calculation are listed in Appendix B.1, Table \ref{tab:dconst2}, and the form factors are given in Figure \ref{fig:BcJpsiFF} there.
\begin{figure}[H]
        \centering
        \includegraphics[width=0.55\textwidth]{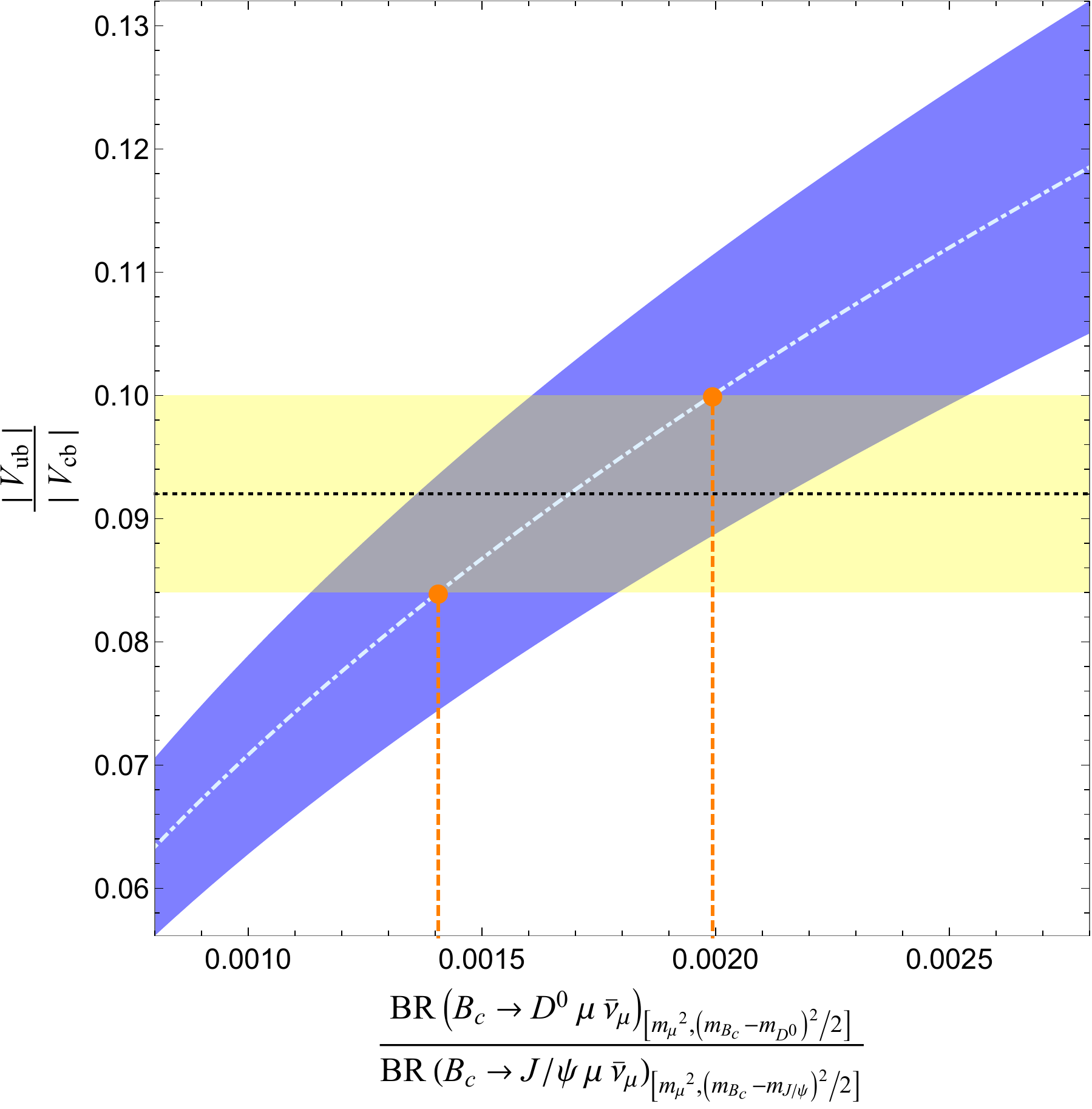}
        \caption{ The prospect for $|V_{ub}|/|V_{cb}|$ determination from  $\mathcal{R}_{D^0 J/\psi}$; the yellow band represents the PDG value of $|V_{ub}|/|V_{cb}|$, Eq.(\ref{eq:CKM3}).}
        \label{fig:VubVcbRAT}
\end{figure}
Also experimentally, due to the very short lifespan of the $B_c$ and a huge background stemming from $B$ decays, measuring the $|V_{ub}|$ quark coupling strength directly is highly challenging \cite{Aaij:2015bfa, LHCb_book}. Therefore, not surprisingly, it turns out that measuring it through the ratio defined above has some benefits from the experimental standpoint as well (such as canceling the production rate uncertainty). Although challenging, the prospects of using Run 1 + Run 2 data at LHCb are indicating  that one could come around 10-20$\%$ uncertainties in $|V_{ub}|/|V_{cb}|$ measurement in an analysis looking at the ${\cal B}(B_c \to  D^0 \mu \bar{\nu}_\mu)/{\cal B}(B_c \to J/\psi \mu \bar{\nu}_\mu)$ ratio~\cite{Marta}. Therefore, the input of the form factors calculated in Sec.2 and the decay rates predicted in a particular $q^2$ region, which we provide in Table \ref{tab:RJDbin} in Appendix C.2, could be valuable information. One can see that by combining our predictions for ${\cal R}_{D^0 J/\psi}$ with future measurements one can achieve the most precise determination of $|V_{ub}|/|V_{cb}|$  in the low-$q^2$ kinematic region in $B_c\to D^0$ transition and moderate-$q^2$ region in $B_c\to J/\psi$ transition.

In Figure \ref{fig:VubVcbRAT} we plot this ratio in the bin defined through $q^2_{\mathrm{min}_1}\!=q^2_{\mathrm{min}_2}\!=m_{\mu}^2$, $q^2_{\mathrm{max}_1} = (m_{B_c}-m_{D^0})^2/2$, and $q^2_{\mathrm{max}_2} = (m_{B_c}-m_{J/\psi})^2/2$, which is approximately the kinematic region in which the QCDSR turn to be most reliable. The current experimental world average of $|V_{ub}|/|V_{cb}|$ is also shown on the plot for comparison of theoretical and future experimental predictions for the ratio of branching fractions directly with the present limit on $|V_{ub}|/|V_{cb}|$.

One should keep in mind that here the differential decay widths are integrated in the lower half of the $q^2$ region of both decay channels and that the form factors used to produce the plot are the result of fits to uncorrelated pseudo-data points. In line with our estimate of the contribution of the correlation among the points to the error budget discussed in  Sec.\ref{sec:corr}, one should assign a further 10\% uncertainty to $|V_{ub}|/|V_{cb}|$ not shown in the plot.
\subsection{The $|V_{ub}|/|V_{cs}|$ ratio}

In \cite{Jenkins:1992nb} the authors propose to use the differential ratio of $BR(B_c \to D^0 l \nu)$ and $BR(B_c \to B_s l \nu)$ at zero recoil defined as
\begin{eqnarray}
 \mathcal{R}_{D^0 B_s}^{\rm max} = \frac{1}{\frac{|V_{ub}|^2}{|V_{cs}|^2}} \frac{\dd\Gamma(B_c^+ \to D^0 e^+ \nu_{e})/\dd q^2|_{q^2 \simeq q^2_{\rm max}}}{ \dd \Gamma(B_c^+ \to B_s e^+ \nu_e)/\dd q^2|_{q^2 \simeq q^2_{\rm max}}} 
 \label{eq:ratioR}
\end{eqnarray} to determine the $|V_{ub}|/|V_{cs}|$ ratio. Experimentally, to measure the ratio $|V_{ub}|/|V_{cs}|$ would be more challenging than $|V_{ub}|/|V_{cb}|$, since the experimental systematics do not nicely cancel in this ratio \cite{Marta}. However, since it was proposed that the theoretical uncertainties coming from the form factors should cancel near the zero recoil point, this could be as well an interesting possibility in the future, which we examine here. 
In that region the heavy quark spin symmetry reduces the number of the form factors of above decays to just one and the differential decay rate ratio in that limit becomes insensitive to the detailed form of $B_c$ wave function and proportional just to a ratio of the final meson masses and decay constants.
Namely, in the heavy quark effective theory (HQET) the following parametrization is valid:
    \begin{equation}
    \begin{split}
		\bra{D^0(v,q^{\prime})}V_{\mu}(q^2)\ket{B_c(v)} & = 2\sqrt{m_{B_c}m_{D^0}}\big[\Sigma_1(a_0q^{\prime})v_{\mu}+a_0\Sigma_2(a_0q^{\prime})q^{\prime}_{\mu}\big],\\
		\bra{B_s(v,q^{\prime})}V_{\mu}(q^2)\ket{B_c(v)} & = 2\sqrt{m_{B_c}m_{B_s}}\big[\Omega^s_1(a_0q^{\prime})v_{\mu}+a_0\Omega^s_2(a_0q^{\prime})q^{\prime}_{\mu}\big],
		\end{split}
	\end{equation}
where $v$ is the velocity of the $B_c$ meson, and $q^{\prime}$ is a small residual velocity carried by the final state meson (denoted such as to avoid confusion with $q$, the momentum carried by the lepton pair system), so that
\begin{equation}
    \begin{split}
        p_{1\mu}=m_{B_c}v_{\mu}; \quad p_{2\mu}=m_{f}v_{\mu} + q^{\prime}_{\mu},
    \end{split}
\end{equation}
 with $m_f=[m_{D^0},m_{B_s}]$. The parameter $a_0$ is connected to the Bohr radius of the $B_c$ meson and will not be discussed here. The form factors $\Sigma_2(a_0q^{\prime})$ and $\Omega^s_2(a_0q^{\prime})$ are irrelevant for this discussion, as they do not contribute around the zero-recoil region $(q^{\prime 2}=0)$, so in principle one could deduce about the differential branching fractions near zero recoil just from the form factors $\Sigma_1(a_0q^{\prime})$ and $\Omega^s_1(a_0q^{\prime})$. In \cite{Jenkins:1992nb} it is argued that, owing to this fact, and by considering the heavy-quark spin symmetry for the remained form factors which one can write as
 \begin{equation}
 \label{eq:HQETFF}
    \begin{split}
        \Sigma_1(a_0q^{\prime})&=\frac{1}{\sqrt{2}}f_{D^0}\sqrt{m_{D^0}}\int\dd^3x\,\mathrm{e}^{\ii\vec{q}^{\prime}\cdot\vec{x}}\Psi(x),\\
        \Omega^s_1(a_0q^{\prime})&=\frac{1}{\sqrt{2}}f_{B_s}\sqrt{m_{B_s}}\int\dd^3x\,\mathrm{e}^{\ii\vec{q}^{\prime}\cdot\vec{x}}\Psi(x),
    \end{split}
\end{equation}
where $\Psi(x)$ is the $B_c$ meson wave function, the ratio at the zero-recoil, 
\begin{equation}
    R_{\mathrm{FF}}=\frac{\Sigma_1(a_0q^{\prime}\approx 0)}{\Omega^s_1(a_0q^{\prime}\approx 0)}
    \label{eq:RFF}
\end{equation}
should in principle very weakly depend on the particular shape of the wave function, due to its cancellation, so that in the heavy quark limit
\begin{equation}
    \label{eq:HQETFF2}
    R_{\mathrm{FF}}^{\mathrm{HQ}}\approx \frac{f_{D^0}}{f_{B_s}}\sqrt{\frac{m_{D^0}}{m_{B_s}}}\approx 0.53.
\end{equation}
We provide in Appendix B.1 some details on  calculation of $B_c\to B_s$ form factors
obtained analogously to the ones of the $B_c\to D^0$ transition 
and by using them we obtain
\begin{equation}
   \label{eq:HQETFF2our}
    R_{\mathrm{FF}}^{\mathrm{our}} = 0.8\pm 0.3,
\end{equation}
which can be also compared with the result from \cite{Colangelo:1999zn}, where the wave functions have been calculated in the framework of a HQET-inspired quark model, explicitly,  
\begin{equation}
   \label{eq:HQETFF2deF}
    R_{\mathrm{FF}}= 0.89 \,.
\end{equation}
This is indeed very close to our result and we can conclude that the heavy quark spin symmetry relations are obeyed in our calculation. This also agrees well with values extracted from other quark models \cite{Ebert:2003cn, Nobes:2000pm, Wang:2008xt}. However, the error in our calculation is quite large, since the form factors in the two decays are not very correlated, and because one can reliably use the sum rules for the $B_c\to B_s$ case only very close to the maximum recoil region, as can be noticed from Figure \ref{fig:BcBsFF}, due to the occurrence of non-Landau singularities. Lattice input might prove to be useful here in order to extrapolate to higher order $z$ terms for both decays and 
with more theoretical input the extraction of the $|V_{ub}|/|V_{cs}|$ ratio from (\ref{eq:ratioR}) could be viable.   

\section{Conclusion}

In this paper we have discussed semileptonic $B_c \to D^{(\ast)}$ decays and examined the possibility to extract the CKM parameter $|V_{ub}|$ from these decays based on the future LHCb experimental data. It was shown that specially from the $B_c \to D^0 \mu \bar{\nu}_{\mu}$ decay the competitive extraction of $|V_{ub}|$ matrix element might be possible.    

For the extraction of $|V_{ub}|$, it is important to know precisely the $B_c \to D^{0}$ form factors since the predictions for the light leptons in the final state come out essentially proportional to  $|f_+(q^2)|^2 |V_{ub}|^2$.
We have calculated the $B_c \to D^0$ form factors $f_+(q^2)$ and $f_0(q^2)$ and the $B_c \to D^{\ast}$ form factors $V(q^2),A_1(q^2),A_0(q^2)$ and $A_2(q^2)$ using the three-point QCD sum rules. The form factors are then theoretically confined in the region of $q^2 \leq 10$ GeV$^2$. The extrapolation to higher $q^2$ values is discussed for the BGL and BCL $z$-series and final predictions are 
given for the BCL parametrization of form factors, summarized in Table \ref{tabular:res} and Figures \ref{fig:BCLBGL} and \ref{fig:BCLBGL2}.

We present the $q^2$ differential decay rate distributions (divided by $|V_{ub}|^2$) for both light ($e$ or $\mu$) and $\tau$ lepton in the final state and also give our predictions for various angular observables in the $B_c \to D^{(\ast)}$ semileptonic transitions in Figures~\ref{fig:angularD0} and~\ref{fig:angularDstar}, which can be useful to further  scrutinize the SM predictions for these decays: like the forward-backward asymmetry $A_{FB}^l(q^2)$, the lepton polarization $P_L^l(q^2)$, convexity parameter $C_F^l(q^2)$ and the $D^*$ meson longitudinal polarization fraction, $F_L^{D^*}$ in $B_c \to D^{\ast}$ decays. 
In addition we provide the values of the ratios of branching fractions of the semileptonic decays to a $\tau$ lepton
to the branching fractions to a muon, $R_c(D^0) = 0.64\pm 0.05$ and $R_c(D^{\ast})= 0.55\pm 0.05$, for testing the lepton favour universality violation in these semileptonic $B_c$ channels, with the $q^2$ distributions shown in Figure \ref{fig:ddR}. 

The possibility of determining the $|V_{ub}|$ CKM matrix element from 
$B_c \to D^{(\ast)}$ decays is carefully studied and we have found that the $|V_{ub}|$ can be determined with the uncertainty of 7.5$\%$ from the $B_c \to D^0 \mu \bar{\nu}_{\mu}$ decay. 
Experimentally there are good prospects for this measurement. The $B_c$ decays will be extensively investigated at LHCb in the Upgrade II \cite{LHCb_book}. With approximately 30,000 reconstructed $B_c \to D^0 l \nu$ decays which can be expected with the 300 fb$^{-1}$ Upgrade II dataset, the  competitive extraction  of $|V_{ub}|$ from $B_c \to D^0$ semileptonic decays can be expected. 
By normalizing $B_c \to D^0 \mu \bar{\nu}_{\mu}$ to $B_c \to J/\psi \mu \bar{\nu}_{\mu}$
the ratio $|V_{ub}|/|V_{cb}|$ could be experimentally extracted with $10-20\%$ of uncertainty \cite{Marta}, which could be also theoretically achieved with the calculated from factors, as shown. We give the binned
distributions of the precision observables $\Delta \zeta_{D^0}$ and ${\cal R}_{D^0 J/\psi}$ in Appendix C.2. 

It was further demonstrated that numerically our form factors do obey the behaviour imposed on them by the heavy quark spin symmetry, as dictated by the ratio of $B_c\to D^0$ and $B_c\to B_s$ transitions \cite{Jenkins:1992nb}. Although the precision is still not satisfactory enough, this opens up new possibilities in terms of extraction of the $|V_{ub}|/|V_{cs}|$ ratio, even if experimentally this will be extremely challenging.

We hope the analysis of the semileptonic $B_c \to D^{(*)}$ decays and the perspective for $|V_{ub}|$ measurement in these decays might contribute to the resolution of the problem of the persisting discrepancy among determinations of $|V_{ub}|$ from exclusive and inclusive $b \to u$ transitions. 


\appendix 
\section{Three-point sum rule contributions to form factors}
\label{app:A1}
\subsection{Perturbative contributions}
The perturbative part is calculated by imposing the Cutkosky rules to calculate  simultaneously discontinuities in the $p_1^2$ and the $p_2^2$ channels of amplitudes (Figure \ref{fig:diags}a), ${\rm Im}_{s_1,s_2}\Pi_{i}(s_1,s_2,q^2)$, and then using the double dispersion relation
\begin{equation}
\Pi_{i}(p_1^2,p_2^2,q^2) = -\frac{1}{(2\pi)^2}\int\!\!\!\int\frac{\rho_{i}(s_1,s_2,q^2)}{(s_1-p^2_1)(s_2-p_2^2)}ds_1ds_2\,,
\end{equation}
where $\rho_{i}(s_1,s_2,q^2) = (-4){\rm Im}_{s_1,s_2}\Pi_{i}(s_1,s_2,q^2)$.
The integration is performed after the Borel transformations in both channels 
\begin{equation}
\mathcal{B}_{-p_1^2}(M_1^2) \mathcal{B}_{-p_2^2}(M_2^2) \rightarrow -\frac{M_1^{-2} M_2^{-2}}{(2\pi)^2}\int\!\!\!\int\rho_{i}(s_1,s_2,q^2) e^{-\frac{s_1}{M_1^2}-\frac{s_2}{M_2^2}} ds_1ds_2
\end{equation}
over a phase space up until some effective thresholds $s^0_1$ and $s^0_2$. They are evaluated by requiring that the decay constants, calculated in the QCD sum rule approach reproduce the lattice results. The final expressions for imaginary parts are
\begin{equation}
\begin{split}
&\rho_{P,1} = \frac{3}{\lambda^{3/2}}\bigg[\big(m_c(m_u-m_c)+\frac{s_2+m_c^2-m_u^2}{2}\big)\lambda \\
& +\mathcal{M}(s_1,s_2,q^2)\big(2s_2(s_1+m_c^2-m_b^2)-(s_2+m_c^2-m_u^2)(s_1+s_2-q^2)\big)\bigg],\\
&\rho_{P,2} = \frac{3}{\lambda^{3/2}}\bigg[\big(m_c(m_b-m_c)+\frac{s_1+m_c^2-m_b^2}{2}\big)\lambda \\
& +\mathcal{M}(s_1,s_2,q^2)\big(2s_1(s_2+m_c^2-m_u^2)-(s_1+m_c^2-m_b^2)(s_1+s_2-q^2)\big)\bigg],\\
\end{split}
\label{eq:appA1}
\end{equation}
for the pseudoscalar case, where 
\begin{equation}
\begin{split}
& \lambda \equiv\lambda(s_1,s_2,q^2) = (s_1+s_2+q^2)^2 - 4s_1s_2\,, \\
\mathcal{M}(s_1,s_2,q^2) & = -m_c^2 - m_b m_u + m_b m_c + m_u m_c + (s_1 + s_2 - q^2)/2\,,
\end{split}
\end{equation}
and for the vector case
\begin{eqnarray}
\rho_{0}^A &=& \frac{3}{2\big[\lambda(s_1,s_2,q^2)\big]^{\frac{3}{2}}} \bigg\{ 2(m_b-m_c) \left ( m_c^2\lambda(s_1,s_2,q^2)+s_1\Delta_2^2+s_2\Delta_1^2-\Delta_1\Delta_2 u
\right )\nonumber \\
&& + \lambda(s_1,s_2,q^2) \bigg( 2 m_c^2(m_c-m_u-m_b)+ m_c (u + 2 m_b m_c )\nonumber \\
&&  - (m_c-m_u)\Delta_1 - (m_c-m_b)\Delta_2 \bigg )
\bigg\}\,, \nonumber \\
\frac{1}{2}(\rho_{1}^A+\rho_{2}^A)&=&  \frac{3}{2\big[\lambda(s_1,s_2,q^2)\big]^{\frac{5}{2}}}\bigg\{\big[(m_u-m_c)\big(2s_1\Delta_2-\Delta_1u\big) +(m_b-m_c)\big(2s_2\Delta_1-\Delta_2 u - \nonumber \\
&&  -2m_c^2(u - 2 s_2)\big) - 2 m_c (2s_2\Delta_1-\Delta_2u) \big] \lambda(s_1,s_2,q^2) +
\nonumber \\
&& + 2(m_b-m_c)\big[\Delta_1\Delta_2(2u^2 + 4 s_1 s_2 - 6s_2u) -3u(s_1 \Delta_2^2 + s_2\Delta_1^2) + 6s_2^2\Delta_1^2 + \nonumber \\
&& + 2 s_1 s_2\Delta_2^2 + \Delta_2^2u^2\big] - m_c\big[\lambda(s_1,s_2,q^2)\big]^2 \bigg\} \,, \nonumber \\
\frac{1}{2}(\rho_{1}^A-\rho_{2}^A) &=& \frac{3}{2\big[\lambda(s_1,s_2,q^2)\big]^{\frac{5}{2}}}
\bigg\{\big[(m_u-m_c)\big(2s_1\Delta_2-\Delta_1u\big) +(m_b-m_c)\big(2s_2\Delta_1-\Delta_2 u - \nonumber \\
&& -2m_c^2(u + 2 s_2)\big) +2 m_c (2s_2\Delta_1-\Delta_2u) \big] \lambda(s_1,s_2,q^2) 
+ \nonumber \\
&& + 2(m_b-m_c)\big[\Delta_1\Delta_2(2u^2 + 4 s_1 s_2 + 6s_2u) -3u(s_1 \Delta_2^2 + s_2\Delta_1^2) - 6s_2^2\Delta_1^2 - 
\nonumber \\
&& - 2 s_1 s_2\Delta_2^2 - \Delta_2^2u^2\big] - m_c\big[\lambda(s_1,s_2,q^2)\big]^2 \bigg\}\,,  \nonumber \\
\rho^V &=& \frac{3}{\big[\lambda(s_1,s_2,q^2)\big]^{\frac{3}{2}}}\big\{
(m_c-m_b)(2s_2\Delta_1-\Delta_2u)+ (m_c-m_u)(2s_1\Delta_2-\Delta_2 u) \nonumber \\
&&+ m_c \lambda(s_1,s_2,q^2)\big\}\,.
\label{eq:appA2}
\end{eqnarray}
Above it was introduced	
\begin{eqnarray*}
\Delta_1 &\equiv& s_1 + m_c^2 - m_b^2\,,\\
\Delta_2 & \equiv& s_2 + m_c^2 - m_u^2\,, \\
u & \equiv& s_1 + s_2 - q^2\,.
\end{eqnarray*}
\subsection{Non-local quark-condensate contributions }
We write our results for the nonlocal quark condensate in terms of the integrals
\begin{equation}
    \label{eq:ints0}
    \begin{split}
    I_0^\prime(p_1^2,p_2^2,q^2;a,b) & =\!\int\!\frac{\dd ^4 k}{(2\pi)^4}\frac{1}{[k^2-m_c^2]^a[(k+p_1)^2-m_b^2]^b}\mathrm{e}^{A(k+p_2)^2}\\
                                    & = \tilde{I}_0(p_1^2,p_2^2,q^2;a,b)\,,\\
    I_1^{\prime\mu}(p_1^2,p_2^2,q^2;a,b) & =\!\int\!\frac{\dd ^4 k}{(2\pi)^4}\frac{k^{\mu}}{[k^2-m_c^2]^a[(k+p_1)^2-m_b^2]^b}\mathrm{e}^{A(k+p_2)^2}\,\\
                                        & = \tilde{I}_1(p_1^2,p_2^2,q^2;a,b)\,p_1^{\mu} + \tilde{I}_2(p_1^2,p_2^2,q^2;a,b)\,p_2^{\mu}\,,\\
    I_2^{\prime\mu\nu}(p_1^2,p_2^2,q^2;a,b) & =\!\int\!\frac{\dd ^4 k}{(2\pi)^4}\frac{k^{\mu}k^{\nu}}{[k^2-m_c^2]^a[(k+p_1)^2-m_b^2]^b}\mathrm{e}^{A(k+p_2)^2}\\
                                        & = \tilde{I}_{00}(p_1^2,p_2^2,q^2;a,b)g^{\mu\nu} + \tilde{I}_{11}(p_1^2,p_2^2,q^2;a,b)p_1^{\mu}p_1^{\nu}\\
                                        & + \tilde{I}_{12}(p_1^2,p_2^2,q^2;a,b)p_1^{\mu}p_2^{\nu} + \tilde{I}_{21}(p_1^2,p_2^2,q^2;a,b)p_2^{\mu}p_1^{\nu}\\
                                        & + \tilde{I}_{22}(p_1^2,p_2^2,q^2;a,b)p_2^{\mu}p_2^{\nu}\,,
    \end{split}
\end{equation}
where, for brevity $A=4/m_0^2$.  After symbolically denoting the operation of Borel transformation of independent tensor structures by the letter $\mathcal{B}$, we can write
\begin{equation}
    \begin{split}
        \mathcal{B}_{-p_{1}^2}(M_1^2)\mathcal{B}_{-p_{2}^2}(M_2^2) \tilde{I}_{[\mathrm{in}]}(p_1^2,p_2^2,q^2;a,b) & =I_{[\mathrm{in}]}(M_1^2,M_2^2,q^2;a,b)\,,
    \end{split}
\end{equation}
where "[in]" stands for any of the indices from Eq.~(\ref{eq:ints0}), and
\begin{equation}
    \label{eq:ints}
    \begin{split}
        I_0(M_1^2,\, & M_2^2,q^2;a,b) \\
        = & \frac{(-1)^{a+b}}{(a-1)!(b-1)!}\frac{\ii}{(4\pi)^2}\frac{1}{M_1^2}\bigg(\frac{A(M_1^2+M_2^2)}{M_1^2(AM_2^2-1)}\bigg)^{a-2}\bigg(A\frac{M_2^2}{M_1^2}\bigg)^{b-1}\mathcal{F}(M_1^2,M_2^2;q^2),\\
        I_1(M_1^2,\, & M_2^2,q^2;a,b) = \frac{M_2^2}{M_1^2} I_2(M_1^2, M_2^2,q^2;a,b)\\
        = & \frac{(-1)^{a+b+1}}{(a-1)!(b-1)!}\frac{\ii}{(4\pi)^2}\frac{1}{M_1^4}\bigg(\frac{A(M_1^2+M_2^2)}{M_1^2(AM_2^2-1)}\bigg)^{a-3}\bigg(A\frac{M_2^2}{M_1^2}\bigg)^{b-1}\mathcal{F}(M_1^2,M_2^2;q^2),\\
        I_{00}(M_1^2,\, & M_2^2,q^2;a,b) \\
        = & \frac{(-1)^{a+b+1}}{(a-1)!(b-1)!}\frac{\ii}{(4\pi)^2}\frac{1}{2M_1^4}\bigg(\frac{A(M_1^2+M_2^2)}{M_1^2(AM_2^2-1)}\bigg)^{a-3}\bigg(A\frac{M_2^2}{M_1^2}\bigg)^{b-2}\mathcal{F}(M_1^2,M_2^2;q^2),\\
        I_{11}(M_1^2,\, & M_2^2,q^2;a,b) \\
        = & \frac{(-1)^{a+b}}{(a-1)!(b-1)!}\frac{\ii}{(4\pi)^2}\frac{1}{M_1^6}\bigg(\frac{A(M_1^2+M_2^2)}{M_1^2(AM_2^2-1)}\bigg)^{a-2}\bigg(A\frac{M_2^2}{M_1^2}\bigg)^{b-3}\mathcal{F}(M_1^2,M_2^2;q^2),\\
        I_{12}(M_1^2,\, & M_2^2,q^2;a,b) = I_{21}(M_1^2,M_2^2,q^2;a,b)\\
        = & \frac{(-1)^{a+b}}{(a-1)!(b-1)!}\frac{\ii}{(4\pi)^2}\frac{1}{M_1^4 M_2^2}\bigg(\frac{A(M_1^2+M_2^2)}{M_1^2(AM_2^2-1)}\bigg)^{a-3}\bigg(A\frac{M_2^2}{M_1^2}\bigg)^{b-2}\mathcal{F}(M_1^2,M_2^2;q^2),\\
        I_{22}(M_1^2,\, & M_2^2,q^2;a,b) \\
        = & \frac{(-1)^{a+b}}{(a-1)!(b-1)!}\frac{\ii}{(4\pi)^2}\frac{1}{M_1^2M_2^4}\bigg(\frac{A(M_1^2+M_2^2)}{M_1^2(AM_2^2-1)}\bigg)^{a-4}\bigg(A\frac{M_2^2}{M_1^2}\bigg)^{b-1}\mathcal{F}(M_1^2,M_2^2;q^2)\\
    \end{split}
\end{equation}
and
\begin{equation}
    \mathcal{F}(M_1^2,M_2^2;q^2)\equiv\frac{1}{AM_2^2-1} \exp[-A\frac{M_1^2+M_2^2}{M_1^2(AM_2^2-1)}m_c^2-A\frac{M_2^2}{M_1^2}m_b^2+\frac{AM_2^2-1}{M_1^2+M_2^2}q^2].
\end{equation}
From above expressions one can deduce the nonlocal quark-condensate contribution to a particular form factor given in the following subsections.

\subsubsection{$B_c\to P$ transition}
The quark condensate contribution to the $B_c\to P$ correlation function is
\begin{equation}
\Pi^{(3)\mu}_P(p_{1},p_2)  = -\frac{3\,i^4}{12}\int\!\!\!\int \dd^4x\,\dd^4y\,\mathrm{e}^{-i(p_1 x-p_2y)} \langle \bar{u}_a(0) u_a(y) \rangle \mathrm{Tr}
\big[S_c(x,y)\gamma_5\gamma^{\mu}(1-\gamma_5)S_b(0,x)\gamma_5 \big],
\end{equation}
where the color trace has been taken. By expanding the $\bar{q}q$ operator one gets
\begin{equation}
    \langle \bar{q}(0)q(y) \rangle \approx \langle \bar{q}q \rangle -g\frac{y^2}{16} \langle \bar{q}\sigma^{\mu\nu}G_{\mu\nu}q \rangle +\dots,
\end{equation}
which, in order to model the nonlocal effects is then substituted with
\begin{equation}
    \langle \bar{q}(0)\exp \Bigg \{i g\int\displaylimits_0^{\infty}dy_{\mu}A^{\mu}(y) \Bigg\} q(y) \rangle  = \langle \bar{q}q \rangle f(y^2).
\end{equation}
After the Fourier transforming of propagators and evaluating the trace we obtain
\begin{equation}
    \Pi^{(3)\mu}_P(p_1,p_2)  = \langle \bar{q}q \rangle \int\!\frac{d^4k}{(2\pi)^4}\frac{(m_b-m_c)k^{\mu}-m_c p_1^{\mu}}{(k^2-m_c^2)((k+p_1)^2-m_b^2)}\tilde{f}(k+p_2), 
\end{equation}
where $\tilde{f}(k+p_2)$ is the Fourier transform of the chosen model function in coordinate space $f(y^2)$. Then it's easy to express the Borel-transformed contribution to the form factors as
\begin{equation}
    \begin{split}
    \mathcal{B}_{-p_1^2}&(M_1^2)\mathcal{B}_{-p_2^2}(M_2^2)\Pi_{P,1}^{(3)}(q^2) = \\
    & 4\pi^2\,i\expval{\bar{q}q} \bigg(\frac{4}{m_0^2}\bigg)^2\bigg[m_c I_0(M_1^2, M_2^2,q^2;1,1) + (m_c-m_b)I_1(M_1^2, M_2^2,q^2;1,1)\bigg],\\
    \mathcal{B}_{-p_1^2}&(M_1^2)\mathcal{B}_{-p_2^2}(M_2^2)\Pi_{P,2}^{(3)}(q^2) = \\
    & 4\pi^2\,i\expval{\bar{q}q} \bigg(\frac{4}{m_0^2}\bigg)^2\bigg[(m_c-m_b)I_2(M_1^2, M_2^2,q^2;1,1)\bigg].\\
    \end{split}
\end{equation}
The mixed quark-gluon condensate contribution amounts to
\begin{equation}
    \begin{split}
        & \Pi^{(5)\mu}_P(p_{1},p_2)  = - i^4\frac{g}{2\cdot 192}\int\!\!\!\int \dd^4x\,\dd^4y\,\dd^4z\,\mathrm{e}^{-i(p_1 x-p_2y)} z^{\alpha}\\
        &  \langle \bar{u}(0) (\sigma\!\cdot\! G)^{\vphantom{y}} u(y) \rangle \mathrm{Tr}_c \big[t^c t^c \big ] \bigg( \mathrm{Tr}\big[S_c(x,z)\gamma^{\beta}S_c(z,y)\gamma_5 \sigma_{\alpha\beta}\gamma^{\mu}(1-\gamma_5)S_b(0,x)\gamma_5\big]\\
        & \qquad\qquad\qquad\qquad\quad\,\,\,\,\,+ \mathrm{Tr}\big[S_c(x,y)\gamma_5\sigma_{\alpha\beta}\gamma^{\mu}(1-\gamma_5)S_b(0,z)\gamma^{\beta}S_b(z,x)\gamma_5\big]\bigg).
    \end{split}
\end{equation}
The quark-gluon condensate can be approximated in terms of the quark condensate as \cite{Va78}
\begin{equation}
g \langle\bar{q}(0)(\sigma\!\cdot\! G)q(y) \rangle \approx m_0^2 
\langle \bar{q}(0)q(y)\rangle .
\end{equation}
After the Fourier transformation the amplitude becomes
\begin{equation}
\begin{split}
& \,\Pi^{(5)\mu}_P  (p_{1},p_2) =i m_0^2\frac{\langle \bar{q}q \rangle}{96} \int\!\frac{d^4k}{(2\pi)^4}\frac{\partial}{\partial q_{\alpha}}\\
& \,\bigg\{\mathrm{Tr}\big[\frac{\slashed{k}+\slashed{q}+m_c}{(k+q)^2-m_c^2}\gamma^{\beta}\frac{\slashed{k}+m_c}{k^2-m_c^2}\gamma_5\sigma_{\alpha\beta}\gamma^{\mu}(1-\gamma_5)\frac{\slashed{k}+\slashed{q}+\slashed{p}_1+m_b}{(k+q+p_1)^2-m_b^2}\gamma_5\big]\\
&\,+\mathrm{Tr}\big[\frac{\slashed{k}+\slashed{q}+\slashed{p}_1+m_b}{(k+q+p_1)^2-m_b^2}\gamma^{\beta}\frac{\slashed{k}+\slashed{p}_1+m_b}{(k+p_1)^2-m_b^2}\gamma_5\frac{\slashed{k}+m_c}{k^2-m_c^2}\gamma_5\sigma_{\alpha\beta}\gamma^{\mu}(1-\gamma_5)\big]\bigg\}_{\!q=0}\!\!\cdot \tilde{f}(k+p_2), \\		
\end{split}
\end{equation}
or, after differentiating
\begin{equation}
\begin{split}
& \,\Pi^{(5)\mu}_P  (p_{1},p_2) =-i\,m_0^2\frac{\langle \bar{q}q \rangle}{96} \int\!\frac{d^4k}{(2\pi)^4}\tilde{f}(k+p_2)\\
& \,\bigg\{\mathrm{Tr}\big[\frac{\slashed{k}+m_c}{k^2-m_c^2}\gamma^\alpha\frac{\slashed{k}+m_c}{k^2-m_c^2}\gamma^{\beta}\frac{\slashed{k}+m_c}{k^2-m_c^2}\gamma_5\sigma_{\alpha\beta}\gamma^{\mu}(1-\gamma_5)\frac{\slashed{k}+\slashed{p}_1+m_b}{(k+p_1)^2-m_b^2}\gamma_5\big]\\
& \,+\mathrm{Tr}\big[\frac{\slashed{k}+m_c}{k^2-m_c^2}\gamma^{\beta}\frac{\slashed{k}+m_c}{k^2-m_c^2}\gamma_5\sigma_{\alpha\beta}\gamma^{\mu}(1-\gamma_5)\frac{\slashed{k}+\slashed{p}_1+m_b}{(k+p_1)^2-m_b^2}\gamma^\alpha\frac{\slashed{k}+\slashed{p}_1+m_b}{(k+p_1)^2-m_b^2}\gamma_5\big]\\
&\,+\mathrm{Tr}\big[\frac{\slashed{k}+\slashed{p}_1+m_b}{(k+p_1)^2-m_b^2}\gamma^\alpha\frac{\slashed{k}+\slashed{p}_1+m_b}{(k+p_1)^2-m_b^2}\gamma^{\beta}\frac{\slashed{k}+\slashed{p}_1+m_b}{(k+p_1)^2-m_b^2}\gamma_5\frac{\slashed{k}+m_c}{k^2-m_c^2}\gamma_5\sigma_{\alpha\beta}\gamma^{\mu}(1-\gamma_5)\big]\bigg\} . \\		
\end{split}
\end{equation}
The integrals are done with complete analogy to the previous case, with one difference. Now, due to the differentiation, additional powers of squares of external momenta are present in the trace, so one needs to Borel-transform according to the rule 
\begin{equation}
    \begin{split}
        I_{[\mathrm{in}]}^{(m)}(M_1^2,M_2^2,q^2;a,b)&\equiv\mathcal{B}_{-p^2}(M^2)\big[(-p^2)^m \tilde{I}_{[\mathrm{in}]}(p_1^2,p_2^2,q^2;a,b)\big]\\
        & =(M^2)^m\bigg(\frac{\partial}{\partial M^2}\bigg)^{\!m}[(M^2)^m \mathcal{B}_{-p^2}(M^2)\,I_{[\mathrm{in}]}(M_1^2,M_2^2,q^2;a,b)],\\
    \end{split}
    \label{eq:pBorel}
\end{equation}
where again, "[in]" stands for any of the indices of the integrals in Eq.~(\ref{eq:ints0}), so that finally, for the Borel-transformed quark-gluon contribution to the correlation function we have 
\begin{equation}
    \begin{split}
    \mathcal{B}_{-p_1^2}&(M_1^2)\mathcal{B}_{-p_2^2}(M_2^2)\Pi_{P,1}^{(5)}(q^2) = \\
    & \frac{\ii\pi^2}{6}\expval{\bar{q}q} \bigg(\frac{4}{m_0^2}\bigg)\bigg\{-40m_cI_0(M_1^2, M_2^2,q^2;2,1) + 32(m_b-m_c)I_1(M_1^2, M_2^2,q^2;2,1)\\
    & +8(m_b-2m_c)I_0(M_1^2, M_2^2,q^2;1,2)+16(m_b-m_c)I_1(M_1^2, M_2^2,q^2;1,2)\\
    & +8m_c\big(m_b-m_c\big)^2I_0(M_1^2, M_2^2,q^2;2,2)\\
    & +8\big[(m_b+m_c)^3-4m_bm_c(m_b+m_c)\big]I_1(M_1^2, M_2^2,q^2;2,2)\\
    & -8m_cI_0^{(1)}(M_1^2, M_2^2,q^2;2,2)-8(m_b+m_c)I_1^{(1)}(M_1^2, M_2^2,q^2;2,2)\bigg\},\\
    \mathcal{B}_{-p_1^2}&(M_1^2)\mathcal{B}_{-p_2^2}(M_2^2)\Pi_{P,2}^{(5)}(q^2) = \\
    & \frac{\ii\pi^2}{6}\expval{\bar{q}q} \bigg(\frac{4}{m_0^2}\bigg)\bigg\{32(m_b-m_c)I_2(M_1^2, M_2^2,q^2;2,1)+16(m_b-m_c)I_2(M_1^2, M_2^2,q^2;1,2)\\
    & +8\big[(m_b+m_c)^3-4m_bm_c(m_b+m_c)\big]I_2(M_1^2, M_2^2,q^2;2,2)\\
    & -8(m_b+m_c)I_2^{(1)}(M_1^2, M_2^2,q^2;2,2)\bigg\}.\\
    \end{split}
\end{equation}
\subsubsection{$B_c\to V$ transition}
The quark condensate contribution to the $B_c\to V$ correlation function is
\begin{equation}
\Pi^{(3)\mu\nu}_V(p_{1},p_2)  = \frac{3\,i^5}{12}\int\!\!\!\int \dd^4x\,\dd^4y\,\mathrm{e}^{-i(p_1 x-p_2y)} \langle \bar{u}_a(0) u_a(y) \rangle \mathrm{Tr}
\big[S_c(x,y)\gamma^{\nu}\gamma^{\mu}(1-\gamma_5)S_b(0,x)\gamma_5 \big],
\end{equation}
so that, after Fourier transforming the propagators, and evaluating the trace we obtain
\begin{equation}
    \begin{split}
    &\Pi^{(3)\mu\nu}_V(p_1,p_2)  = \\
    & \quad i\langle \bar{q}q \rangle \int\!\frac{d^4k}{(2\pi)^4}\frac{\big(k^2+(k\cdot p_1)-m_bm_c\big)g^{\mu\nu}+k^{\nu}p_1^{\mu}-p_1^{\nu}k^{\mu}+\ii \epsilon^{\mu\nu\alpha\beta}k_{\alpha}p_{1\beta}}{(k^2-m_c^2)((k+p_1)^2-m_b^2)}\tilde{f}(k+p_2).
    \end{split}
\end{equation}
Then it's easy to express the Borel-transformed contribution to the form factors as
\begin{equation*}
    \begin{split}
    \mathcal{B}_{-p_1^2}&(M_1^2)\mathcal{B}_{-p_2^2}(M_2^2)\Pi_{V,0}^{(3)}(q^2) = \\
    & 4\pi^2\,i\expval{\bar{q}q} \bigg(\frac{4}{m_0^2}\bigg)^2\bigg[m_b(m_b-m_c) I_0(M_1^2, M_2^2,q^2;1,1) - I_0^{(1)}(M_1^2, M_2^2,q^2;1,1)\bigg],\\
    \mathcal{B}_{-p_1^2}&(M_1^2)\mathcal{B}_{-p_2^2}(M_2^2)\Pi_{V,1}^{(3)}(q^2) = -4\pi^2\,i\expval{\bar{q}q} \bigg(\frac{4}{m_0^2}\bigg)^2I_2(M_1^2, M_2^2,q^2;1,1),\\
    \mathcal{B}_{-p_1^2}&(M_1^2)\mathcal{B}_{-p_2^2}(M_2^2)\Pi_{V,2}^{(3)}(q^2) = 0,\\
    \mathcal{B}_{-p_1^2}&(M_1^2)\mathcal{B}_{-p_2^2}(M_2^2)\Pi_{V,v}^{(3)}(q^2) = 4\pi^2\,i\expval{\bar{q}q} \bigg(\frac{4}{m_0^2}\bigg)^2I_2(M_1^2, M_2^2,q^2;1,1).\\
    \end{split}
\end{equation*}
The mixed quark-gluon condensate contribution amounts to
\begin{equation}
    \begin{split}
        & \Pi^{(5)\mu\nu}_V(p_{1},p_2)  = i^5\frac{g}{2\cdot 192}\int\!\!\!\int \dd^4x\,\dd^4y\,\dd^4z\,\mathrm{e}^{-i(p_1 x-p_2y)} z^{\alpha}\\
        &  \langle \bar{u}(0) (\sigma\!\cdot\! G)^{\vphantom{y}} u(y) \rangle \mathrm{Tr}_c \big[t^c t^c \big ] \bigg( \mathrm{Tr}\big[S_c(x,z)\gamma^{\beta}S_c(z,y)\gamma^{\nu} \sigma_{\alpha\beta}\gamma^{\mu}(1-\gamma_5)S_b(0,x)\gamma_5\big]\\
        & \qquad\qquad\qquad\qquad\qquad\,\,+ \mathrm{Tr}\big[S_c(x,y)\gamma^\nu\sigma_{\alpha\beta}\gamma^{\mu}(1-\gamma_5)S_b(0,z)\gamma^{\beta}S_b(z,x)\gamma_5\big]\bigg).
    \end{split}
\end{equation}
Or, completely analogously to the previous case, by differentiating we get
\begin{equation}
\begin{split}
& \,\Pi^{(5)\mu\nu}_V  (p_{1},p_2) =-\,m_0^2\frac{\langle \bar{q}q \rangle}{96} \int\!\frac{d^4k}{(2\pi)^4}\tilde{f}(k+p_2)\\
& \,\bigg\{\mathrm{Tr}\big[\frac{\slashed{k}+m_c}{k^2-m_c^2}\gamma^\alpha\frac{\slashed{k}+m_c}{k^2-m_c^2}\gamma^{\beta}\frac{\slashed{k}+m_c}{k^2-m_c^2}\gamma^{\nu}\sigma_{\alpha\beta}\gamma^{\mu}(1-\gamma_5)\frac{\slashed{k}+\slashed{p}_1+m_b}{(k+p_1)^2-m_b^2}\gamma_5\big]\\
& \,+\mathrm{Tr}\big[\frac{\slashed{k}+m_c}{k^2-m_c^2}\gamma^{\beta}\frac{\slashed{k}+m_c}{k^2-m_c^2}\gamma^{\nu}\sigma_{\alpha\beta}\gamma^{\mu}(1-\gamma_5)\frac{\slashed{k}+\slashed{p}_1+m_b}{(k+p_1)^2-m_b^2}\gamma^\alpha\frac{\slashed{k}+\slashed{p}_1+m_b}{(k+p_1)^2-m_b^2}\gamma_5\big]\\
&\,+\mathrm{Tr}\big[\frac{\slashed{k}+\slashed{p}_1+m_b}{(k+p_1)^2-m_b^2}\gamma^\alpha\frac{\slashed{k}+\slashed{p}_1+m_b}{(k+p_1)^2-m_b^2}\gamma^{\beta}\frac{\slashed{k}+\slashed{p}_1+m_b}{(k+p_1)^2-m_b^2}\gamma_5\frac{\slashed{k}+m_c}{k^2-m_c^2}\gamma^\nu\sigma_{\alpha\beta}\gamma^{\mu}(1-\gamma_5)\big]\bigg\} . \\		
\end{split}
\end{equation}
The final contributions are then
\begin{equation*}
    \begin{split}
    \mathcal{B}_{-p_1^2}&(M_1^2)\mathcal{B}_{-p_2^2}(M_2^2)\Pi_{V,0}^{(5)}(q^2) = \\
    & \frac{4\ii\pi^2}{3}\expval{\bar{q}q} \bigg(\frac{4}{m_0^2}\bigg)\bigg\{I_0(M_1^2, M_2^2,q^2;1,1) + \frac{1}{2}(m_b^2-m_c^2)^2I_0(M_1^2, M_2^2,q^2;2,2)\\
    & -(m_b^2+m_c^2)I_0^{(1)}(M_1^2, M_2^2,q^2;2,2)+\frac{1}{2}I_0^{(2)}(M_1^2, M_2^2,q^2;2,2)\bigg\},\\
    \mathcal{B}_{-p_1^2}&(M_1^2)\mathcal{B}_{-p_2^2}(M_2^2)\Pi_{V,1}^{(5)}(q^2) = \\
    & \frac{4\ii\pi^2}{3}\expval{\bar{q}q} \bigg(\frac{4}{m_0^2}\bigg)\bigg\{\big[4m_c^2-(m_b+m_c)^2\big]I_2(M_1^2, M_2^2,q^2;2,2) + I_2^{(1)}(M_1^2, M_2^2,q^2;2,2)\\
    & +4I_{21}^{(1)}(M_1^2, M_2^2,q^2;2,2)\bigg\},\\
    \mathcal{B}_{-p_1^2}&(M_1^2)\mathcal{B}_{-p_2^2}(M_2^2)\Pi_{V,2}^{(5)}(q^2) = \\
    & \frac{16\ii\pi^2}{3}\expval{\bar{q}q} \bigg(\frac{4}{m_0^2}\bigg)\bigg\{\big(m_c^2-m_b^2\big)I_1(M_1^2, M_2^2,q^2;2,2) + I_1^{(1)}(M_1^2, M_2^2,q^2;2,2)\\
    & +I_{11}^{(1)}(M_1^2, M_2^2,q^2;2,2)\bigg\},\\
    \mathcal{B}_{-p_1^2}&(M_1^2)\mathcal{B}_{-p_2^2}(M_2^2)\Pi_{V,v}^{(5)}(q^2) = \\
    & \frac{4\ii\pi^2}{3}\expval{\bar{q}q} \bigg(\frac{4}{m_0^2}\bigg)\bigg\{\big(m_b-m_c\big)^2I_2(M_1^2, M_2^2,q^2;2,2) - I_2^{(1)}(M_1^2, M_2^2,q^2;2,2)\bigg\}.\\
    \end{split}
\end{equation*}
\subsection{Gluon condensate contributions}

Here we only briefly sketch the method of the calculation of gluon condensate contributions to the correlation functions (\ref{eq:corr1}) and (\ref{eq:corr2}).  

There are altogether 6 diagrams of the type shown in Figure \ref{fig:diags}d. 
The calculation is done in the Fock-Schwinger fixed-point gauge following the method of the excellent review \cite{Shifman}.  

In the process of evaluation of the diagrams we have encountered integrals of the type
\begin{eqnarray}
I_{\mu_1,\mu_2,...\mu_i}  = \int \frac{d^D k}{(2 \pi)^4} \frac{k_{\mu_1} k_{\mu_2}...k_{\mu_i}}{[k^2 -m_3^2]^a [(k+p_1)^2 - m_1^2]^b [(k+p_2)^2 - m_2^2]^c }
\end{eqnarray}
Although the integrals are finite, to simplify the calculation we have worked out the scalar integral $I_0$ integral in $D$-dimension so that
we can write for the main integrals \cite{Duplancic}
\begin{eqnarray}
I^{\mu}(4, \{a,b,c\}) &=& \int \frac{d^4 k}{(2 \pi)^4} \frac{k_{\mu}}{[k^2 -m_3^2]^a [(k+p_1)^2 - m_1^2]^b [(k+p_2)^2 - m_2^2]^c } \nonumber \\
&=& p_1^{\mu} \frac{\Gamma(b+1)}{\Gamma(b)} I_0(6, \{a,b+1,c\}) + p_2^{\mu} \frac{\Gamma(c+1)}{\Gamma(c)} I_0(6, \{a,b,c+1\}),
\end{eqnarray}
whereas
\begin{eqnarray}
I^{\mu\nu}(4, \{a,b,c\}) &=& \int \frac{d^4 k}{(2 \pi)^4} \frac{k_{\mu}k_{\nu}}{[k^2 -m_3^2]^a [(k+p_1)^2 - m_1^2]^b [(k+p_2)^2 - m_2^2]^c } \nonumber \\
&=& - \frac{1}{2} g^{\mu\nu} I_0 ( 6, \{ a,b,c\}) \nonumber \\
&& + p_1^{\mu} p_1^{\nu} \frac{\Gamma(b+2)}{\Gamma(b)} I_0(8, \{a,b+2,c\})  + p_2^{\mu} p_2^{\nu} \frac{\Gamma(c+2)}{\Gamma(c)} I_0(8, \{a,b,c+2 \}) 
\nonumber \\
&& + \left (  p_1^{\mu} p_2^{\nu} +  p_1^{\nu} p_1^{\mu}\right ) \frac{\Gamma(b+1)\Gamma(c+1)}{\Gamma(b)\Gamma(c)} I_0(8, \{a,b+1,c+1\}). 
\end{eqnarray}
To Borel-transform the integrals the following expression for $I_0(D, \{a,b,c\})$ is useful:
\begin{eqnarray}
I_0(D, \{a,b,c\}) &=& \frac{(-1)^{a+b+c} i }{(4\pi)^{D/2} \Gamma(a)\Gamma(b)\Gamma(c)} \int_0^{\infty} d\alpha 
d\beta d\gamma \frac{\alpha^{a-1} \beta^{b-1}\gamma^{c-1}}{(\alpha +\beta+\gamma)^{D/2}} e^{-\alpha m_3^2 -\beta m_1^2 -\gamma m_2^2} e^{\beta\gamma Q^2/(\alpha+\beta+\gamma)}
\nonumber \\
&& \cdot e^{-p_{1E}^2 \left (\beta - \frac{\beta(\beta + \gamma)}{\alpha+\beta+\gamma} \right )} e^{-p_{2E}^2 \left ( \gamma - \frac{\gamma(\beta + \gamma)}{\alpha+\beta+\gamma} \right )}
\end{eqnarray}
which can be then Borelized by applying 
\begin{eqnarray}
\mathcal{B}_{p_{iE}^2}&(M_i^2) e^{-p_{iE}^2 X_i} = \delta (1 - X_i M_i^2 ). 
\end{eqnarray}
All Borel-transformed integrals are then easily calculated by using the analogous expression of (\ref{eq:pBorel}) and 
\begin{eqnarray}
\mathcal{B}_{-p_{1}^2}(M_1^2)\mathcal{B}_{-p_{2}^2}(M_2^2) \, I_0 (D, \{a,b,c\}) && 
\nonumber \\
&& \hspace*{-6cm} = \frac{(-1)^{a+b+c} i }{(4\pi)^{D/2} \Gamma(a)\Gamma(b)\Gamma(c)} \left (M_1^2\right )^{D/2-a-b} \left (M_2^2\right )^{D/2 - a-c} 
\int_0^\infty dy \,( y + M_1^2 + M_2^2 )^{a+b+c-D} y^{D/2 -1 -b-c} 
\nonumber \\
&& \hspace*{-3cm} e^{- \frac{-Q^2}{y}} e^{-m_1^2 \left ( \frac{y + M_1^2 + M_2^2}{y M_1^2}  \right )-m_2^2 \left ( \frac{y + M_1^2 + M_2^2}{y M_2^2}  \right )-m_3^2 \left ( \frac{y + M_1^2 + M_2^2}{ M_1^2 M_2^2 }  \right )}, 
\end{eqnarray}
where for $B_c \to D^{(\ast)}$ transition one has to take $m_1 =m_b, m_2 = 0$ and $m_3 = m_c$. 

\section{QCDSR parameters in $B_c \to D^{(\ast)}$ and $B_c \to J/\psi, B_c \to B_s$ form factor calculations and discussion about their $q^2$ dependence}

\subsection{Parameters in the $B_c\to J/\psi,B_s$ 3ptSR calculations}
\begin{figure}[h!]
    \centering
    \includegraphics[width=0.47\textwidth]{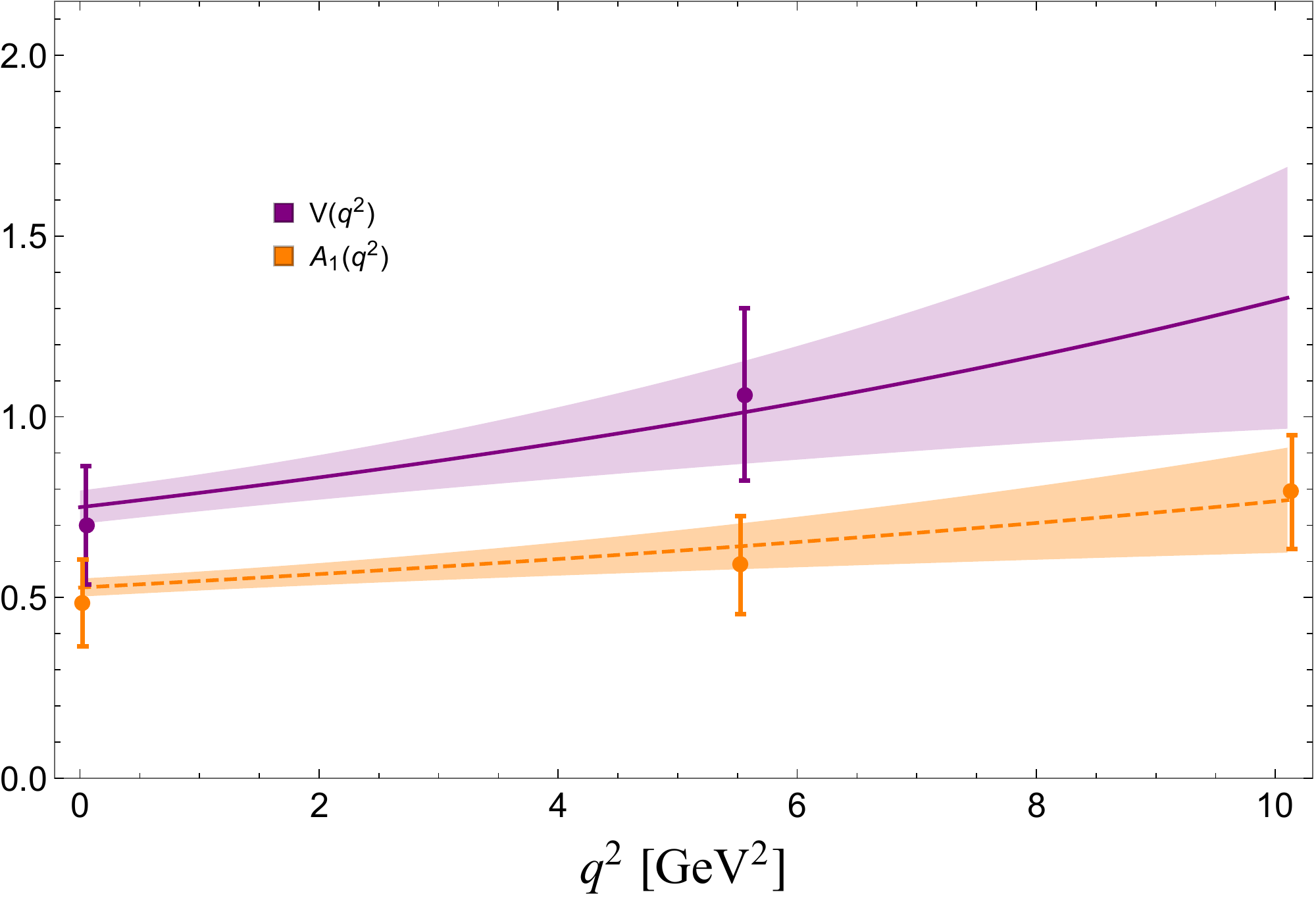}
        \hspace*{0.2cm}
    \includegraphics[width=0.47\textwidth]{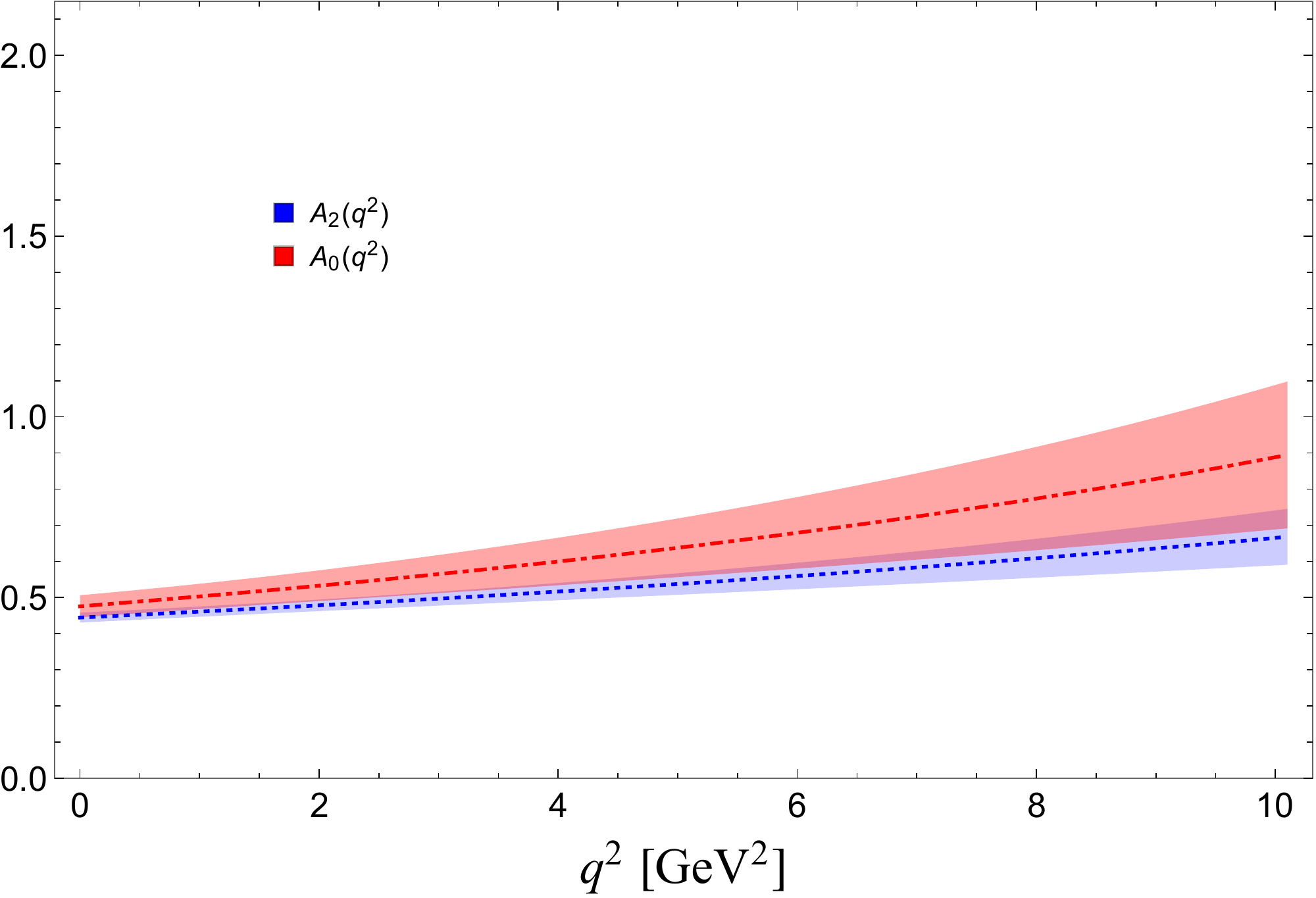}
    \caption{Final predictions for $B_c \to J/\psi$ form factors obtained by extrapolating 3ptSR results to higher $q^2$ regions using the BCL parametrization.}
    \label{fig:BcJpsiFF}
\end{figure}

\begin{figure}[t]
    \centering
    \includegraphics[width=0.47\textwidth]{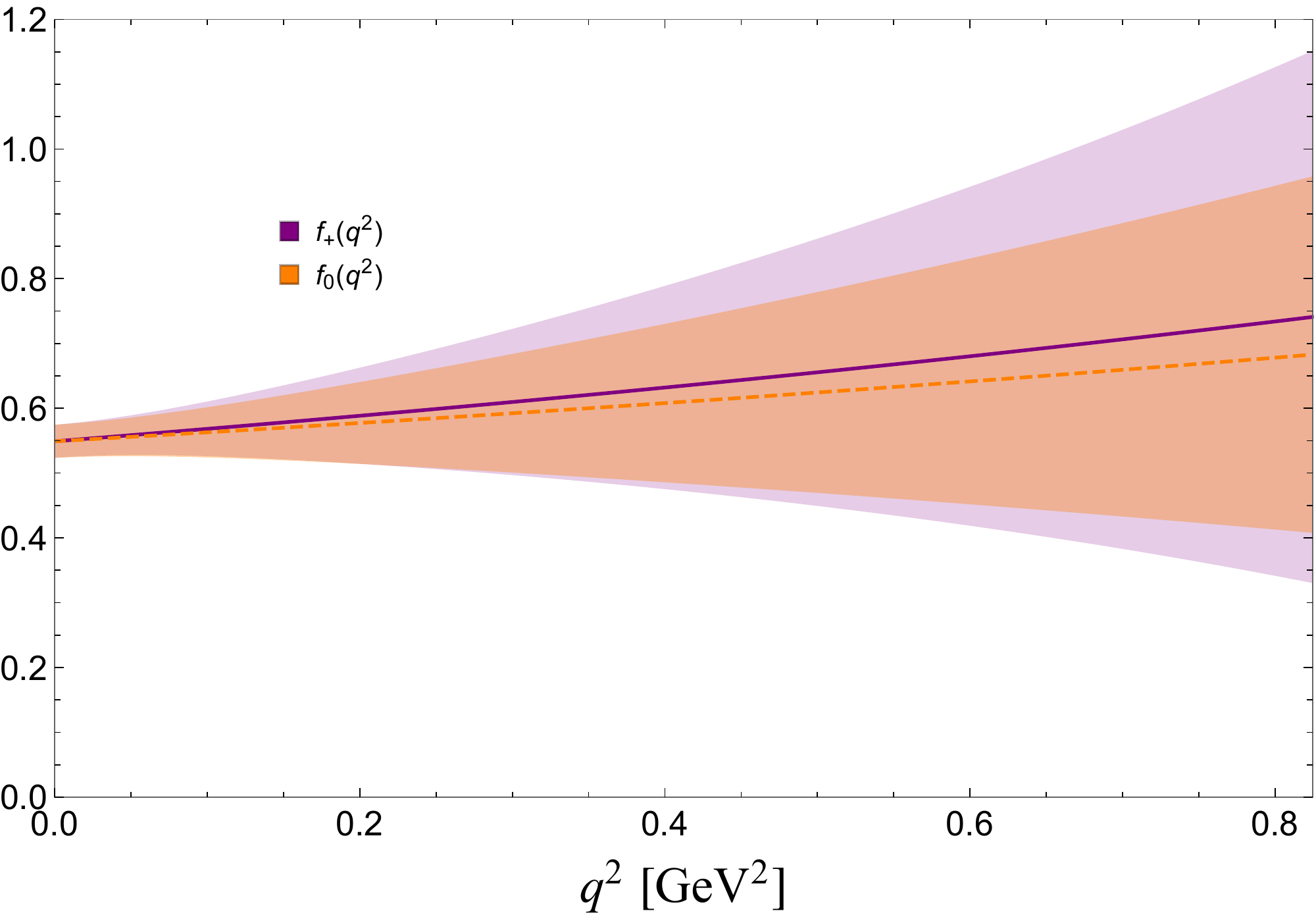}
    \caption{Final predictions for $B_c \to B_s$ form factors obtained by extrapolating 3ptSR results to higher $q^2$ regions using the BCL parametrization.}
    \label{fig:BcBsFF}
\end{figure}

When fitting the threshold parameters for the $B_c\to J/\psi$ and $B_c\to B_s$ transitions to those obtained in the calculation of $f_{J/\psi}$ and $f_{B_s}$ decay constants using the same criteria as for $B_c \to D^{(\ast)}$ explained in the text, in Sec.2, we obtain parameters listed in Table~\ref{tab:dconst2}. 
\begin{center}
\addtolength{\tabcolsep}{-3pt}
\renewcommand{\arraystretch}{1.5}
 \begin{tabular}{|| c | c c c c ||}
 \hline
 Meson & lattice [MeV] & our value [MeV] & $s_0^{\mathrm{eff}}$ [GeV$^2$] & $M^2$ [GeV$^2$]\\ [0.5ex]
 \hline\hline
 $f_{J/\psi}$ & 405$\pm$6~\cite{Donald:2012ga}, 399$\pm 6$~\cite{Bailas:2018car} & $394\pm 17$ & 16-17 & 10-15 \\
 \hline
 $f_{B_s}$ & 224$\pm$5~\cite{Colquhoun:2015oha}, 229$\pm$5 \cite{Bussone:2016iua} & 225$\pm$16 & 37.5-39.5 & 20-35\\
 \hline
\end{tabular}
\captionof{table}{Decay constants of mesons with 3ptSR parameters.}
\label{tab:dconst2} 
\end{center}
As for the Borel mass window in the three-point calculation, the approximate relation from~(\ref{eq:3pt2ptrel}) holds, and we have $M^2_{3\mathrm{pt},J/\psi}\approx 20-25$ GeV$^2$, and $M^2_{3\mathrm{pt},B_s}\approx 30-60$ GeV$^2$.
The plot of the form factors for the $B_c\to J/\psi$ transition obtained using the latter parameters is given in Figure~\ref{fig:BcJpsiFF} together with the lattice points given by the HPQCD Collaboration~\cite{Colquhoun:2016osw}. Once again excellent agreement can be seen between the lattice result and our form factors, but one can also notice that we also agree extraordinarily well in all of the form factors with the LFQM from~\cite{Wang:2008xt}.
We also include the plot of the $B_c\to B_s$ form factors, Figure \ref{fig:BcBsFF}, which show a large uncertainty appearing due to the inability of utilizing QCDSR deeper in the high $q^2$ region. One should also always keep in mind that these uncertainties do not include the truncation error, which is always of the order of 20-30\% in our calculations, since we extrapolate only linearly in $z(q^2)$. 
Finally, the Table~\ref{tabular:res2} contains the results of these two fits. The unitarity threshold used in each case is listed in the same table under the column "threshold". Note that when fitting the $B_c\to J/\psi$ form factors we exclude two of the poles appearing beneath the $B^*D$ threshold, since numerically their value is very close to the threshold itself. Namely these poles are $M(1P_1)\approx M(1P_1^\prime)\approx 7.14$ GeV, and are very close to $\sqrt{t_*}\approx 7.2$ GeV. This is done in order to keep the monotonic behaviour of $A_1(q^2)$ and $A_2(q^2)$ around $q^2_{\mathrm{max}}$, and it does not significantly alter their numerical value. A more nuanced discussion on the impact of near-threshold poles in pole fits one can find in e.g.~\cite{Ball:2004rg}. The poles for the $B_c\to J/\psi$ case are taken from~\cite{Eichten:1994gt}, whereas for the $B_c\to B_s$ case the needed pole masses are known from experiments~\cite{PDG}.
\begin{center}
    \addtolength{\tabcolsep}{-3pt}
    \renewcommand{\arraystretch}{1.5}
     \begin{tabular}{|| c c c || c c c | c ||}
     \hline
        $J^P$ & threshold & $m_R$  [GeV] & BCL: & $b_0$ & $b_1$ & $\chi^2[10^{-4}]$ \\ [0.5ex]
        \hline
        $1^-$ & $ BD $ & 6.34, 6.90, 7.01 & $V^{B_c\to J/\psi}$ & 0.69 & 2 & 23\\
        $1^+$ & $B^*D$ & 6.73, 6.74 & $A_1^{B_c\to J/\psi}$ & 0.50 & 1 & 45\\
        $1^+$ & $B^*D$ & 6.73, 6.74 & $A_2^{B_c\to J/\psi}$ & 0.42 & 1 & 102\\
        $0^-$ & $B^*D$ & 6.28, 6.84 & $A_0^{B_c\to J/\psi}$ & 0.50 & -1 & 13\\
        \hline
        \hline
        $1^-$ & $D_s \eta^\prime$ & 2.11, 2.71, 2.86 & $f_+^{B_c\to B_s}$ & 0.5 & 3 & 0.7\\
        $0^+$ & $D_s \eta^\prime$ & 2.138 & $f_0^{B_c\to B_s}$ & 0.6 & -1 & 0.3\\
        \hline
    \end{tabular}
        \captionof{table}{Summary of the fits for $B_c \to J/\psi$ and $B_c \to B_s$ form factors.}
    \label{tabular:res2}
\end{center}
\subsection{Covariance matrices}
For the $B_c\to D$ transition the form factors are related to each other at the maximum recoil point, so the fit is done simultaneously, and the error covariance matrix, defined as 
\begin{equation}
    \big(\mathcal{M}_{f_{+,0}}\big)_{ij} = \mathrm{cov}[\theta_i,\theta_j]
\end{equation}
with vectors in this case being $\vec{\theta} = (b_0^{f_+},b_1^{f_+},b_0^{f_0})^{T}$, is
\begin{equation}
    \mathcal{M}_{f_{+,0}^{B_c\to D}} = \begin{pmatrix*}[r]
        0.000531195	& -0.00492472 &	1.48099\cross 10^{-7} \\[6pt]
        -0.00492472	& 0.0536992 & -0.0000114013 \\[6pt]
        1.48099\cross 10^{-7} & -0.0000114013 & 0.000313175 \\[6pt]
    \end{pmatrix*}, 
\end{equation}
where the fourth expansion parameter has been fixed using the fact that $f_+(0) = f_0(0)$ as
\begin{equation}
    b^{f_0}_1 = \frac{b^{f_+}_0 - b^{f_0}_0}{z(0)} + b^{f_+}_1 \,.
\end{equation}
In the case of $B_c\to D^*$ transition the form factors aren't related so that the vectors entering the covariance look like $\vec{\theta} = (b_0^{\mathrm{F}},b_1^{\mathrm{F}})^{T}$, with $\mathrm{F} = \{ V, A_{\{1,2,0\}} \}$, and 
\begin{equation}
    \begin{split}
        \mathcal{M}_{V^{B_c\to D^*}} = \begin{pmatrix*}[r]
            0.00329643 & -0.0359087 \\[6pt]
            -0.0359087 & 0.909345 \\[6pt]
        \end{pmatrix*}, & \quad
        \mathcal{M}_{A_1^{B_c\to D^*}} = \begin{pmatrix*}[r]
            0.000366363	& -0.00330984 \\[6pt]
            -0.00330984 & 0.165033, \\[6pt]
        \end{pmatrix*} \\
        \mathcal{M}_{A_2^{B_c\to D^*}} = \begin{pmatrix*}[r]
            0.000161295 & -0.000675167 \\[6pt]
            -0.000675167 & 0.100573 \\[6pt]
        \end{pmatrix*}, & \qquad
        \mathcal{M}_{A_0^{B_c\to D^*}} = \begin{pmatrix*}[r]
            0.00134904	& -0.0145026 \\[6pt]
            -0.0145026 & 0.336261 \\[6pt]
        \end{pmatrix*}. \\
    \end{split}
\end{equation}

\subsection{$z$-series fits}
The outer functions required for the fit of the $B_c\to D^0$ form factors to the BGL parametrization in Eqs.(\ref{eq:fits},\ref{eq:fitsF}) are obtained from \cite{Boyd2} and are given as  
\begin{equation}
    \begin{split}
        \phi_+(z) & = \frac{8N_0^{\frac{5}{4}}}{m_{B_c}}\sqrt{\frac{1}{3\pi\chi^T(u)}}\frac{r^2\big[(\sqrt{N_0}-1)z+(\sqrt{N_0}+1)\big]^{\frac{3}{2}}(1+z)^2\sqrt{1-z}}{\big[(1+r)(1-z)+2\sqrt{N_0r}(1+z)\big]^5}\,,\\
        \phi_0(z) & = \frac{2N_0^{\frac{3}{4}}}{m_{B_c}^2}\sqrt{\frac{1}{\pi\chi^L(u)}}\frac{r\big[(\sqrt{N_0}-1)z+(\sqrt{N_0}+1)\big]^{\frac{1}{2}}\sqrt{1-z}(1-z^2)}{\big[(1+r)(1-z)+2\sqrt{N_0r}(1+z)\big]^4}\,,
    \end{split}
\end{equation}
where a useful mass ratio $r=m_{D^0}/m_{B_c}$ has been defined, and also $N_0=\frac{t_*-t_0}{t_*-t_-}$. Notice as well that in some analyses $t_0$ is chosen to be $t_-$, which would further simplify the expressions. For the $B_c\to D^*$ case, we have
\begin{equation}
    \begin{split}
        \phi_g(z) & = 4N_0^{\frac{5}{4}}\sqrt{\frac{2}{3\pi\chi^T(u)}}\frac{r^2\big[(\sqrt{N_0}-1)z+(\sqrt{N_0}+1)\big]^{\frac{3}{2}}(1+z)^2}{\sqrt{1-z}\big[(1+r)(1-z)+2\sqrt{N_0r}(1+z)\big]^4}\,,\\
        \phi_f(z) & = 2\frac{N_0^{\frac{3}{4}}}{m_{B_c}^2}\sqrt{\frac{2}{3\pi\chi^T(-u)}}\frac{r\big[(\sqrt{N_0}-1)z+(\sqrt{N_0}+1)\big]^{\frac{1}{2}}\sqrt{1-z}(1-z^2)}{\big[(1+r)(1-z)+2\sqrt{N_0r}(1+z)\big]^4}\,,\\
        \phi_1(z) & = 2\frac{N_0^{\frac{3}{4}}}{m_{B_c}^3}\sqrt{\frac{1}{3\pi\chi^T(-u)}}\frac{r\big[(\sqrt{N_0}-1)z+(\sqrt{N_0}+1)\big]^{\frac{1}{2}}(1+z)(1-z)^{\frac{5}{2}}}{\big[(1+r)(1-z)+2\sqrt{N_0r}(1+z)\big]^5}\,,\\
        \phi_2(z) & = 4N_0^{\frac{5}{4}}\sqrt{\frac{1}{\pi\chi^L(-u)}}\frac{r^2\big[(\sqrt{N_0}-1)z+(\sqrt{N_0}+1)\big]^{\frac{3}{2}}(1+z)^2}{\sqrt{1-z}\big[(1+r)(1-z)+2\sqrt{N_0r}(1+z)\big]^4}\,,\\
    \end{split}
\end{equation}
where now the mass ratio is $r=m_{D^*}/m_{B_c}$. The functions $\chi_{T,L}(\pm u)$ are calculated perturbatively in QCD, and 
\begin{equation}
    \begin{split}
        m_b^2\big[\chi^T_{\mathrm{pert}}(u)\big]^{\mathrm{LO}} = \frac{1}{32\pi^2(1-u^2)^5}&\big[(1-u^2)(3+4u-21u^2+40u^3-21u^4+4u^5+3u^6)\\
        & +12u^3(2-3u+2u^2)\ln{u^2}\big],
    \end{split}
\end{equation}
\begin{equation}
    \begin{split}
        m_b^2\big[\chi^T_{\mathrm{pert}}(u)&\big]^{\mathrm{NLO}} =  \frac{\alpha_s}{576\pi^3(1-u^2)^6}\\ &\big[(1-u^2)^2(75+360u-1031u^2+1776u^3-1031u^4+360u^5+75u^6)\\
        & +4u(1-u^2)(18-99u+732u^2-1010u^3+732u^4-99u^5+18u^6)\ln{u^2}\\
        &+4u^3(108-324u+648u^2-456u^3+132u^4+59u^5-12u^6-9u^7)\ln^2{u^2}\\
        &+8(1-u^2)^3(9+12u-32u^2+12u^3+9u^4)\mathrm{Li}_2(1-u^2)
        \big],
    \end{split}
\end{equation}
\begin{equation}
    \begin{split}
        \big[\chi^L_{\mathrm{pert}}(u)\big]^{\mathrm{LO}} = \frac{1}{8\pi^2(1-u^2)^3}&\big[(1-u^2)(1+u+u^2)(1-4u+u^2)-6u^3\ln{u^2}\big]
    \end{split}
\end{equation}
\begin{equation}
    \begin{split}
        \big[\chi^L_{\mathrm{pert}}(u)&\big]^{\mathrm{NLO}} =  \frac{\alpha_s}{48\pi^3(1-u^2)^4}\\ &\big[(1-u^2)^2(1-36u-22u^2-36u^3+u^4)\\
        &-2u(1-u^2)(9+4u+66u^2+4u^3+9u^4)\ln{u^2}\\
        &-4u^3(9+18u^2-2u^3-3u^4+u^5)\ln^2{u^2}+8(1-u^2)^3(1-3u+u^2)\mathrm{Li}_2(1-u^2)
        \big],
    \end{split}
\end{equation}
where $\chi^{T,L}_{\mathrm{pert}}(u)=\big[\chi^{T,L}_{\mathrm{pert}}(u)\big]^{\mathrm{LO}}+\big[\chi^{T,L}_{\mathrm{pert}}(u)\big]^{\mathrm{NLO}}$. The non-perturbative corrections form condensates are 
\begin{equation}
    \begin{split}
        \chi^T_{\mathrm{cond}}(u) = & -\expval{\bar{q}q}\frac{2-3u+2u^2}{2m_b^5(1-u^2)^5}\\
        & -\expval{\frac{\alpha_s}{\pi}G^2}\frac{1}{24m_b^6(1-u^2)^7}\\
        &\cross\big[(1-u^2)(2-104u+148u^2-270u^3+145u^4-104u^5+5u^6-2u^7)\\
        &-12u\ln{u^2}(3-5u+17u^2-15u^3+17u^4-5u^5+3u^6)\big],
    \end{split}
\end{equation}
and 
\begin{equation}
    \begin{split}
        \chi^L_{\mathrm{cond}}(u) = & -\expval{\bar{q}q}\frac{1}{m_b^3(1-u^2)^3}\\
        & +\expval{\frac{\alpha_s}{\pi}G^2}\frac{1}{12m_b^4(1-u^2)^5}\big[(1-u^2)(1-21u+10u^2-20u^3+u^4-u^5)\\
        &-3u\ln{u^2}(3-2u+8u^2-2u^3+3u^4)\big].
    \end{split}
\end{equation}
\section{Differential $q^2$ distributions and $|V_{ub}|$ extraction}
\subsection{Differential $(q^2,\cos \theta_l)$ distributions of semileptonic $B_c \to D^{(*)}$ decays}
For the case of $B_c\to D^0$ the two-fold functions in  Eq.(\ref{eq:doubleGamma}) are
\begin{equation}
    \begin{split}
        a_{\theta_l}^D & = \frac{G_F^2 |V_{ub}|^2q^2}{256\pi^3m_{B_c}^3}\sqrt{\lambda(m_{B_c}^2,m_{D^*}^2,q^2)}\bigg(1-\frac{m_l^2}{q^2}\bigg)^2\big[|h_0(q^2)|^2+\frac{m_l^2}{q^2}|h_t(q^2)|^2\big],\\
        b_{\theta_l}^D & = \frac{G_F^2 |V_{ub}|^2q^2}{128\pi^3m_{B_c}^3}\sqrt{\lambda(m_{B_c}^2,m_{D^*}^2,q^2)}\bigg(1-\frac{m_l^2}{q^2}\bigg)^2\frac{m_l^2}{q^2}\mathrm{Re}\big[h_0(q^2)h_t^*(q^2)\big],\\
        c_{\theta_l}^D & = - \frac{G_F^2 |V_{ub}|^2q^2}{256\pi^3m_{B_c}^3}\sqrt{\lambda(m_{B_c}^2,m_{D^*}^2,q^2)}\bigg(1-\frac{m_l^2}{q^2}\bigg)^3|h_0(q^2)|^2\,,\\
    \end{split}
\end{equation}
whereas for the $B_c\to D^*$ we have
\begin{equation}
    \begin{split}
        a_{\theta_l}^{D^*} & = \frac{G_F^2 |V_{ub}|^2q^2}{512\pi^3m_{B_c}^3}\sqrt{\lambda(m_{B_c}^2,m_{D^*}^2,q^2)}\bigg(1-\frac{m_l^2}{q^2}\bigg)^2\\
        &\qquad\bigg[\big(|H_+(q^2)|^2+|H_-(q^2)|^2\big)\bigg(1+\frac{m_l^2}{q^2}\bigg)+2\big(|H_0(q^2)|^2+\frac{m_l^2}{q^2}|H_t(q^2)|^2\big)\bigg],\\
        b_{\theta_l}^{D^*} & = \frac{G_F^2 |V_{ub}|^2q^2}{256\pi^3m_{B_c}^3}\sqrt{\lambda(m_{B_c}^2,m_{D^*}^2,q^2)}\bigg(1-\frac{m_l^2}{q^2}\bigg)^2\\
        &\qquad\bigg[|H_-(q^2)|^2-|H_+(q^2)|^2+2\frac{m_l^2}{q^2}\mathrm{Re}\big[H_0(q^2)H_t^*(q^2)\big]\bigg],\\
        c_{\theta_l}^{D^*} & = \frac{G_F^2 |V_{ub}|^2q^2}{512\pi^3m_{B_c}^3}\sqrt{\lambda(m_{B_c}^2,m_{D^*}^2,q^2)}\bigg(1-\frac{m_l^2}{q^2}\bigg)^3\bigg[|H_+(q^2)|^2+|H_-(q^2)|^2-2|H_0(q^2)|^2\bigg].\\
    \end{split}
\end{equation}
\subsection{Binned $q^2$ distributions of $\Delta \zeta_{D^0}$ and ${\cal R}_{D^0 J/\psi}$} \label{binneddel}
Bins of $\Delta \zeta_{D^0}(q_1^2,q_2^2)$, Eq.(\ref{eq:zetaD0}),  which can be used together with future experimental data to determine $V_{ub}$ from $B_c\rightarrow D^0 \mu \bar{\nu}_{\mu}$ decays are given in Table~\ref{tab:deltabin}. One should keep in mind, though, that in this table we do not include the errors stemming from our estimate of the correlation between the pseudo-data points. In order to account for this one would need to add a further 10\% uncertainty in the first 4 bins (up to $8$ GeV$^2$), as discussed in Sec. \ref{sec:corr}.
\begin{center}
\addtolength{\tabcolsep}{-3pt}
\renewcommand{\arraystretch}{1.5}
    \begin{adjustbox}{center}
    \begin{tabular}{||c | c c c c c c c c c c||}
        \hline
        $[q_1^2-q_2^2]\,[$GeV$^2$] & [$m_{\mu}^2-2$] & [2-4] & [4-6] & [6-8] & [8-10] & [10-12] & [12-14] & [14-16] & [16-18] & [18-$q^2_{\mathrm{max}}$]\\
        \hline\hline
        $\Delta \zeta_{D^0}(q_1^2,q_2^2) \, [10^{-4}$ eV] & $2.2 $ & $2.4 $ & $2.5$ & $2.5$ & $2.6$ & $2.5$ & $2.3$ & $1.9$ & $1.1$ & $0.21$\\
        & $\pm 0.2$ & $\pm 0.2$ & $\pm 0.3$ & $\pm 0.3$ & $\pm 0.4$ & $\pm 0.5$ & $\pm 0.5$ & $\pm 0.5$ & $\pm 0.3$ & $\pm 0.06$\\
        \hline
\end{tabular}
\end{adjustbox}
\captionof{table}{Distribution of $\Delta \zeta_{D^0}(q_1^2,q_2^2)$ placed in bins spaced by $2$ GeV apart.}
\label{tab:deltabin} 
\end{center}
In Table~\ref{tab:RJDbin} one can find the ratio defined in Eq.~(\ref{eq:ratiodef}) calculated for equally spaced bins, denoted here as bin$_{[D^0]}$ for the bin values of the numerator integral, and bin$_{[J/\psi]}$ for the bin values of the denominator integral, in GeV$^2$. Again, here the form factors used stem from the uncorrelated fits.
\begin{center}
\addtolength{\tabcolsep}{-4pt}
\renewcommand{\arraystretch}{1.5}
    \begin{adjustbox}{center}
    \begin{tabular}{||c || c c c c c c c c c c||}
        \hline
        \diagbox{bin$_{[D^0]}$}{bin$_{[J/\psi]}$} & [$m_{\mu}^2$-1] & [1-2] & [2-3] & [3-4] & [4-5] & [5-6] & [6-7] & [7-8] & [8-9] & [9-$q^2_{\mathrm{max}}$]\\
        \hline\hline
        [$m_{\mu}^2$-2] & \begin{tabular}{@{}c@{}}0.26\\ $\pm$ 0.06 \end{tabular} & \begin{tabular}{@{}c@{}}0.21\\ $\pm$ 0.05 \end{tabular} & \begin{tabular}{@{}c@{}}0.19\\ $\pm$ 0.04 \end{tabular} & \begin{tabular}{@{}c@{}}0.17\\ $\pm$ 0.04 \end{tabular} & \begin{tabular}{@{}c@{}}0.15\\ $\pm$ 0.03 \end{tabular} & \begin{tabular}{@{}c@{}}0.15\\ $\pm$ 0.03 \end{tabular} & \begin{tabular}{@{}c@{}}0.14\\ $\pm$ 0.03 \end{tabular} & \begin{tabular}{@{}c@{}}0.15\\ $\pm$ 0.03 \end{tabular} & \begin{tabular}{@{}c@{}}0.17\\ $\pm$ 0.04 \end{tabular} & \begin{tabular}{@{}c@{}}0.24\\ $\pm$ 0.05 \end{tabular}\\
        \hline
        [2-4] & \begin{tabular}{@{}c@{}}0.28\\ $\pm$ 0.07 \end{tabular} & \begin{tabular}{@{}c@{}}0.23\\ $\pm$ 0.05 \end{tabular} & \begin{tabular}{@{}c@{}}0.20\\ $\pm$ 0.04 \end{tabular} & \begin{tabular}{@{}c@{}}0.18\\ $\pm$ 0.04 \end{tabular} & \begin{tabular}{@{}c@{}}0.16\\ $\pm$ 0.03 \end{tabular} & \begin{tabular}{@{}c@{}}0.16\\ $\pm$ 0.03 \end{tabular} & \begin{tabular}{@{}c@{}}0.15\\ $\pm$ 0.03 \end{tabular} & \begin{tabular}{@{}c@{}}0.16\\ $\pm$ 0.03 \end{tabular} & \begin{tabular}{@{}c@{}}0.18\\ $\pm$ 0.04 \end{tabular} & \begin{tabular}{@{}c@{}}0.26\\ $\pm$ 0.05 \end{tabular}\\
        \hline
        [4-6] & \begin{tabular}{@{}c@{}}0.29\\ $\pm$ 0.07 \end{tabular} & \begin{tabular}{@{}c@{}}0.24\\ $\pm$ 0.05 \end{tabular} & \begin{tabular}{@{}c@{}}0.21\\ $\pm$ 0.05 \end{tabular} & \begin{tabular}{@{}c@{}}0.19\\ $\pm$ 0.04 \end{tabular} & \begin{tabular}{@{}c@{}}0.17\\ $\pm$ 0.04 \end{tabular} & \begin{tabular}{@{}c@{}}0.16\\ $\pm$ 0.03 \end{tabular} & \begin{tabular}{@{}c@{}}0.16\\ $\pm$ 0.03 \end{tabular} & \begin{tabular}{@{}c@{}}0.16\\ $\pm$ 0.03 \end{tabular} & \begin{tabular}{@{}c@{}}0.19\\ $\pm$ 0.04 \end{tabular} & \begin{tabular}{@{}c@{}}0.27\\ $\pm$ 0.06 \end{tabular}\\
        \hline
        [6-8] & \begin{tabular}{@{}c@{}}0.30\\ $\pm$ 0.08 \end{tabular} & \begin{tabular}{@{}c@{}}0.24\\ $\pm$ 0.06 \end{tabular} & \begin{tabular}{@{}c@{}}0.21\\ $\pm$ 0.05 \end{tabular} & \begin{tabular}{@{}c@{}}0.19\\ $\pm$ 0.04 \end{tabular} & \begin{tabular}{@{}c@{}}0.18\\ $\pm$ 0.04 \end{tabular} & \begin{tabular}{@{}c@{}}0.17\\ $\pm$ 0.04 \end{tabular} & \begin{tabular}{@{}c@{}}0.16\\ $\pm$ 0.04 \end{tabular} & \begin{tabular}{@{}c@{}}0.17\\ $\pm$ 0.04 \end{tabular} & \begin{tabular}{@{}c@{}}0.19\\ $\pm$ 0.04 \end{tabular} & \begin{tabular}{@{}c@{}}0.28\\ $\pm$ 0.06 \end{tabular}\\
        \hline
        [8-10] & \begin{tabular}{@{}c@{}}0.30\\ $\pm$ 0.08 \end{tabular} & \begin{tabular}{@{}c@{}}0.24\\ $\pm$ 0.06 \end{tabular} & \begin{tabular}{@{}c@{}}0.21\\ $\pm$ 0.05 \end{tabular} & \begin{tabular}{@{}c@{}}0.19\\ $\pm$ 0.05 \end{tabular} & \begin{tabular}{@{}c@{}}0.18\\ $\pm$ 0.04 \end{tabular} & \begin{tabular}{@{}c@{}}0.17\\ $\pm$ 0.04 \end{tabular} & \begin{tabular}{@{}c@{}}0.17\\ $\pm$ 0.04 \end{tabular} & \begin{tabular}{@{}c@{}}0.17\\ $\pm$ 0.04 \end{tabular} & \begin{tabular}{@{}c@{}}0.19\\ $\pm$ 0.05 \end{tabular} & \begin{tabular}{@{}c@{}}0.28\\ $\pm$ 0.07 \end{tabular}\\
        \hline
        [10-12] & \begin{tabular}{@{}c@{}}0.30\\ $\pm$ 0.09 \end{tabular} & \begin{tabular}{@{}c@{}}0.24\\ $\pm$ 0.07 \end{tabular} & \begin{tabular}{@{}c@{}}0.21\\ $\pm$ 0.06 \end{tabular} & \begin{tabular}{@{}c@{}}0.19\\ $\pm$ 0.05 \end{tabular} & \begin{tabular}{@{}c@{}}0.17\\ $\pm$ 0.05 \end{tabular} & \begin{tabular}{@{}c@{}}0.16\\ $\pm$ 0.04 \end{tabular} & \begin{tabular}{@{}c@{}}0.16\\ $\pm$ 0.04 \end{tabular} & \begin{tabular}{@{}c@{}}0.17\\ $\pm$ 0.04 \end{tabular} & \begin{tabular}{@{}c@{}}0.19\\ $\pm$ 0.05 \end{tabular} & \begin{tabular}{@{}c@{}}0.27\\ $\pm$ 0.07 \end{tabular}\\
        \hline
        [12-14] & \begin{tabular}{@{}c@{}}0.27\\ $\pm$ 0.08 \end{tabular} & \begin{tabular}{@{}c@{}}0.22\\ $\pm$ 0.07 \end{tabular} & \begin{tabular}{@{}c@{}}0.19\\ $\pm$ 0.06 \end{tabular} & \begin{tabular}{@{}c@{}}0.17\\ $\pm$ 0.05 \end{tabular} & \begin{tabular}{@{}c@{}}0.16\\ $\pm$ 0.05 \end{tabular} & \begin{tabular}{@{}c@{}}0.15\\ $\pm$ 0.04 \end{tabular} & \begin{tabular}{@{}c@{}}0.15\\ $\pm$ 0.04 \end{tabular} & \begin{tabular}{@{}c@{}}0.15\\ $\pm$ 0.04 \end{tabular} & \begin{tabular}{@{}c@{}}0.17\\ $\pm$ 0.05 \end{tabular} & \begin{tabular}{@{}c@{}}0.25\\ $\pm$ 0.07 \end{tabular}\\
        \hline
        [14-16] & \begin{tabular}{@{}c@{}}0.22\\ $\pm$ 0.07 \end{tabular} & \begin{tabular}{@{}c@{}}0.18\\ $\pm$ 0.06 \end{tabular} & \begin{tabular}{@{}c@{}}0.16\\ $\pm$ 0.05 \end{tabular} & \begin{tabular}{@{}c@{}}0.14\\ $\pm$ 0.04 \end{tabular} & \begin{tabular}{@{}c@{}}0.13\\ $\pm$ 0.04 \end{tabular} & \begin{tabular}{@{}c@{}}0.12\\ $\pm$ 0.04 \end{tabular} & \begin{tabular}{@{}c@{}}0.12\\ $\pm$ 0.04 \end{tabular} & \begin{tabular}{@{}c@{}}0.12\\ $\pm$ 0.04 \end{tabular} & \begin{tabular}{@{}c@{}}0.14\\ $\pm$ 0.04 \end{tabular} & \begin{tabular}{@{}c@{}}0.20\\ $\pm$ 0.06 \end{tabular}\\
        \hline
        [16-18] & \begin{tabular}{@{}c@{}}0.13\\ $\pm$ 0.05 \end{tabular} & \begin{tabular}{@{}c@{}}0.11\\ $\pm$ 0.04 \end{tabular} & \begin{tabular}{@{}c@{}}0.10\\ $\pm$ 0.03 \end{tabular} & \begin{tabular}{@{}c@{}}0.09\\ $\pm$ 0.03 \end{tabular} & \begin{tabular}{@{}c@{}}0.08\\ $\pm$ 0.03 \end{tabular} & \begin{tabular}{@{}c@{}}0.08\\ $\pm$ 0.02 \end{tabular} & \begin{tabular}{@{}c@{}}0.07\\ $\pm$ 0.02 \end{tabular} & \begin{tabular}{@{}c@{}}0.08\\ $\pm$ 0.02 \end{tabular} & \begin{tabular}{@{}c@{}}0.09\\ $\pm$ 0.03 \end{tabular} & \begin{tabular}{@{}c@{}}0.12\\ $\pm$ 0.04 \end{tabular}\\
        \hline
        [18-$q^2_\mathrm{max}$] & \begin{tabular}{@{}c@{}}0.03\\ $\pm$ 0.01 \end{tabular} & \begin{tabular}{@{}c@{}}0.02\\ $\pm$ 0.01 \end{tabular} & \begin{tabular}{@{}c@{}}0.018\\ $\pm$ 0.006 \end{tabular} & \begin{tabular}{@{}c@{}}0.016\\ $\pm$ 0.006 \end{tabular} & \begin{tabular}{@{}c@{}}0.015\\ $\pm$ 0.005 \end{tabular} & \begin{tabular}{@{}c@{}}0.014\\ $\pm$ 0.005 \end{tabular} & \begin{tabular}{@{}c@{}}0.014\\ $\pm$ 0.005 \end{tabular} & \begin{tabular}{@{}c@{}}0.014\\ $\pm$ 0.005 \end{tabular} & \begin{tabular}{@{}c@{}}0.016\\ $\pm$ 0.006 \end{tabular} & \begin{tabular}{@{}c@{}}0.023\\ $\pm$ 0.008 \end{tabular}\\
        \hline
\end{tabular}
\end{adjustbox}
\captionof{table}{$\mathcal{R}_{D^0 J/\psi}$ calculated in different pairs of bins.}
\label{tab:RJDbin}
\end{center}
\acknowledgments
We thank members of the LHCb Collaboration Marta Calvi and Matthew William Kenzie for discussions on possibility to measure $B_c \to D^{(\ast)}$ semileptonic decays and for sharing with us the status of the LHCb analysis and the experimental constraints. D. L. would also like to thank Goran Duplan\v ci\'c and Danny van Dyk for their very helpful comments and fruitful discussions. 
This project has been supported by the European Union through the European regional Development Fund - the Competitiveness and Cohesion Operational Programme (KK.01.1.1.06).  B.M. would like to acknowledge the support of the Alexander von Humboldt foundation as well as the hospitality of the Institute for Theoretical Physics at Johannes Gutenberg University in Mainz, where this work has started. This research was also partially supported by the Munich Institute for Astro- and Particle Physics (MIAPP) of the DFG cluster of excellence "Origin and Structure of the Universe".
%

\end{document}